\documentclass[cernpreprint,USenglish]{na61doc}
\usepackage[utf8]{inputenc}
\usepackage[T1]{fontenc}
\usepackage{amsmath}
\usepackage{url}
\usepackage{multirow}
\usepackage{colortbl}
\usepackage[colorlinks=true,linkcolor=firebrick,citecolor=darkgreen,urlcolor=darkblue]{hyperref}
\usepackage{mathptmx} % - Times New Roman font (required by EPJC)
\usepackage{enumerate}
\usepackage{lineno}
\usepackage{xspace}

\definecolor{darkred}{rgb}{0.5,0,0}
\definecolor{darkblue}{rgb}{0,0,0.5}
\definecolor{firebrick}{rgb}{0.75,0.125,0.125}
\definecolor{darkgreen}{rgb}{0,0.5,0}

\renewcommand{\NASixtyOne}{\mbox{NA61/SHINE}\xspace}

\newcommand{\GeantThree}{{\scshape Geant3}\xspace}
\newcommand{\Epos}{{\scshape Epos}\xspace}
\newcommand{\EposLong}{{\scshape Epos1.99}\xspace}

\newcommand{\coordinate}[1]{{\fontfamily{lmss}\selectfont#1}}

\newcommand{\MeV}{\mbox{Me\kern-0.1em V}\xspace}
\newcommand{\GeV}{\mbox{Ge\kern-0.1em V}\xspace}
\newcommand{\A}{\textit{A}\xspace}
\newcommand{\GeVc}{\mbox{Ge\kern-0.1em V\kern-0.15em /\kern-0.05em\textit{c}}\xspace}
\newcommand{\MeVc}{\mbox{Me\kern-0.1em V\kern-0.15em /\kern-0.05em\textit{c}}\xspace}

\newcommand{\AGeVc}{\A~\GeVc}
\newcommand{\Ar}{\textsuperscript{40}Ar\xspace}
\newcommand{\Sc}{\textsuperscript{45}Sc\xspace}
\newcommand{\ArSc}{\mbox{\Ar~+~\Sc}\xspace}
\newcommand{\dedx}{\mbox{\ensuremath{\textrm{d}E\!/\!\textrm{d}x}}\xspace}
\newcommand{\sNN}{\ensuremath{\sqrt{s_{\mathrm{NN}}}}\xspace}
\newcommand{\AVG}[1]{\ensuremath{\langle #1\rangle}\xspace}
\newcommand{\chindf}{\mbox{\ensuremath{\chi^{2}\kern-0.3em/\kern-0.1em\text{ndf}}}\xspace}

\ShineTitle{
Search for the critical point of strongly-interacting matter in \ArSc
collisions at 150\AGeVc using scaled factorial moments of protons
}

\PreprintIdNumber{CERN-EP-2023-082}

\ShineJournal{}

\ShineJournalRef{}
\ShineDOI{}

\ShineAbstract{
The critical point of dense, strongly interacting matter is searched for at
the CERN SPS in \ArSc collisions at 150\AGeVc.
The dependence of second-order scaled factorial moments of proton multiplicity distribution
on the number of subdivisions of transverse momentum space is measured.
The intermittency analysis is performed using both transverse momentum and
cumulative transverse momentum.
For the first time, statistically independent data sets are used for each
subdivision number.

The obtained results do not indicate any statistically significant intermittency pattern.
An upper limit on the fraction of critical proton pairs and the power of
the correlation function is obtained based on a comparison with
the Power-law Model developed for this purpose.
}

\begin{document}

\maketitle

%\linenumbers

\section{Introduction}

The experimental results are presented on intermittency analysis using second-order scaled
factorial moments of mid-rapidity protons produced in central \ArSc collisions at 150\AGeVc
beam momentum (\sNN = 16.84~\GeV).
The measurements were performed by the multi-purpose \NASixtyOne~\cite{Abgrall:2014xwa} apparatus at
the CERN Super Proton Synchrotron (SPS). They are part of the strong interactions program of
\NASixtyOne devoted to the study of the properties of strongly interacting matter such as
onset of deconfinement and critical end point (CP). Within this program, a two-dimensional scan in
collision energy and size of colliding nuclei was conducted~\cite{Aduszkiewicz:2642286}.

In the proximity of CP, the fluctuations of the order parameter are self-similar \cite{Antoniou:2006zb},
belonging to the 3D-Ising universality class, and can be detected in transverse momentum space
within the framework of intermittency analysis of proton density fluctuations by use of scaled
factorial moments. This analysis was performed in intervals of transverse momentum and cumulative
transverse momentum distributions. For the first time, statistically independent data sets were used
to obtain results for different number of intervals (at the cost of reducing event statistics).
%Special attention was drawn to defining the experimental results, enabling comparison with model predictions.

The paper is organized as follows. Section~\ref{sec:sfm} introduces quantities exploited for
the CP search using the intermittency analysis. In Sec.~\ref{sec:detector}, the characteristics
of the \NASixtyOne detector, relevant for the current study, are briefly presented.
The details of data selection and the analysis procedure are presented in Sec.~\ref{sec:analysis}.
Results obtained are shown in Sec.~\ref{sec:results} and compared with several models in
Sec.~\ref{sec:models}. A summary in Sec.~\ref{sec:summary} closes the paper.

Throughout this paper, the rapidity, $y = \text{arctanh}\left(\beta_{L}\right)$, is calculated in the
collision center-of-mass frame by
shifting rapidity in laboratory frame by rapidity of the center-of-mass, assuming proton mass.
$\beta_{L} = p_{L}/E$ is the longitudinal (\coordinate{z}) component of the velocity, while $p_{L}$
and $E$ are particle longitudinal momentum and energy in the collision center-of-mass frame.
The transverse component of the momentum is denoted as $p_{T} = \sqrt{p_{x}^{2} + p_{y}^{2}}$, where
$p_{x}$ and $p_{y}$ are its horizontal and vertical components. The azimuthal angle $\phi$ is the
angle between the transverse momentum vector and the horizontal (\coordinate{x}) axis. Total momentum
in the laboratory frame is denoted as $p_{\text{lab}}$. The collision energy per nucleon pair in the
center-of-mass frame is denoted as \sNN.

The \ArSc collisions are selected by requiring a low value of the energy measured by the forward calorimeter,
Projectile Spectator Detector (PSD). This is the energy emitted into the region populated mostly by
projectile spectators. These collisions are referred to as PSD-central collisions and a selection of
collisions based on the PSD energy is called a PSD-centrality selection.
\section{Scaled factorial moments}
\label{sec:sfm}

\subsection{Critical point and intermittency in heavy-ion collisions}
\label{sec:sfm_CP}

% ----------------------
A second-order phase transition leads to the divergence of the correlation length ($\xi$).
The infinite system becomes scale-invariant with the particle density-density correlation function
exhibiting power-law scaling, which induces intermittent behavior of particle multiplicity
fluctuations~\cite{Wosiek:1988}.

% Also, other
% properties of the system should be sensitive to the vicinity of the critical
% point~\cite{Stephanov:1999zu} and these fluctuations have specific
% characteristics~\cite{Wosiek:1988,Bialas:1990xd}.

The maximum CP signal is expected when the freeze-out occurs close to the CP. On the other hand, the
energy density at the freeze-out is lower than at the early stage of the collision. Thus, the
critical point should be experimentally searched for in nuclear collisions at energies higher than
that of the onset of deconfinement -- the beginning of quark-gluon plasma creation.
According to the NA49 results~\cite{Afanasiev:2002mx,Alt:2007aa}, this general condition limits the
critical point search to the collision energies higher than $\sNN \approx 7$~\GeV.

The intermittent multiplicity fluctuations~\cite{Bialas:1985jb}  were
discussed as the signal of CP by Satz~\cite{Satz:1989vj}, Antoniou et al.~\cite{Antoniou:1990vf} and
Bialas, Hwa~\cite{Bialas:1990xd}. This initiated experimental studies of the structure of the phase
transition region via analyses of particle multiplicity fluctuations using scaled factorial
moments~\cite{NA49:2012ebu}. Later, additional measures of fluctuations were also proposed as probes
of the critical behavior ~\cite{Stephanov:1998dy,Stephanov:1999zu}. The \NASixtyOne experiment
has performed a systematic scan in collision energy and system size. The new measurements may answer
the question about the nature of the transition region and, in particular, whether or not the critical
point of strongly interacting matter exists.

The scaled factorial moments $F_r(M)$~\cite{Bialas:1985jb} of order $r$ are defined as:
\begin{equation}
  F_r(M) = \frac
    {\bigg<{\displaystyle{\frac{1}{M^{D}}\sum_{i=1}^{M^{D}}} n_i\;...\;(n_i-r+1) }\bigg>}
    {\bigg<{\displaystyle{\frac{1}{M^{D}}\sum_{i=1}^{M^{D}}} n_i }\bigg>^r }\:,
  \label{eq:scaled-factorial-moments}
\end{equation}
where $M$ is the number of subdivision intervals in each of the $D$ dimensions of the selected range
$\Delta$, $n_{i}$ is the particle multiplicity in a given subinterval and angle brackets denote
averaging over the analyzed events. In the presented analysis, $\Delta$ is divided into two-dimensional
($D=2$) cells in $p_{x}$ and $p_{y}$.

In case the mean particle multiplicity, $n_{i}$, is proportional to the subdivision interval size and
for a poissonian multiplicity distribution, $F_r(M)$ is equal to 1 for all values of $r$ and $M^{D}$.
This condition is satisfied in the configuration space, where the particle density is uniform
throughout the gas volume. The momentum distribution is, in general non-uniform and thus in the
momentum space, it is more convenient to use the so-called cumulative variables~\cite{Bialas:1990dk}
which, for very small cell size, leave a power-law unaffected and at the same time lead to uniformly
distributed particle density. By construction, particle density in the cumulative variables is
uniformly distributed.

If the system at freeze-out is close to the CP, its properties are expected to be very different from
those of an ideal gas. Such a system is a simple fractal and $F_r(M)$ follows a power-law dependence:
\begin{equation}
\label{eq:cp_1}
  F_{r}(M) = F_{r}(\Delta) \cdot (M^{D})^{\phi_{r}}~.
\end{equation}
Moreover, the exponent (intermittency index) $\phi_{r}$ obeys the relation:
\begin{equation}
\label{eq:cp_2}
  \phi_{r} = ( r - 1 ) \cdot (d_{r}/D)\:,
\end{equation}
where the anomalous fractal dimension $d_{r}$ is independent of $r$~\cite{Bialas:1990xd}.
Such behavior is the analogue of the phenomenon of critical opalescence in conventional
matter~\cite{Antoniou:2006zb}.
Importantly the critical properties given by Eqs.~\ref{eq:cp_1} and~\ref{eq:cp_2} are approximately
preserved for very small cell size (large $M$) under transformation to the cumulative
variables~\cite{Bialas:1990dk,Samanta:2021dxq}.

The ideal CP signal, Eqs.~\ref{eq:cp_1} and~\ref{eq:cp_2}, derived for the infinite system in
equilibrium may be distorted by numerous experimental effects present in high-energy collisions.
This includes finite size and evolution time of the system, other dynamical correlations between
particles, limited acceptance and resolution of measurements. Moreover, to experimentally search for
CP in high-energy collisions, the momentum-space region's dimension, interval size and location must
be chosen. Note that unbiased results can be obtained only by analyzing variables and dimensions in
which the singular behavior appears~\cite{Bialas:1990gu,Ochs:1988ky,Ochs:1990mg}. Any other procedure
is likely to distort the critical-fluctuation signal.

Another question is the selection of particle type used in the experimental search for CP. The
QCD-inspired considerations~\cite{Antoniou:2000ms,Stephanov:2004wx} suggest that the order parameter
of the phase transition is the chiral condensate. Suppose a carrier of the critical properties of the
chiral condensate is the isoscalar $\sigma$-field. In that case, the critical behavior can be observed
either directly from its decay products ($\pi^{+}\pi^{-}$ pairs)~\cite{Antoniou:2005am} or by
measuring the fluctuations of the number of protons. The former requires precise reconstruction of
pion pairs. In this case, $d = \phi_2 = 2/3$~\cite{Antoniou:2005am} is expected. The latter is based on
the assumption that the critical fluctuations are transferred to the net-baryon density, which
mixes with the chiral condensate~\cite{Fukushima:2010bq, Hatta:2002sj, Stephanov:2004wx,
Antoniou:2008vv,Karsch:2010ck,Skokov:2010uh,Morita:2012kt}. Thus, the net-baryon density may serve
as an order parameter of the phase transition. Such fluctuations are expected to be present in the
net-proton number and the proton and anti-proton numbers, separately~\cite{Hatta:2003wn}. For protons,
$d = \phi_2 = 5/6$~\cite{Antoniou:2006zb} is expected.

\subsection{Cumulative transformation}
\label{sec:sfm_cumulative}

Scaled factorial moments are sensitive to the shape of the single-particle momentum distribution.
This dependence may bias the signal of critical fluctuations. To remove it, one has two possibilities.
First, to construct a mixed events data set, where each event is constructed using particles from
different experimental events thereby removing all possible dynamical correlations. Then the quantity:
\begin{equation}
    \Delta F_{2}(M) = F_{2}^\text{data}(M) - F_{2}^{\text{mixed}}(M)
\end{equation}
is calculated. It was shown~\cite{NA49:2012ebu} that this procedure removes (approximately) the
dependence of $\Delta F_{2}(M)$ on the shape of single-particle distribution.

The second possibility is to use cumulative transformation~\cite{Bialas:1990dk}, which for a
one-dimensional single-particle distribution $f(x)$ reads:
\begin{equation}
    Q_{x} = \int\limits_{a}^{x} f(x)dx \Bigg/ \int\limits_{a}^{b} f(x)dx~,
\end{equation}
where $a$ and $b$ are lower and upper limits of the variable $x$.
For a two-dimensional distribution $f(x,y)$ and a given $x$ the transformation reads
\begin{equation}
    Q_{y}(x) = \int\limits_{a}^{y} f(x,y)dy \Bigg/ \int\limits_{a}^{b} f(x,y)dy.
\end{equation}
The cumulative transformation transforms any single-particle distribution into a uniform one
ranging from 0 to 1, and therefore it removes the dependence on the shape of the single-particle
distribution for uncorrelated particles.
At the same time, it has been verified that the transformation preserves the critical
behavior~\cite{Samanta:2021dxq} given by Eq.~\ref{eq:cp_1}, at least for the second-order
scaled factorial moments.

\begin{figure}[!ht]
    \centering
    \includegraphics[height=5cm]{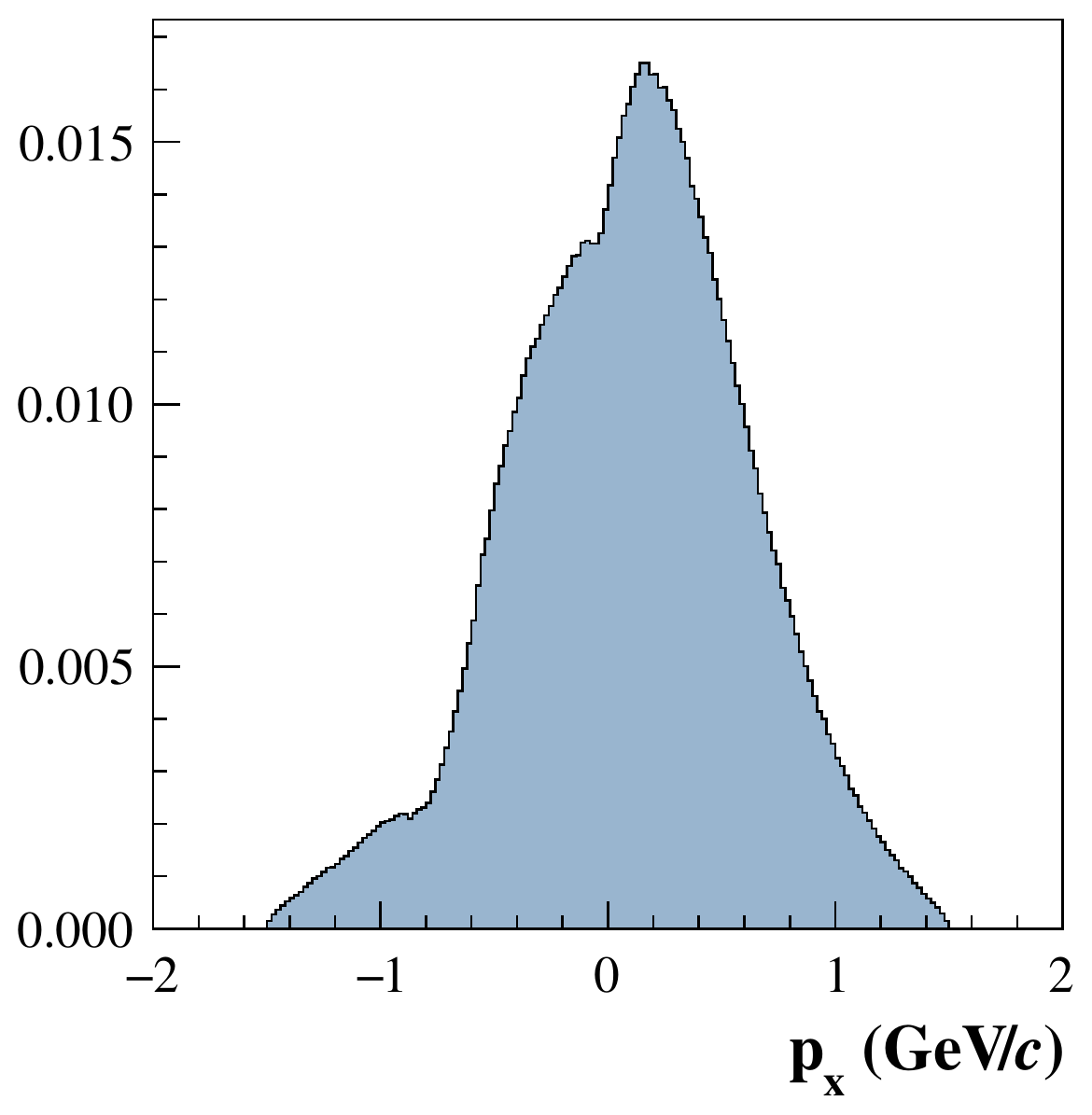}\hfill
    \includegraphics[height=5cm]{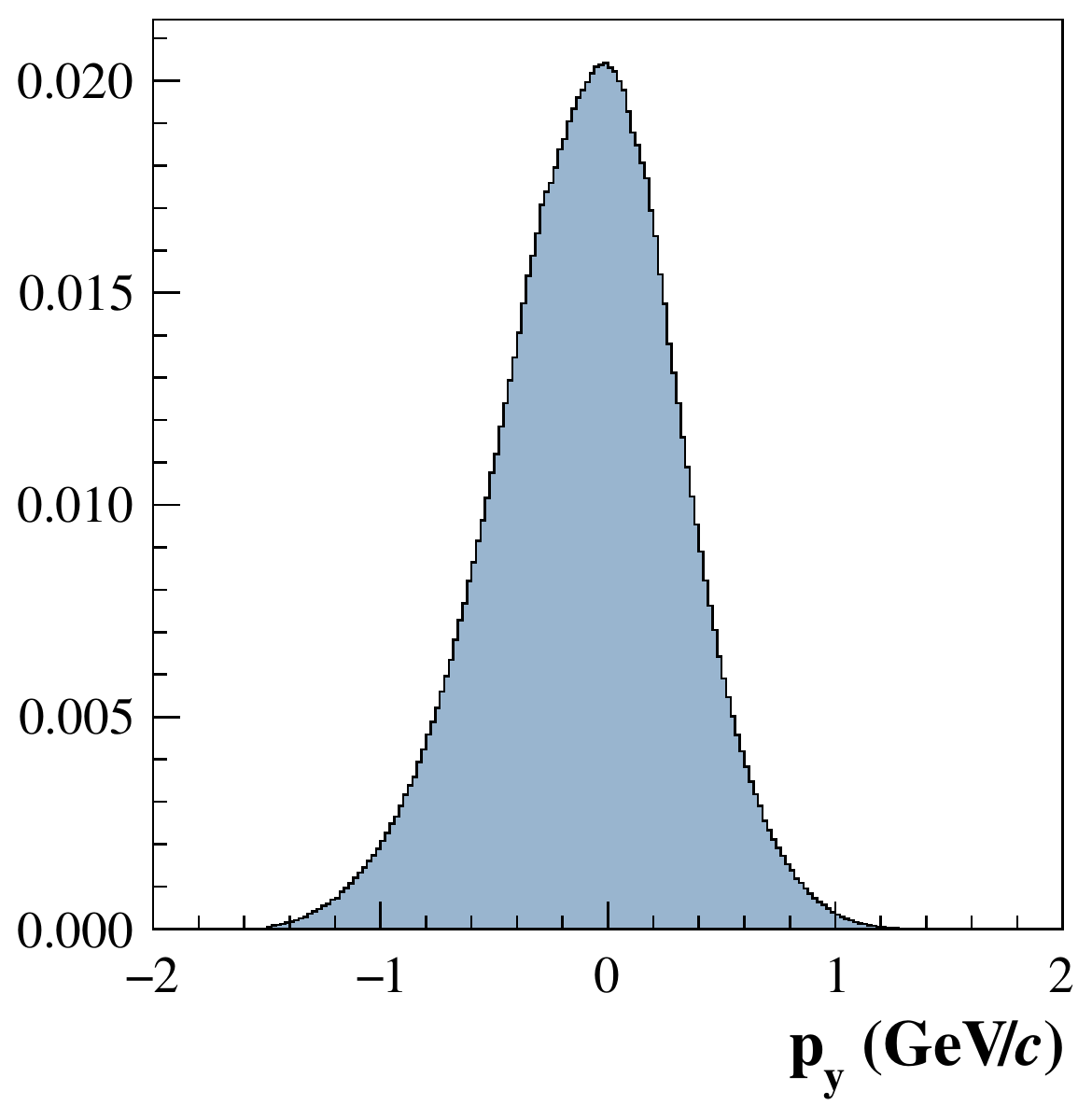}\hfill
    \includegraphics[height=5.2cm]{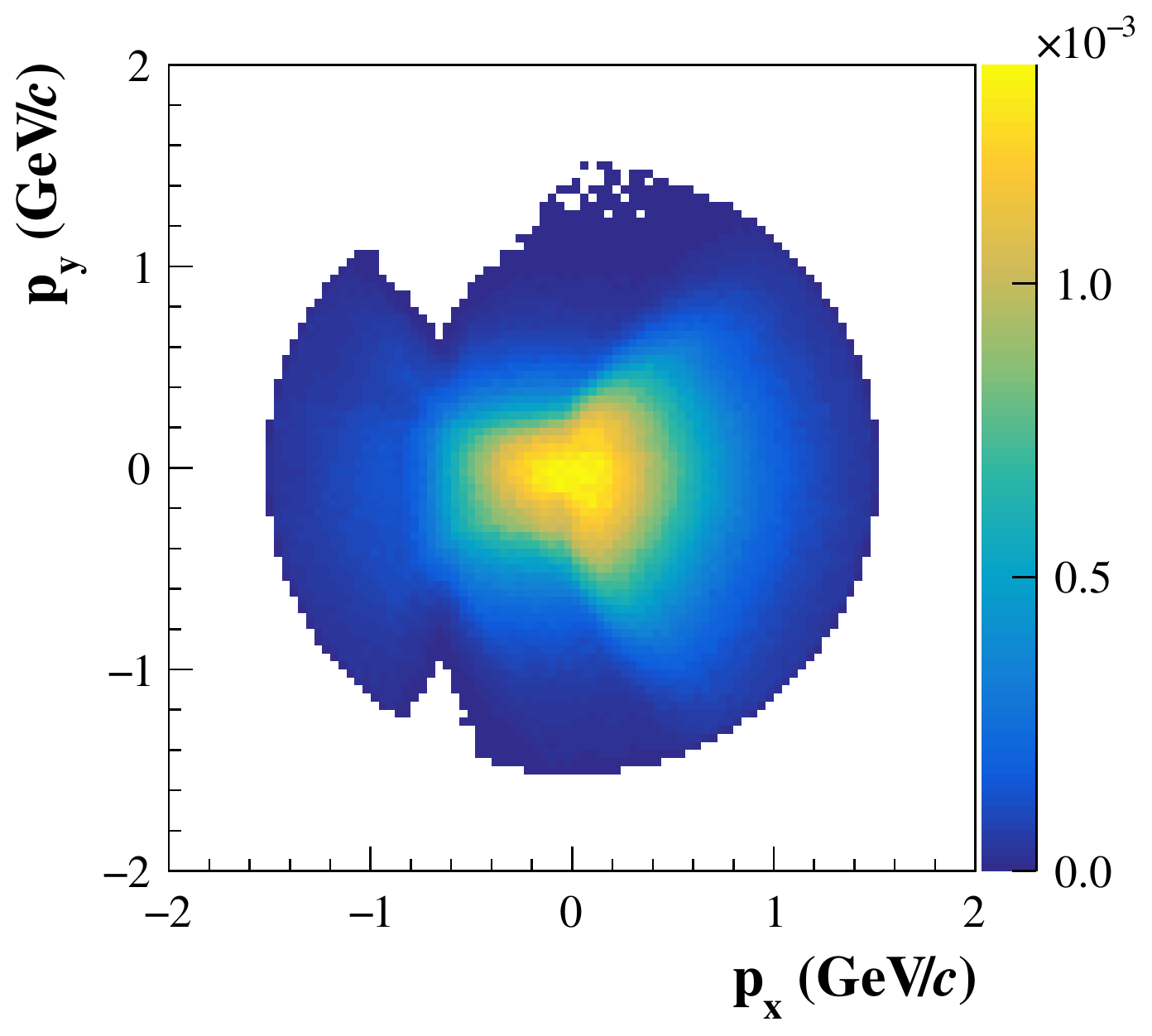}\\
    \includegraphics[height=5cm]{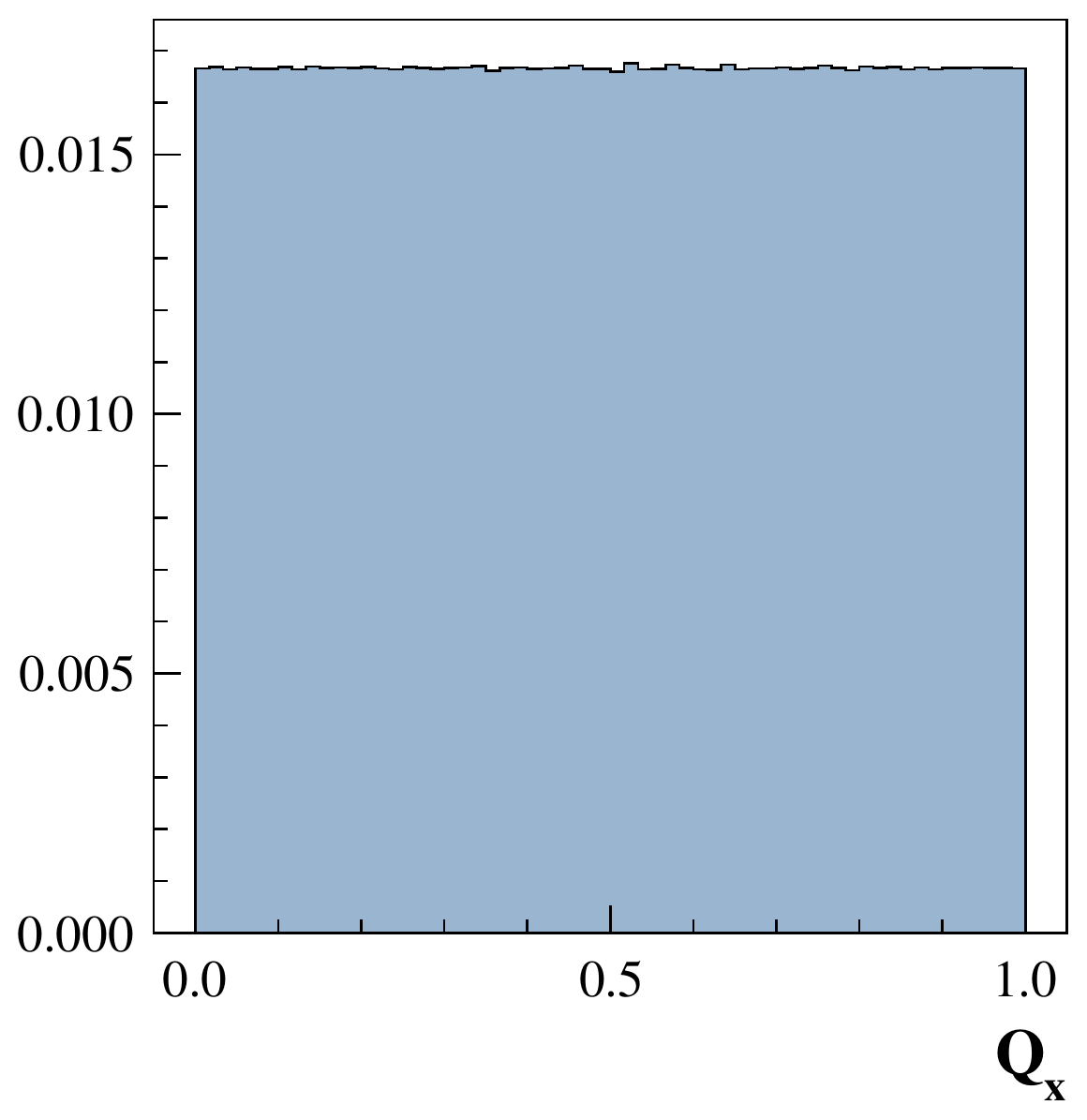}\hfill
    \includegraphics[height=5cm]{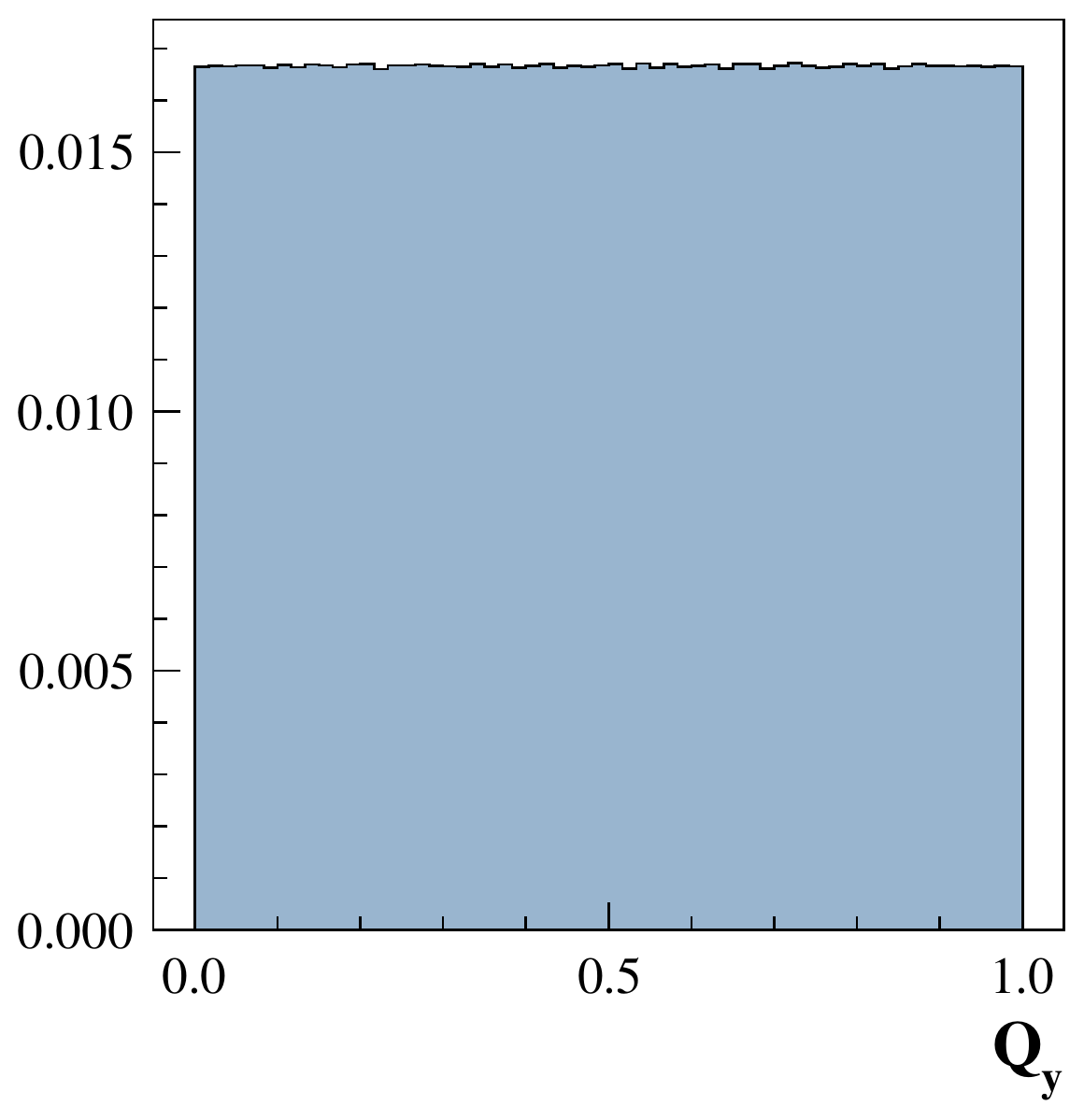}\hfill
    \includegraphics[height=5.2cm]{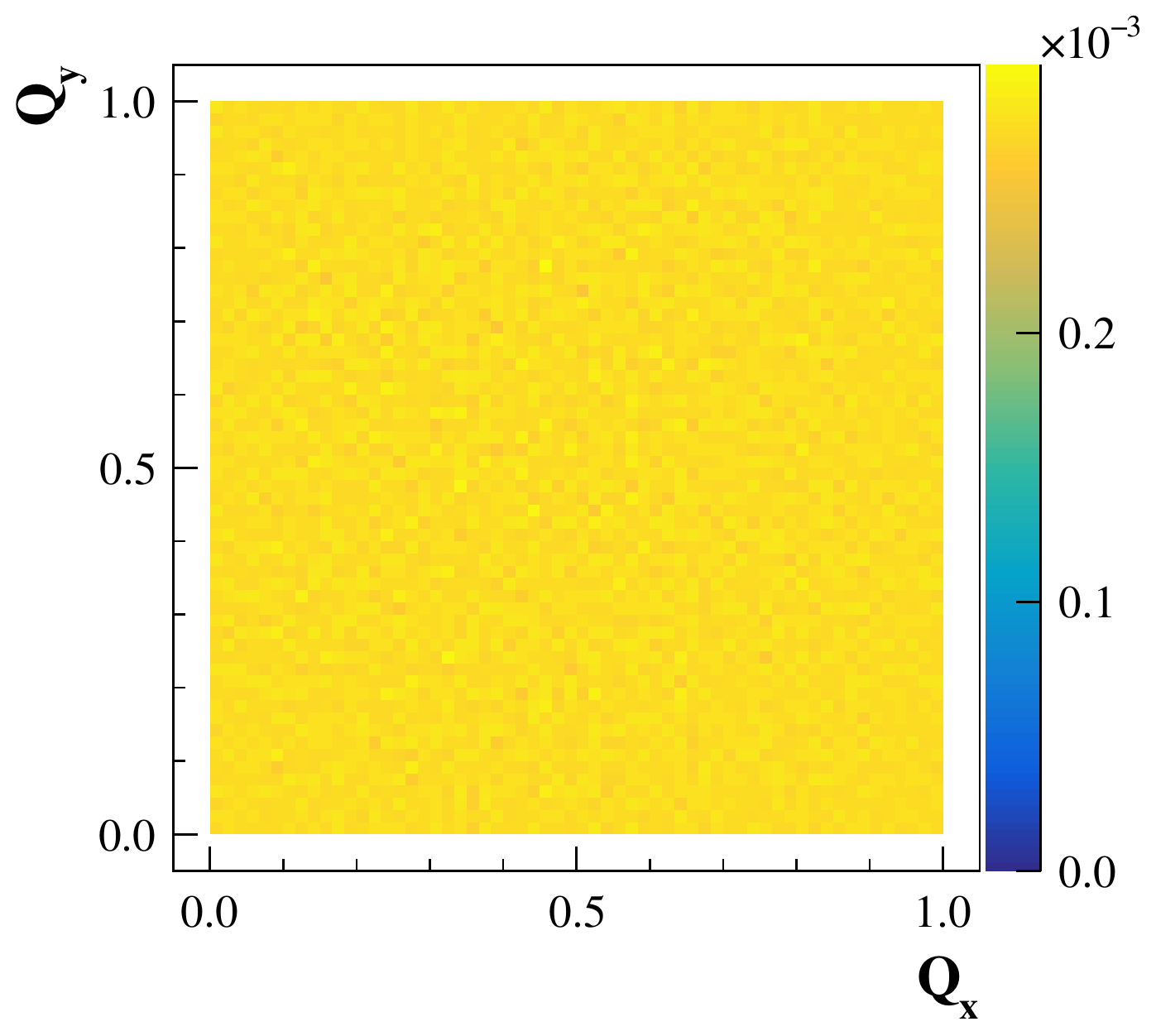}
    \caption{
        Example of the effect of the cumulative transformation of transverse momentum components,
        $p_{x}$ and $p_{y}$ of proton candidates selected for
        intermittency analysis of the \NASixtyOne \ArSc at 150\AGeVc data.
        Distributions before (\emph{top}) and after (\emph{bottom}) the transformation.
    }
    \label{fig:cumulative}
\end{figure}

An example of the transformation of transverse momentum components $p_{x}$ and $p_{y}$ for
protons produced in 5\% most central \ArSc collisions at 150\AGeVc (see next sections for details)
is shown in Fig.~\ref{fig:cumulative}.

Both methods are approximate. Subtracting moments for mixed data set may introduce negative
$\Delta F_{2}(M)$ values~\cite{NA49:2012ebu} and using cumulative quantities
mixes the scales of the momentum differences and therefore may distort eventual
power-law behavior.

\section{The \NASixtyOne detector}
\label{sec:detector}

\begin{figure*}[ht]
  \centering
  \includegraphics[width=\textwidth]{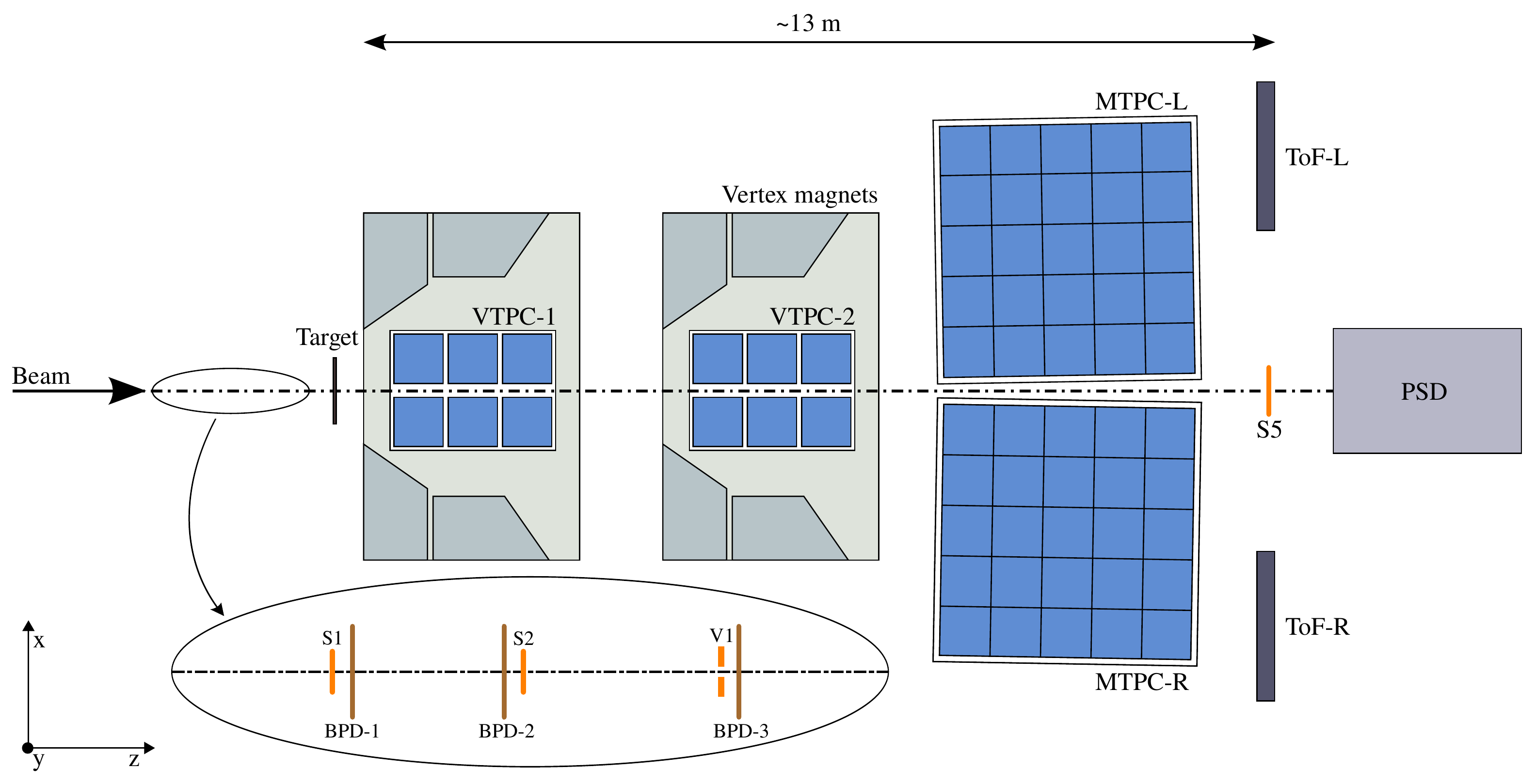}
  \caption{
    The schematic layout of the \NASixtyOne experiment at the CERN SPS ~\cite{Abgrall:2014xwa}
    showing the components used for the \ArSc energy scan (horizontal cut, not to scale).
    The detector configuration upstream of the target is shown in the inset.
    Alignment of the chosen coordinate system is shown on the plot; its origin
    (\coordinate{x}=\coordinate{y}=\coordinate{z}=0) lies in the middle of VTPC-2,
    on the beam axis. The nominal beam direction is along the \coordinate{z}-axis.
    Target is placed at \mbox{\coordinate{z} $\approx -580$~cm}.
    The magnetic field bends charged particle trajectories in the \coordinate{x}--\coordinate{z}
    (horizontal) plane. The drift direction in the TPCs is along the (vertical) \coordinate{y}-axis.
  }
  \label{fig:setup}
\end{figure*}

The \NASixtyOne detector (see Fig.~\ref{fig:setup}) is a large-acceptance hadron spectrometer
situated in the North Area H2 beam-line of the CERN SPS~\cite{Abgrall:2014xwa}.
The main components of the detection system used in the analysis are four large-volume Time Projection Chambers (TPC).
Two of them, called Vertex TPCs (VTPC-1/2), are located downstream of the target inside superconducting magnets
with maximum combined bending power of 9~Tm, which was set for the data collection at 150\AGeVc.
The main TPCs (MTPC-L/R) and two walls of pixel Time-of-Flight (ToF-L/R) detectors are placed symmetrically on either
side of the beamline downstream of the magnets.
The TPCs are filled with Ar:CO\textsubscript{2} gas mixtures in proportions 90:10 for the VTPCs and 95:5 for the MTPCs.
The Projectile Spectator Detector (PSD), a zero-degree hadronic calorimeter, is positioned 16.7 m downstream of the MTPCs,
centered in the transverse plane on the deflected position of the beam.
It consists of 44 modules that cover a transverse area of almost 2.5~m\textsuperscript{2}. The central part of the PSD consists of
16 small modules with transverse dimensions of 10~x~10~cm\textsuperscript{2} and its outer part consists of
28 large 20~x~20~cm\textsuperscript{2} modules.
Moreover, a brass cylinder of 10~cm length and 5~cm diameter (degrader) was placed in front of the center of the
PSD in order to reduce electronic saturation effects and shower leakage from the downstream side caused by the
Ar beam and its heavy fragments.

Primary beams of fully ionized \Ar nuclei were extracted from the SPS accelerator at 150\AGeVc beam momentum.
Two scintillation counters, S1 and S2, provide beam definition, together with a veto counter V1 with a 1~cm diameter hole,
which defines the beam before the target. The S1 counter also provides the timing reference (start time for all counters).
Beam particles are selected by the trigger system requiring the coincidence
\mbox{$\textrm{T1} = \textrm{S1}\wedge\textrm{S2} \wedge\overline{\textrm{V1}}$}.
Individual beam particle trajectories are precisely measured by the three beam position detectors (BPDs) placed
upstream of the target~\cite{Abgrall:2014xwa}. Collimators in the beam line were adjusted to obtain beam rates
of~$\approx 10^4$/s during the~$\approx$~10~s spill and a super-cycle time of 32.4~s.

The target was a stack of 2.5~x~2.5~cm\textsuperscript{2} area and 1~mm thick \Sc plates of 6~mm total thickness
placed~$\approx$~80~cm upstream of VTPC-1. Impurities due to other isotopes and elements were measured to be
0.3\%~\cite{Banas:2018sak}. No correction was applied for this negligible contamination.

Interactions in the target are selected with the trigger system by requiring an incoming \Ar ion and a signal
below that of beam ions from S5, a small 2~cm diameter scintillation counter placed on the beam trajectory
behind the MTPCs. This minimum bias trigger is based on the breakup of the beam ion due to interactions in
and downstream of the target. In addition, central collisions were selected by requiring an energy signal below
a set threshold from the 16 central modules of the PSD, which measure mainly the energy carried by projectile
spectators. The cut was set to retain only the events with the $\approx$~30\% smallest energies in the PSD.
The event trigger condition thus was \mbox{$\textrm{T2} = \textrm{T1}\wedge\overline{\textrm{S5}}\wedge\overline{\textrm{PSD}}$}.
The statistics of recorded events at 150\AGeVc are summarized in Table~\ref{tab:statbeam}.

\begin{table}
	\caption{
      Basic beam properties and number of events recorded and used in the analysis of \ArSc interactions at
      incident momentum of 150\AGeVc.
    }
	\vspace{0.5cm}
	\centering
	\begin{tabular}{ c | c | c | c }
  	    $p_{beam}$ (\GeVc) &
  	    \sNN (\GeV) &
  	    Recorded central triggers &
  	    Number of selected events \\
    \hline
    150\A & 16.84 & $1.7\cdot10^{6}$ & $1.1\cdot10^{6}$
  	\end{tabular}
	\label{tab:statbeam}
\end{table}
\section{Analysis}
\label{sec:analysis}

The goal of the analysis was to search for the critical point of the strongly interacting matter
by measuring the second-order scaled factorial moments for a selection of protons produced in central
\ArSc interactions at 150\AGeVc, using statistically independent points and cumulative variables.

\subsection{Event selection}
\label{sec:event-selection}

\NASixtyOne detector recorded over 1.7 million collisions using 150\AGeVc \Ar beam impinging
on a stationary \Sc target.
However, not all of those events contain well-reconstructed central Ar+Sc interactions.
Therefore the following criteria were used to select data for further analysis.
\begin{enumerate}[(i)]
    \item no off-time beam particle detected within a time window of $\pm 4 \mu$s around the trigger
      particle,
    \item no interaction-event trigger detected within a time window of $\pm 25 \mu$s around the
      trigger particle,
    \item beam particle detected in at least two planes out of four of BPD-1 and BPD-2 and in both
      planes of BPD-3,
    \item T2 trigger (set to select central and semi-central collisions),
    \item a high-precision interaction vertex with \coordinate{z} position (fitted using  the beam
      trajectory and TPC tracks) no further than 10~cm away from the center of the Sc target (the cut
      removes less than 0.4\% of T2 trigger ($E_{PSD}$) selected interactions),
    \item energy in small PSD modules should be less than 2800~\GeV,
    \item energy in large PSD modules should be in the range between 800~\GeV and 5000~\GeV,
    \item if the number of tracks in the vertex fit is less than 50, then the ratio of tracks in fit
      to all tracks must be at least 0.25.
\end{enumerate}

After applying the selection criteria, about 1.1 million events remain for further analysis.

\subsection{Centrality selection}

\begin{figure}[!ht]
    \centering
    \includegraphics[width=.42\textwidth]{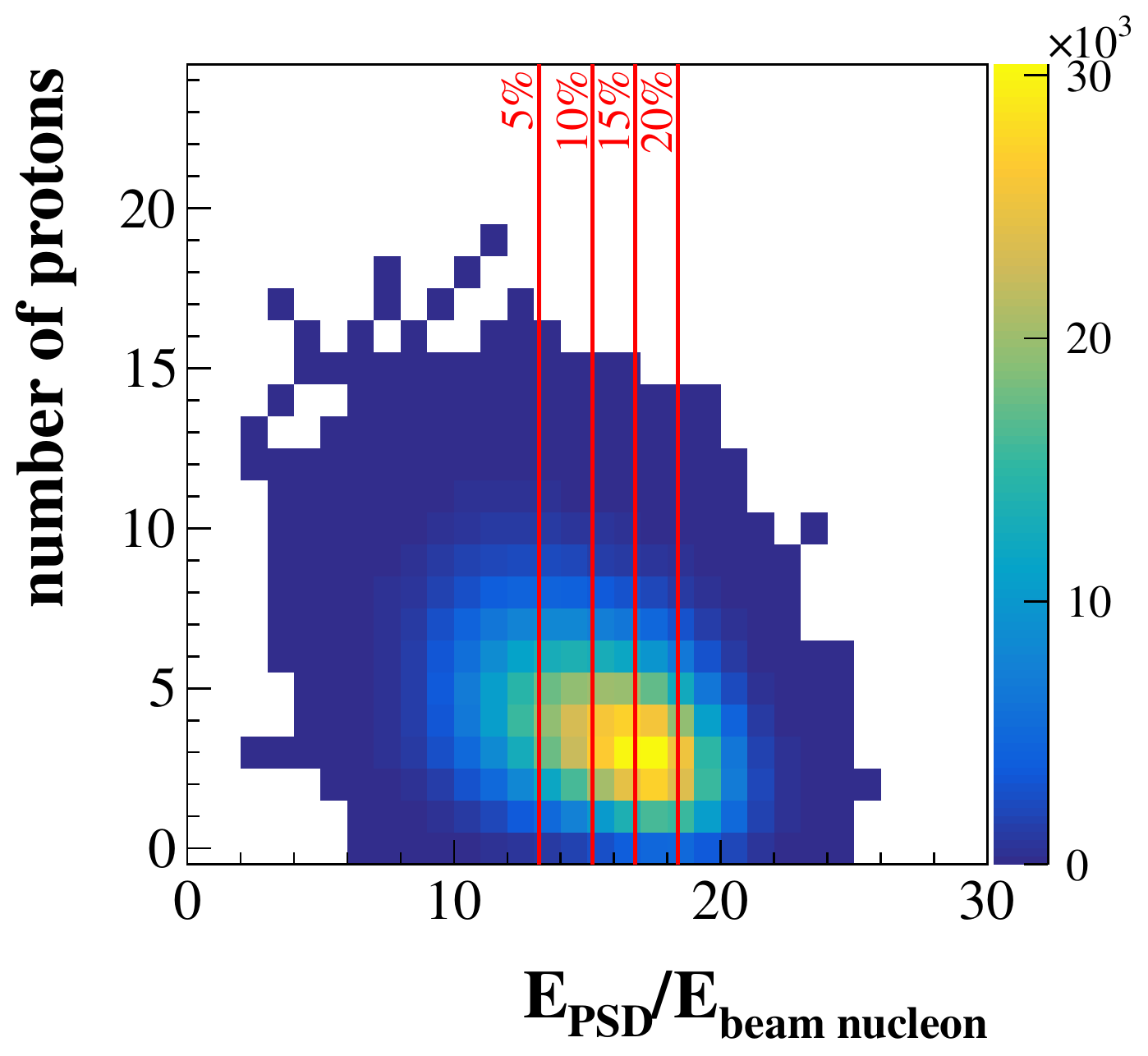}\qquad
    \includegraphics[width=.4\textwidth]{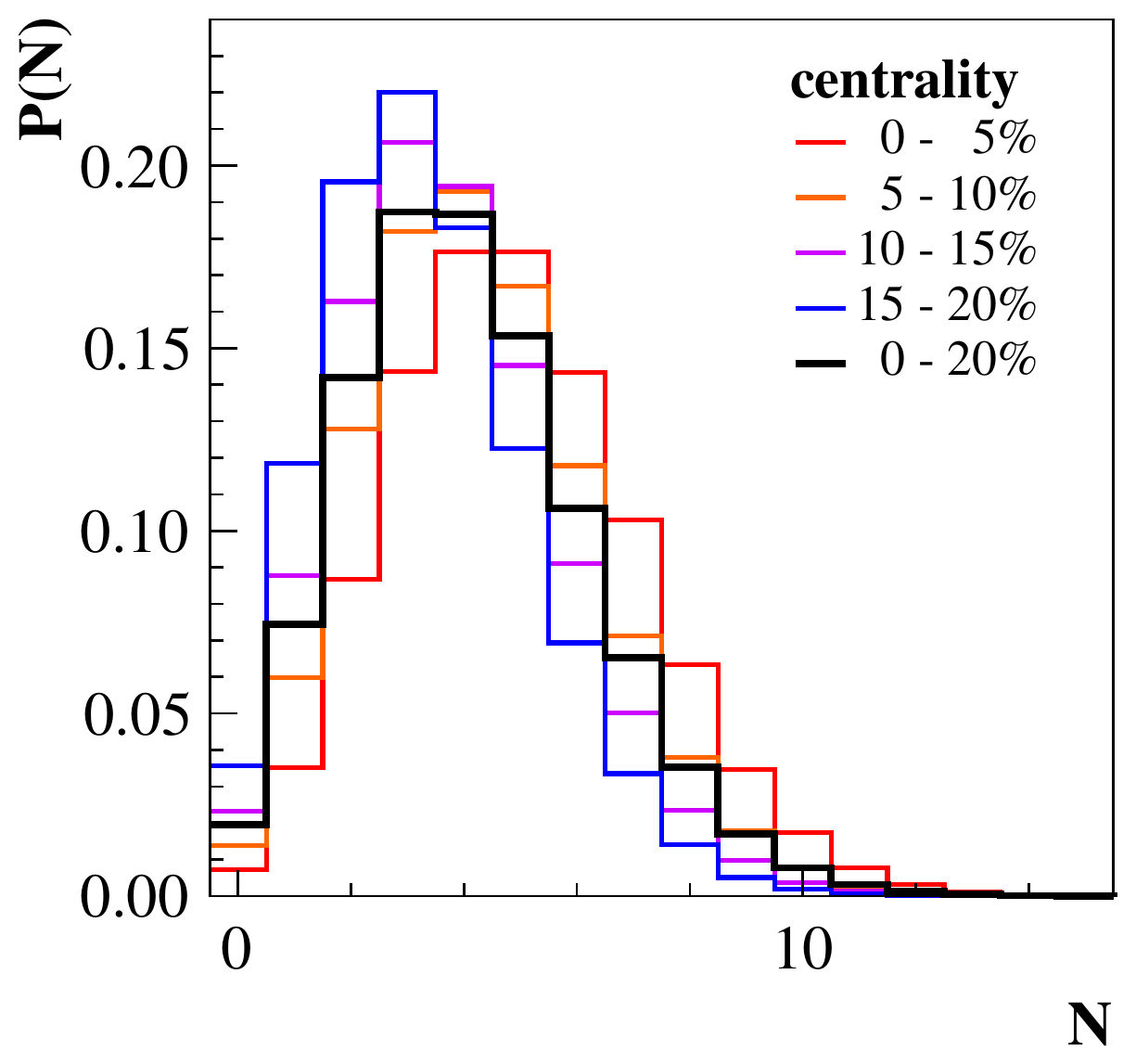}
    \caption{
        \emph{Left:}
        Distribution of selected events in the number of accepted proton candidates and the
        energy registered by the Particle Spectator Detector with PSD-energy values for
        centrality selection marked as red vertical lines.
        \emph{Right:}
        Multiplicity distributions of proton candidates for different centrality selections
        (for all events selected for the analysis and after all track cuts described in
        Sec.~\ref{sec:track_selection}).
    }
    \label{fig:centrality}
\end{figure}

The analysis was performed in several centrality intervals (0-5\%, 5-10\%, 10-15\%, 15-20\% and 0-20\%).
Centrality is determined using the energy deposited in the PSD forward calorimeter, $E_{PSD}$ (see
Ref.~\cite{NA61SHINE:2020ggt} for details). Figure~\ref{fig:centrality} shows the proton candidate
multiplicity distributions for each of the studied centrality classes and the distributions of the
number of accepted proton candidates for different selections of energy deposited in PSD.

Table \ref{tab:events} presents number of events in each of the chosen centrality intervals
selected for the analysis.
\begin{table}[!ht]
    \centering
    \caption{
        Number of events selected for the analysis.
    }
    \vspace{1ex}
    \begin{tabular}{ l | c | c | c | c || c }
        \multicolumn{1}{c|}{\multirow[c]{2}{*}{data set}} & \multicolumn{5}{c}{number of events in centrality intervals} \\
        \cline{2-6}
        & 0 -- 5\% & 5 -- 10\% & 10 -- 15\% & 15 -- 20\% & 0 -- 20\% \\
        \hline
        experimental data & 237k & 235k & 236k & 216k & 924k \\
        \EposLong         & 323k & 323k & 323k & 324k & 1293k \\
    \end{tabular}
    \label{tab:events}
\end{table}

\subsection{Single-track selection}
\label{sec:track_selection}

To select tracks of primary charged hadrons and to reduce the contamination by particles from secondary
interactions, weak decays and off-time interactions, the following track selection criteria were applied:
\begin{enumerate}[(i)]
    \item track momentum fit including the interaction vertex should have converged,
    \item total number of reconstructed points on the track should be greater than 30,
    \item sum of the number of reconstructed points in VTPC-1 and VTPC-2 should be greater than 15,
    \item the ratio of the number of reconstructed points to the potential (maximum possible) number
      of reconstructed points should be greater than 0.5 and less than 1.1,
    \item number of points used to calculate energy loss (\dedx) should be greater than 30,
    \item the distance between the track extrapolated to the interaction plane and the vertex (track
      impact parameter) should be smaller than 4~cm in the horizontal (bending) plane and 2~cm in the
      vertical (drift) plane.
\end{enumerate}

As the analysis concerns mid-rapidity protons, only particles with center-of-mass rapidity (assuming
proton mass) greater than -0.75 and less than 0.75 were considered.

Only particles with transverse momentum components, $p_{x}$ and $p_{y}$, absolute values less than
1.5~\GeVc were accepted for the analysis.

\subsubsection{Proton selection}

\begin{figure}[!ht]
    \centering
    \includegraphics[width=.5\textwidth]{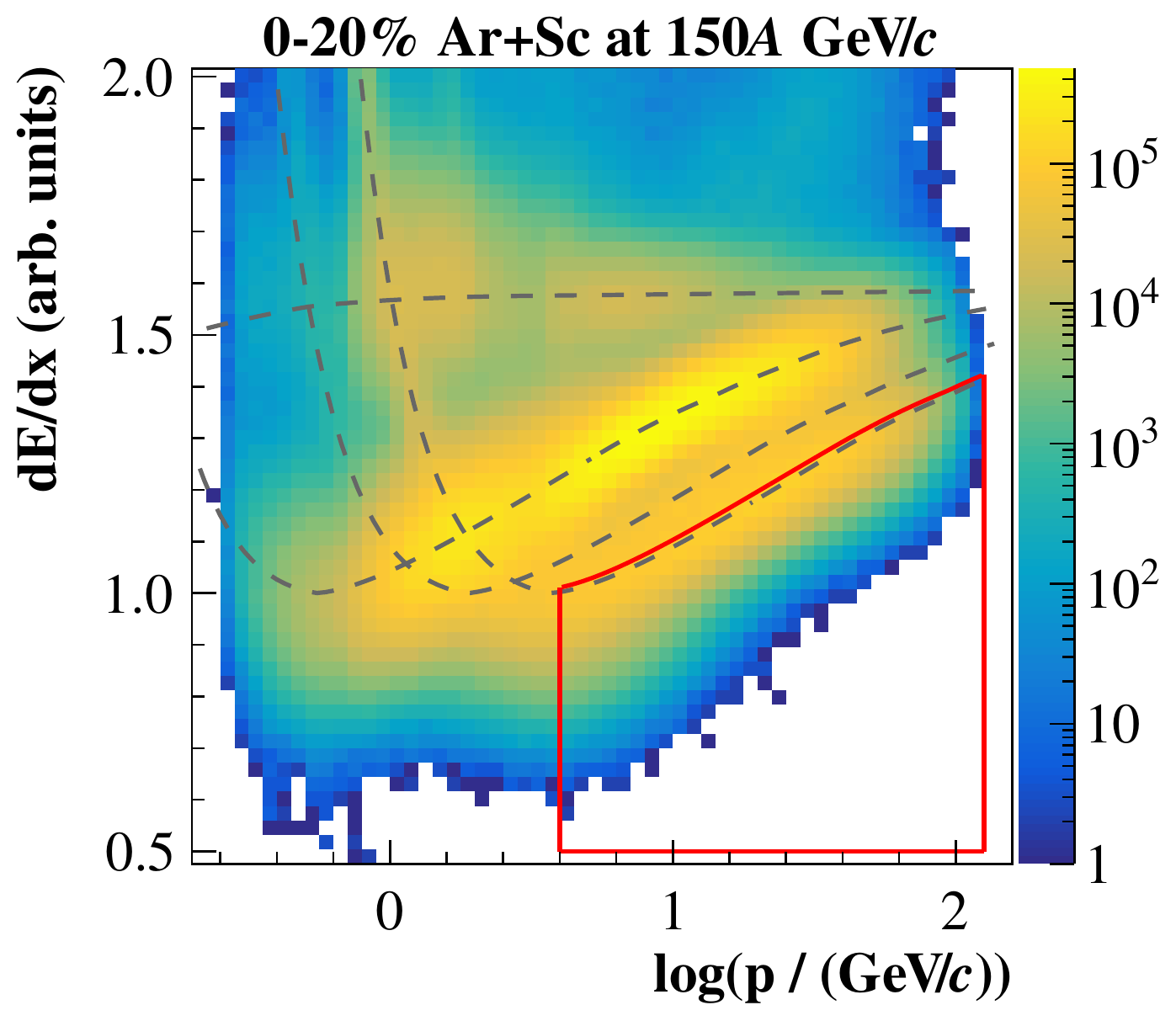}
    \caption{
        Energy loss vs total momentum of positively charged particles measured with the \NASixtyOne
        Time Projection Chambers in the selected \ArSc events at 150\AGeVc.
        Dashed lines represent the nominal Bethe-Bloch values.
        The graphical cut to select proton candidates is marked with a red line.
    }
    \label{fig:dEdx}
\end{figure}

To identify proton candidates, positively charged particles were selected. Their ionization energy
loss in TPCs is taken to be greater than 0.5 and less than the proton Bethe-Bloch value increased
by the 15\% difference between the values for kaons and protons while the momentum is in the
relativistic-rise region (from 4 to 125~\GeVc). The \dedx distribution for selected positive particles
is shown in Fig.~\ref{fig:dEdx}. The selected region is marked with a red line.

The selection was found to select, on average, approximately 60\% of protons and leave, on average,
less than 4\% of kaon contamination. The corresponding random proton losses do not bias the final
results in case of independent production of protons in the transverse momentum space.
The results for correlated protons will be biased by the selection (see Sec.~\ref{sec:results} for
an example), thus the random proton selection should be considered when calculating model predictions.

\subsection{Acceptance maps}
\label{sec:maps}

\subsubsection{Single-particle acceptance map}

A three-dimensional (in $p_{x}$, $p_{y}$ and center-of-mass rapidity) acceptance map
\cite{na61AccMapProtinIntermittencyArSc150} was created to describe the momentum region selected for
this analysis. The map was created by comparing the number of Monte Carlo-generated mid-rapidity
protons before and after detector simulation and reconstruction. Only particles from the regions with
at least 70\% reconstructed tracks are analyzed. The single-particle acceptance maps should be used
for calculating model predictions.

\subsubsection{Two-particle acceptance map}
\label{sec:track-pair-selection}

Time Projection Chambers (the main tracking devices of \NASixtyOne) are not capable of distinguishing
tracks that are too close to each other in space. At a small distance, their clusters overlap, and
signals are merged.

The mixed data set is constructed by randomly swapping particles from different events so that each
particle in each mixed event comes from different recorded events.

\begin{figure}[!ht]
    \centering
    \includegraphics[width=.4\textwidth]{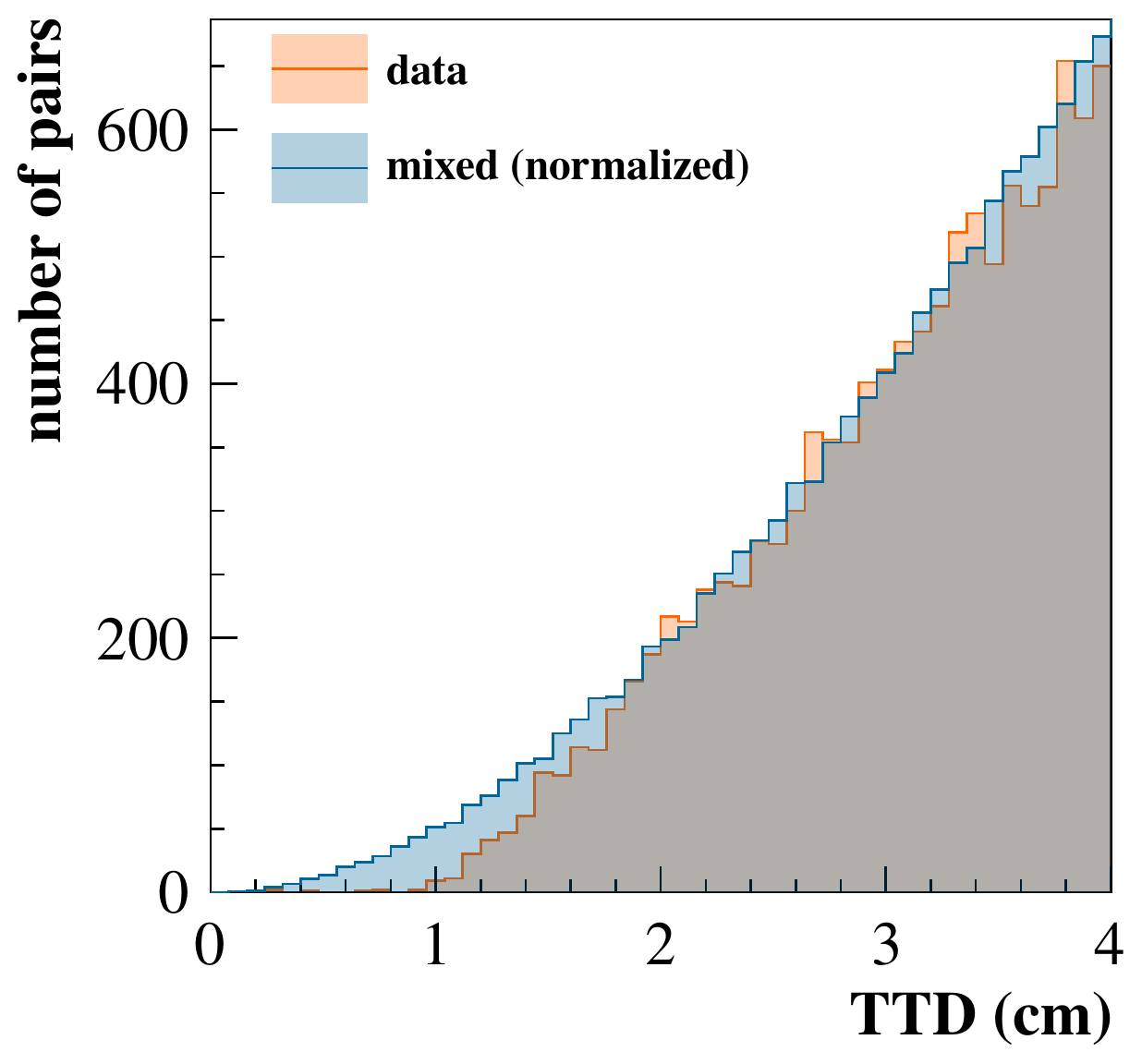}\qquad
    \includegraphics[width=.4\textwidth]{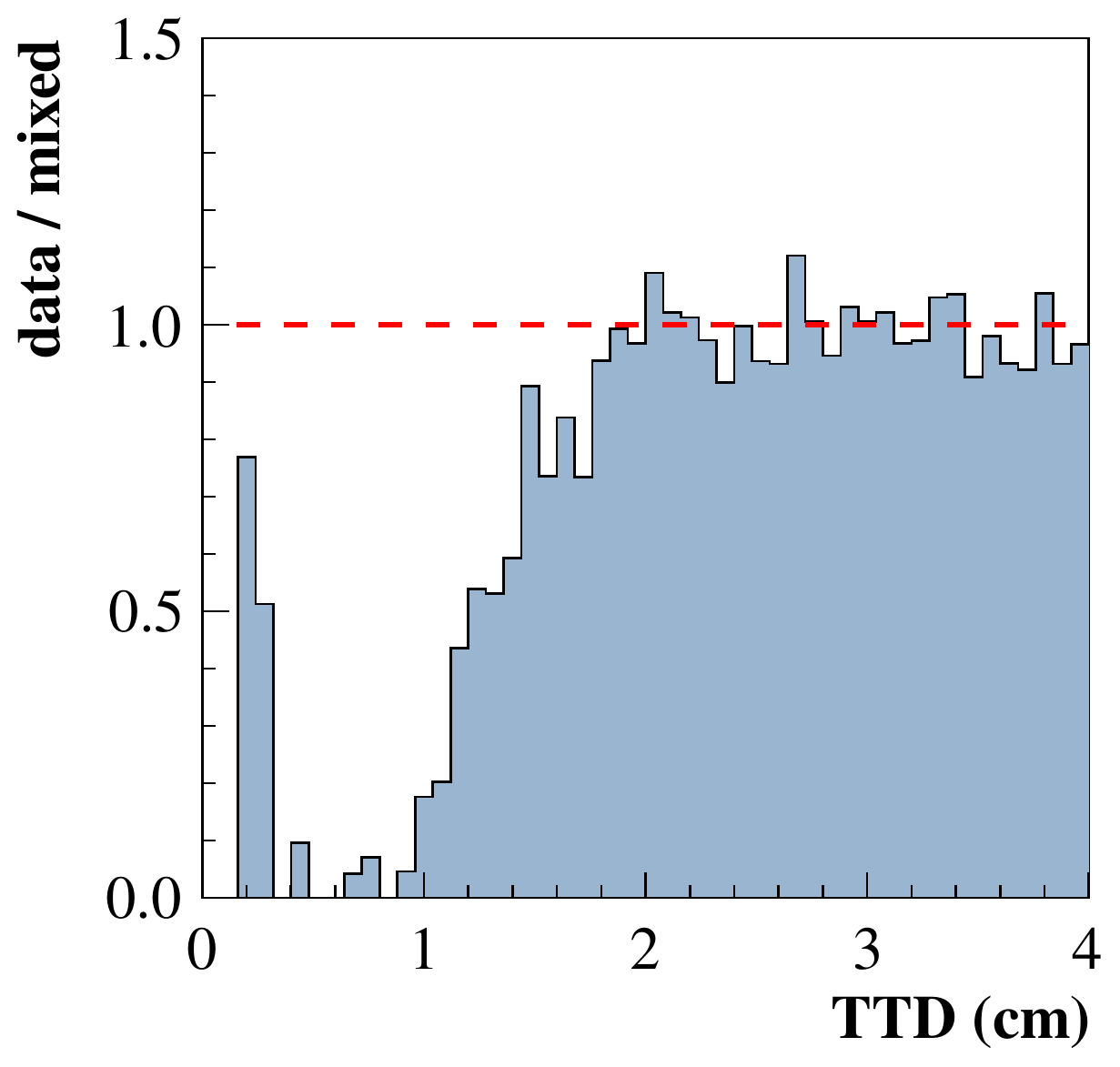}
    \caption{
        Distributions of Two-Track Distance for experimental and mixed data (\emph{left})
        and their ratio (\emph{right}).
    }
    \label{fig:ttd}
\end{figure}

For each pair of particles in both recorded and mixed events, a Two-Track Distance (TTD) is calculated.
It is an average distance of their tracks in $p_{x}$-$p_{y}$ plane at eight different $z$ planes
(-506, -255, -201, -171, -125, 125, 352 and 742~cm).
Figure~\ref{fig:ttd} presents TTD distributions for both data sets (\emph{left}) and their ratio
(\emph{right}). The TPC's limitation to recognizing close tracks is clearly visible for TTD < 2~cm.

Calculating TTD requires knowledge of the \NASixtyOne detector geometry and magnetic field. Hence it
is restricted to the Collaboration members. Therefore, a momentum-based Two-Track Distance (mTTD) cut
was introduced to allow for a meaningful comparison with models.

The magnetic field bends the trajectory of charged particles in the \coordinate{x}-\coordinate{z}
plane. Thus, it is most convenient to express the momentum of each positive particle in both recorded
and mixed data sets in the following coordinates:
\begin{align*}
    s_{x} & = p_{x}/p_{xz},\\
    s_{y} & = p_{y}/p_{xz},\\
    \rho & = 1/p_{xz},
\end{align*}
where $p_{xz} = \sqrt{p_{x}^{2} + p_{z}^{2}}$.
For each pair of positively charged particles, a difference in these coordinates is calculated as
\begin{align*}
    \Delta s_{x} & = s_{x,2} - s_{x,1},\\
    \Delta s_{y} & = s_{y,2} - s_{y,1},\\
    \Delta \rho & = \rho_{2} - \rho_{1}.
\end{align*}

\begin{figure}[!ht]
    \centering
    \includegraphics[width=.4\textwidth]{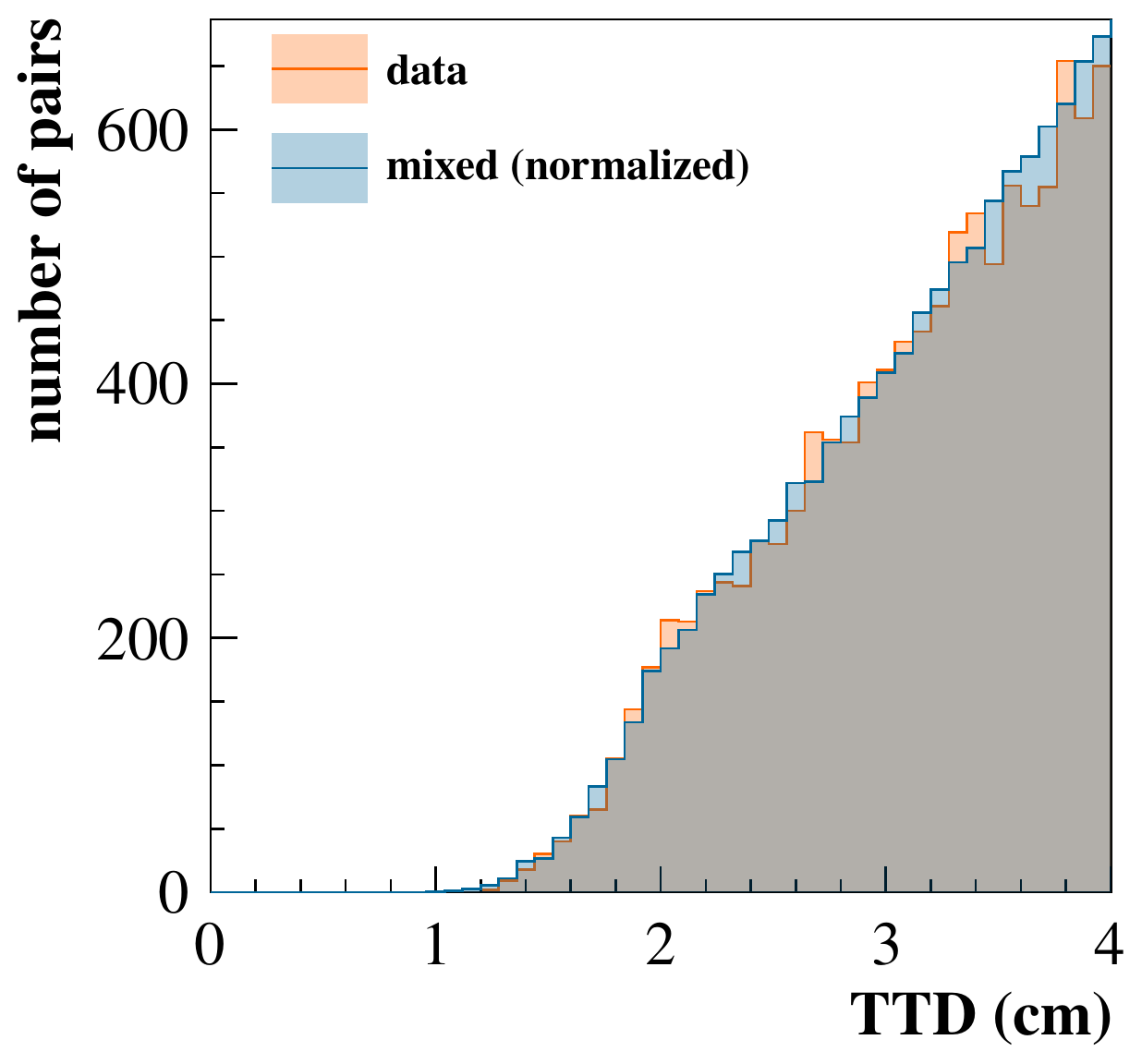}\qquad
    \includegraphics[width=.4\textwidth]{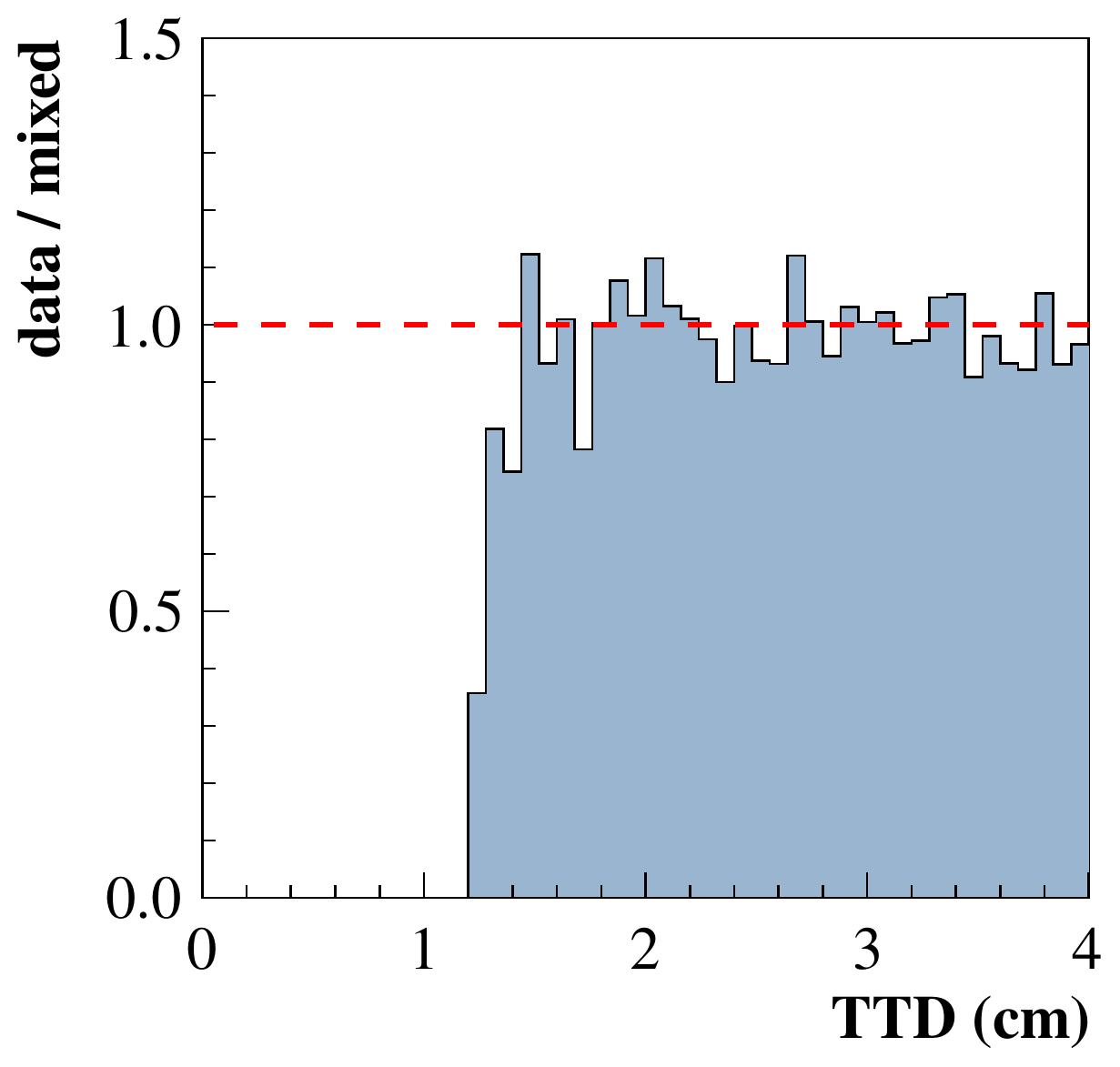}
    \caption{
        Distributions of Two-Track Distance for experimental and mixed data (\emph{left}) and their
        ratio (\emph{right}) after applying the momentum-based Two-Track Distance cut.
    }
    \label{fig:mTTD}
\end{figure}

The distribution of particle pairs' momentum difference for pairs with TTD < 2~cm is parametrized
with ellipses in the new coordinates. Such elliptic cuts are applied to recorded and mixed events.
Their distributions and their ratio are shown in Fig.~\ref{fig:mTTD}.

The mTTD cut defines the two-particle acceptance used in the analyses. The explicit formulas with
numerical values of parameters are:
\begin{align*}
    \left(\frac{\Delta \rho}{0.0105}\right)^{2}  + \left(\frac{\Delta s_{y}}{0.0018}\right)^{2} & \leq 1,\\
    \left(\frac{\Delta s_{x}}{0.0080}\right)^{2} + \left(\frac{\Delta s_{y}}{0.0018}\right)^{2} & \leq 1,\\
    \left(\frac{\Delta \rho\cos(\text{51}^{\circ}) - \Delta s_{x}\sin(\text{51}^{\circ})}{0.0200}\right)^{2} +
    \left(\frac{\Delta \rho\sin(\text{51}^{\circ}) + \Delta s_{x}\cos(\text{51}^{\circ})}{0.0023}\right)^{2} & \leq 1.
\end{align*}
Particle pairs with momenta inside all the ellipses are rejected. The two-particle acceptance maps
should be used for calculating model predictions.

\subsection{Statistically-independent data points}

The intermittency analysis yields the dependence of scaled factorial moments on the number of
subdivisions of transverse momentum and cumulative transverse momentum intervals. In the past, the
same data set was used for the analysis performed for different subdivision numbers. This resulted
in statistically-correlated data points uncertainties, therefore the full covariance matrix is required
for proper statistical treatment of the results. The latter may be numerically not
trivial~\cite{Wosiek:1990kw}. Here, for the first time, statistically-independent data subsets were
used to obtain results for each subdivision number. In this case, the results for different
subdivision numbers are statistically independent. Only diagonal elements of the covariance matrix
are non-zero, and thus the complete relevant information needed to interpret the results is easy to
present graphically and use in the statistical tests. However, the procedure significantly decreases
the number of events used to calculate each data point increasing statistical uncertainties and
therefore forcing to reduce the number of the data points to 10.

Number of events used in each subset was selected to obtain similar magnitudes of the statistical
uncertainties of results for different subsets. Table~\ref{tab:fraction-of-events} presents the
fractions of all available events used to calculate each of the 10 points.
\begin{table}[!ht]
    \centering
    \caption{
        Fraction of the total number of analyzed events for each centrality interval used to calculate
        second-order scaled factorial moments for the chosen number of cumulative momentum cells.
    }
    \vspace{1ex}
    \begin{tabular}{ l | c | c | c | c | c | c | c | c | c | c }
        number of cells $M^{2}$    & $1^{2}$   & $50^{2}$ & $70^{2}$ & $86^{2}$ & $100^{2}$ & $111^{2}$ & $122^{2}$ & $132^{2}$ & $141^{2}$ & $150^{2}$ \\ \hline
        fraction of all events (\%) & 0.5 &  3.0 &  5.0 &  7.0 &   9.0 &  11.0 &  13.0 & 15.5 & 17.0 &  19.0
    \end{tabular}
    \label{tab:fraction-of-events}
\end{table}

\subsection{Uncertainties and biases}

The standard expression for the scaled factorial moments,
Eq.~\ref{eq:scaled-factorial-moments} can be rewritten as
\begin{equation}
\label{eq:sfm}
    F_{2}(M) = 2M^{2}\frac{\AVG{N_{2}(M)}}{\AVG{N}^{2}},
\end{equation}
where $N_{2}(M)$ denotes the total number of particle pairs in all of the $M^{2}$ bins in an event.
Then the statistical uncertainties can be calculated using the standard error propagation:
\begin{equation}
    \frac{\sigma_{F_{2}}}{|F_{2}|} = \sqrt{\frac{(\sigma_{N_{2}})^{2}}{\AVG{N_{2}}^{2}}
    + 4\frac{(\sigma_{N})^{2}}{\AVG{N}^{2}}
    - 4\frac{(\sigma_{N_{2}N})^{2}}{\AVG{N}\AVG{N_{2}}}}.
\end{equation}

\begin{figure}[!ht]
    \centering
    \includegraphics[width=.4\textwidth]{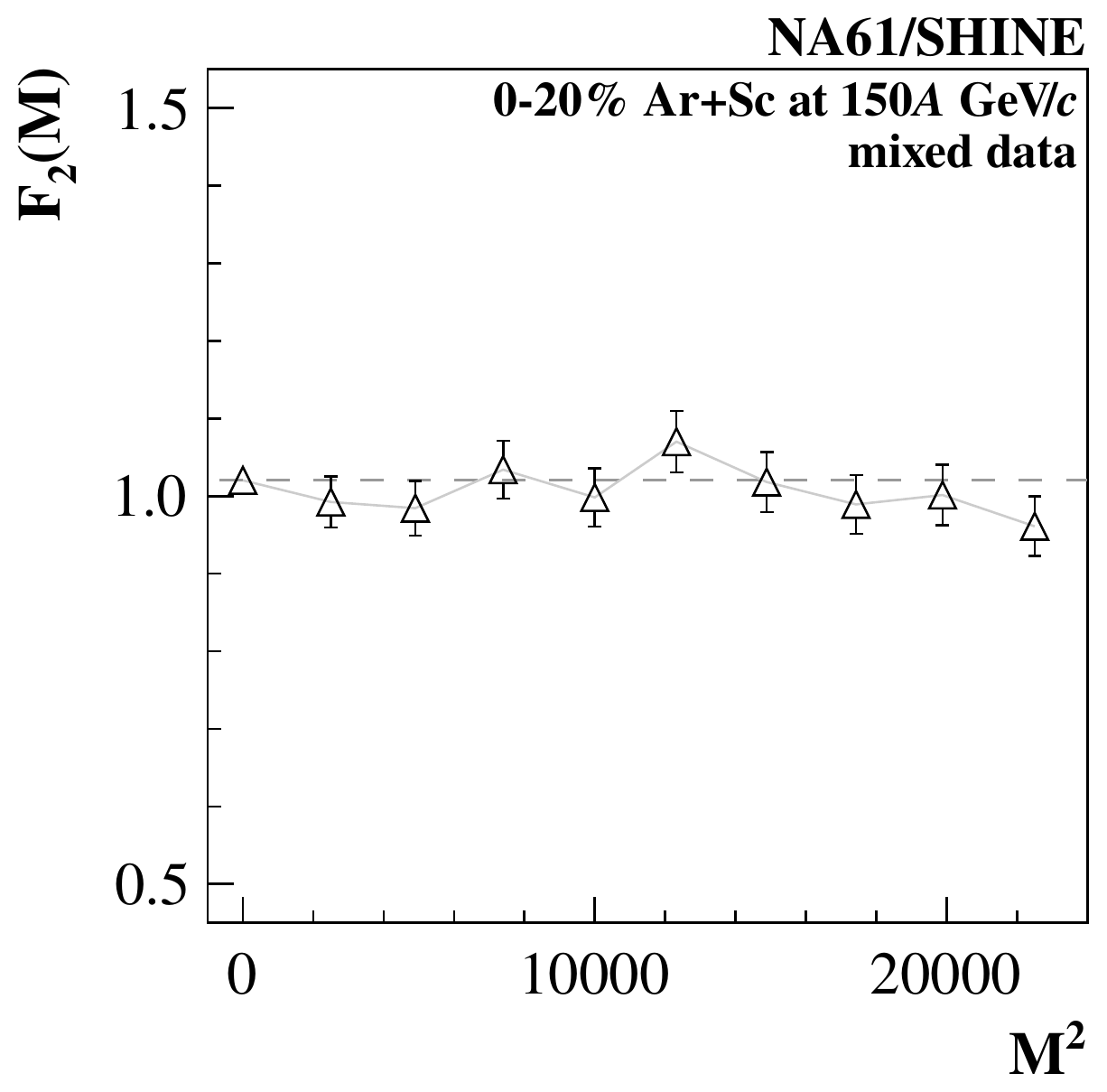}\qquad
    \includegraphics[width=.4\textwidth]{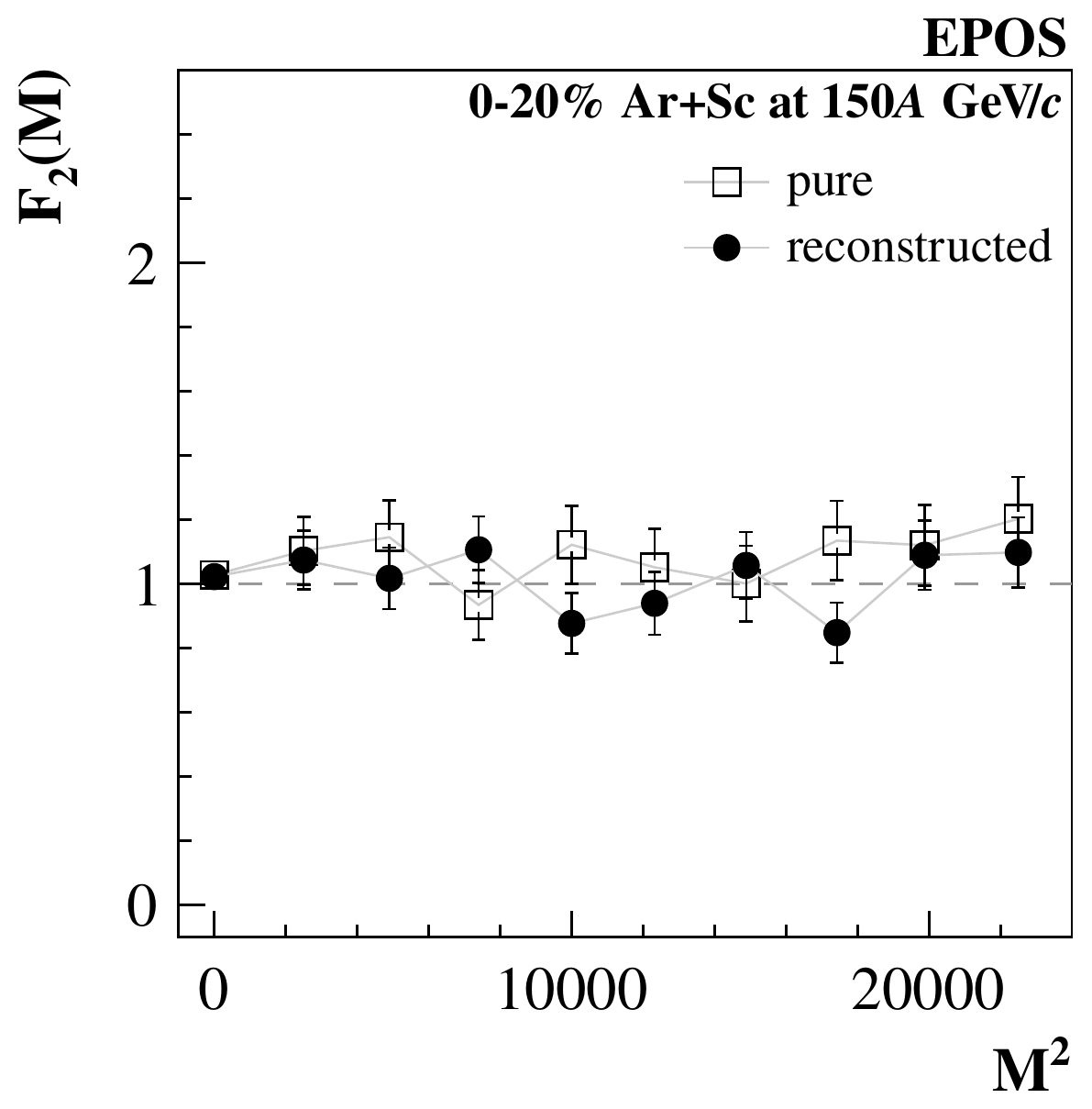}
    \caption{
        Results on the dependence of the scaled factorial moment of proton
        multiplicity distribution on the number of subdivisions in cumulative
        transverse momentum space for mixed events (\emph{left}) as well as
        events generated with \EposLong model before and after reconstruction
        (\emph{right}).
    }
    \label{fig:uncert}
\end{figure}

The \emph{left} plot in Fig.~\ref{fig:uncert} shows $F_{2}(M)$ results for the
mixed data set (see Sec.~\ref{sec:results} for details). As expected,
the $F_{2}$ values are independent of $M^{2}$. Deviation of the points from
the value for the first point (marked with the dashed line) is approximately
\chindf = 7.7/9, which validates the values of statistical uncertainties.

Final results presented in Sec.~\ref{sec:results} are not corrected for
possible biases. Systematic uncertainty was estimated by comparing
results for pure \EposLong and \EposLong subjected to the detector simulation,
reconstruction and data-like analysis as shown in Fig.~\ref{fig:uncert} (\emph{right}).
Their differences are significantly smaller than statistical uncertainties
(\chindf = 9.7/9) of the experimental data and increase with $M^{2}$
up to the order of 0.1 at large $M^{2}$ values.
Note that protons generated by \EposLong do not show significant correlation
in the transverse momentum space, see Sec.~\ref{sec:models}. In this case,
the momentum resolution does not affect the results significantly.

In the case of the critical correlations, the impact of the momentum resolution
may be significant, see Ref.~\cite{Samanta:2021usk} and Sec.~\ref{sec:models} for detail.
Thus a comparison with models including short-range correlations in the
transverse momentum space requires smearing of their momenta according to the
the experimental resolution, which can be approximately parametrized as:
\begin{equation}\label{eq:smearing}
\begin{split}
    p_{x}^{\text{smeared}} &= p_{x}^{\text{original}} + \delta p,\\
    p_{y}^{\text{smeared}} &= p_{y}^{\text{original}} + \delta p,
\end{split}
\end{equation}
where $\delta p$ is randomly drawn from a Gaussian distribution with $\sigma = 3$~\MeVc.

Uncertainties on final results presented in Sec.~\ref{sec:results} correspond to
statistical uncertainties.

\section{Results}
\label{sec:results}

This section presents results on second-order scaled factorial moments
(Eq.~\ref{eq:scaled-factorial-moments}) of~$\approx60\%$ randomly selected
protons (losses due to proton misidentification) with momentum smeared due
to reconstruction resolution (Eq.~\ref{eq:smearing}) produced within the
acceptance maps defined in Sec.~\ref{sec:maps} by strong and electromagnetic
processes in 0-5\%, 5-10\%, 10-15\%, 15-20\% and
0-20\% most central \ArSc collisions at 150\AGeVc.
The results are shown as a function of the number of subdivisions in
transverse momentum space -- the so-called intermittency analysis.
The analysis was performed for
cumulative and original transverse momentum components.
Independent data sets were used to calculate results for each subdivision.

Uncertainties correspond to statistical ones.
Biases estimated using the \EposLong~\cite{Werner:2008zza} model (see Sec.~\ref{sec:models})
are significantly smaller than statistical uncertainties of the experimental data.

\subsection{Subdivisions in cumulative transverse momentum space}
\label{sec:results_cumulative}

\begin{figure}[!ht]
    \centering
    \includegraphics[width=.31\textwidth]{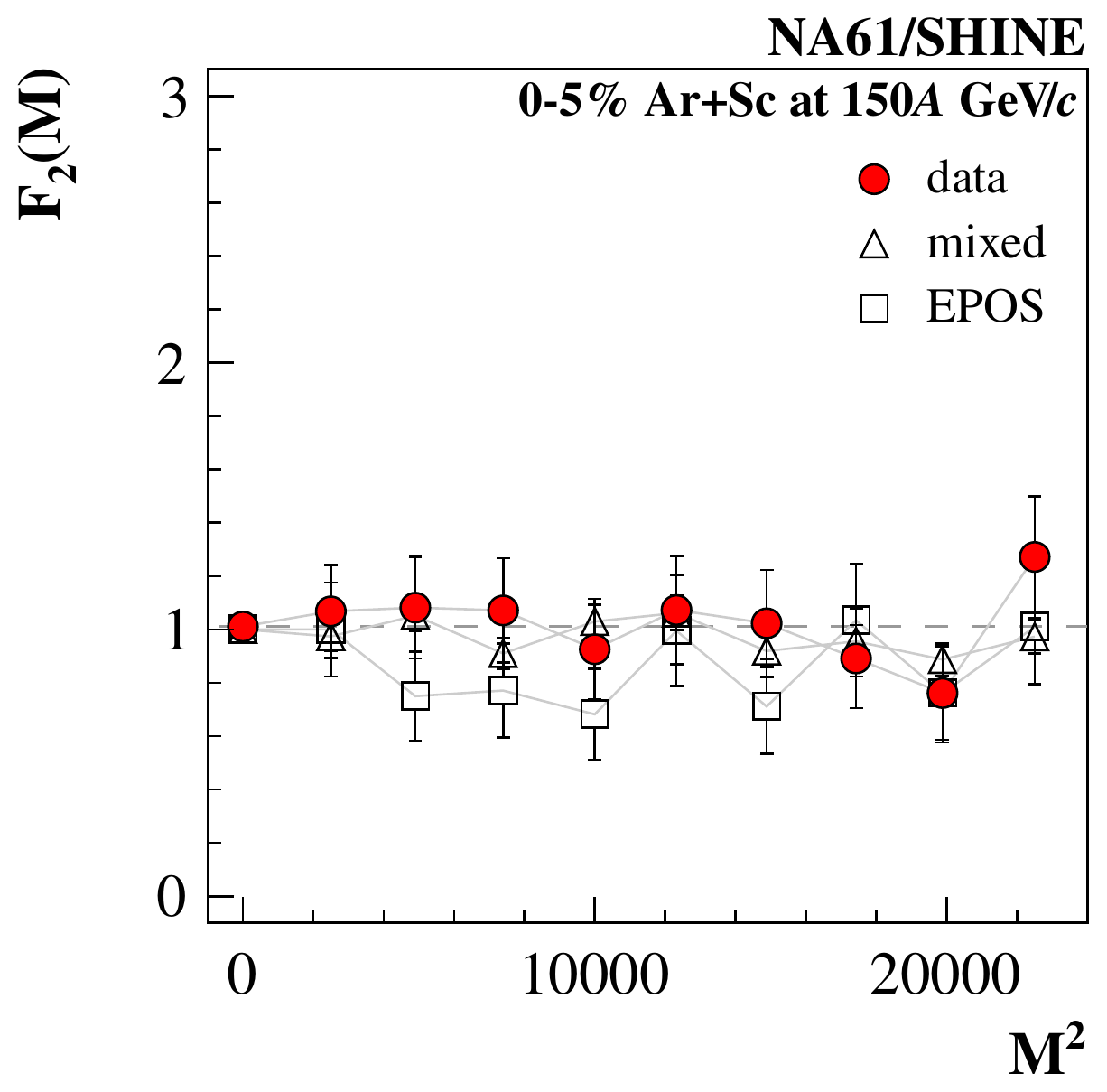}\hfill
    \includegraphics[width=.31\textwidth]{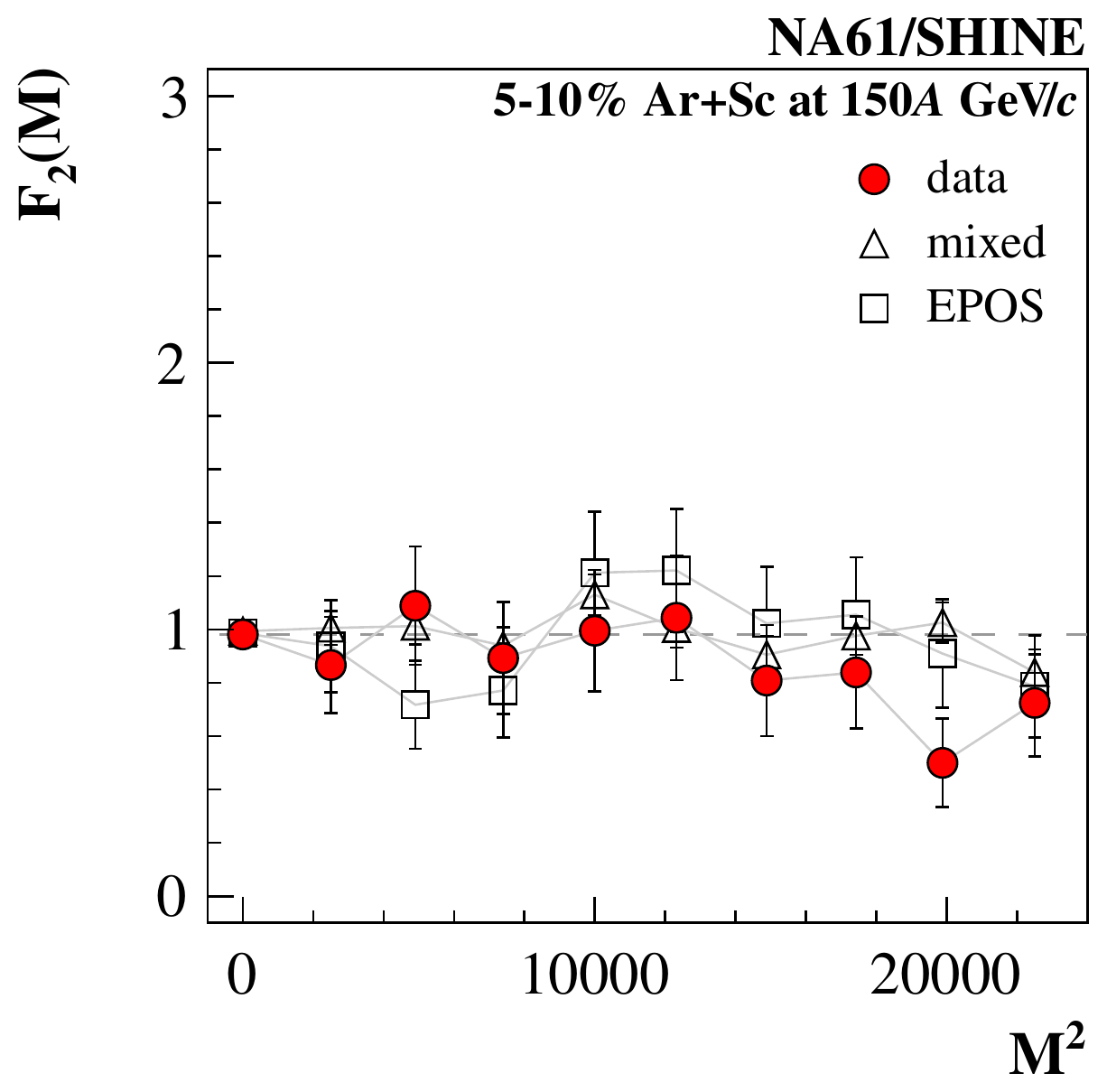}\hfill
    \includegraphics[width=.31\textwidth]{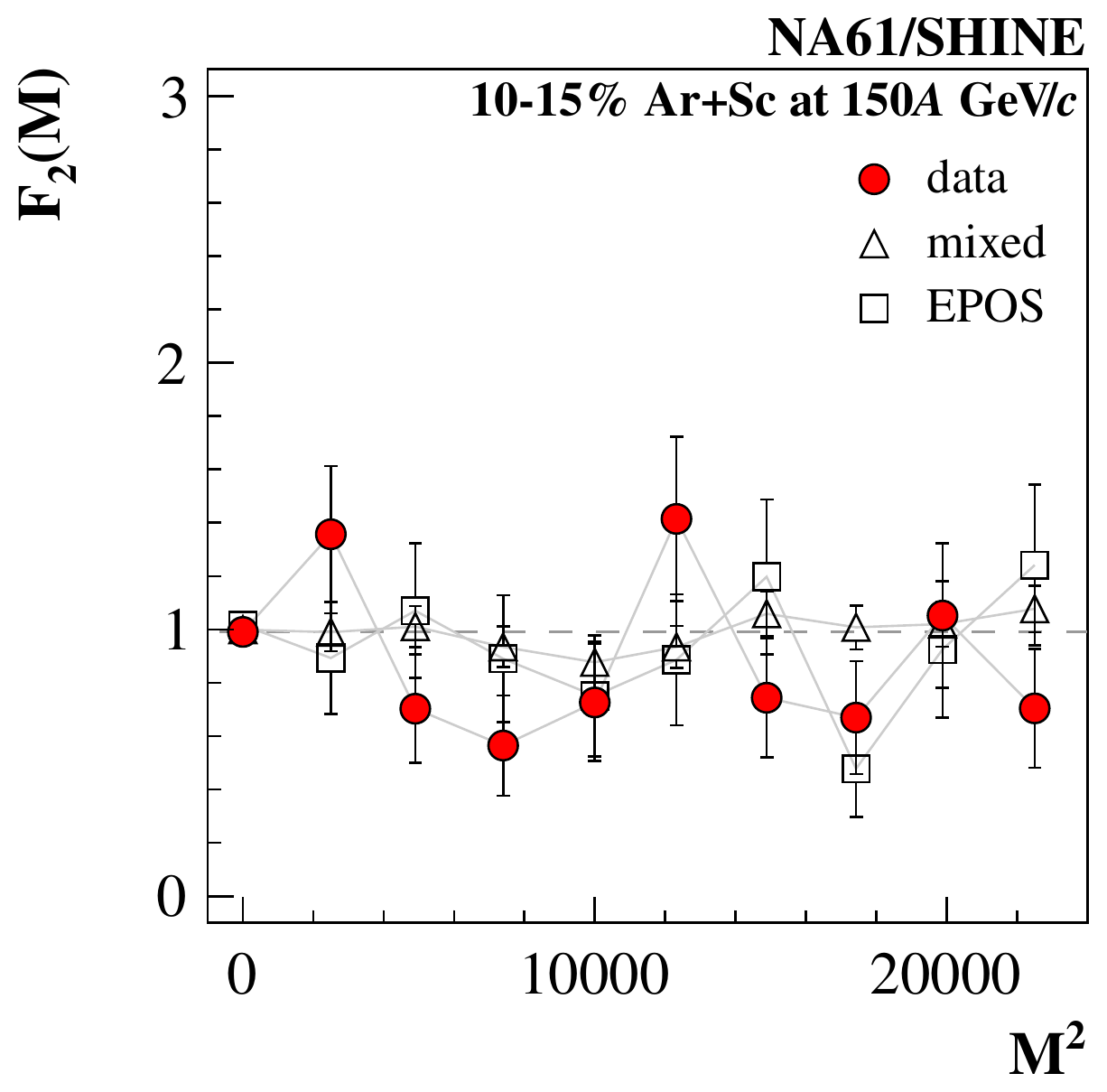}\\
    \includegraphics[width=.31\textwidth]{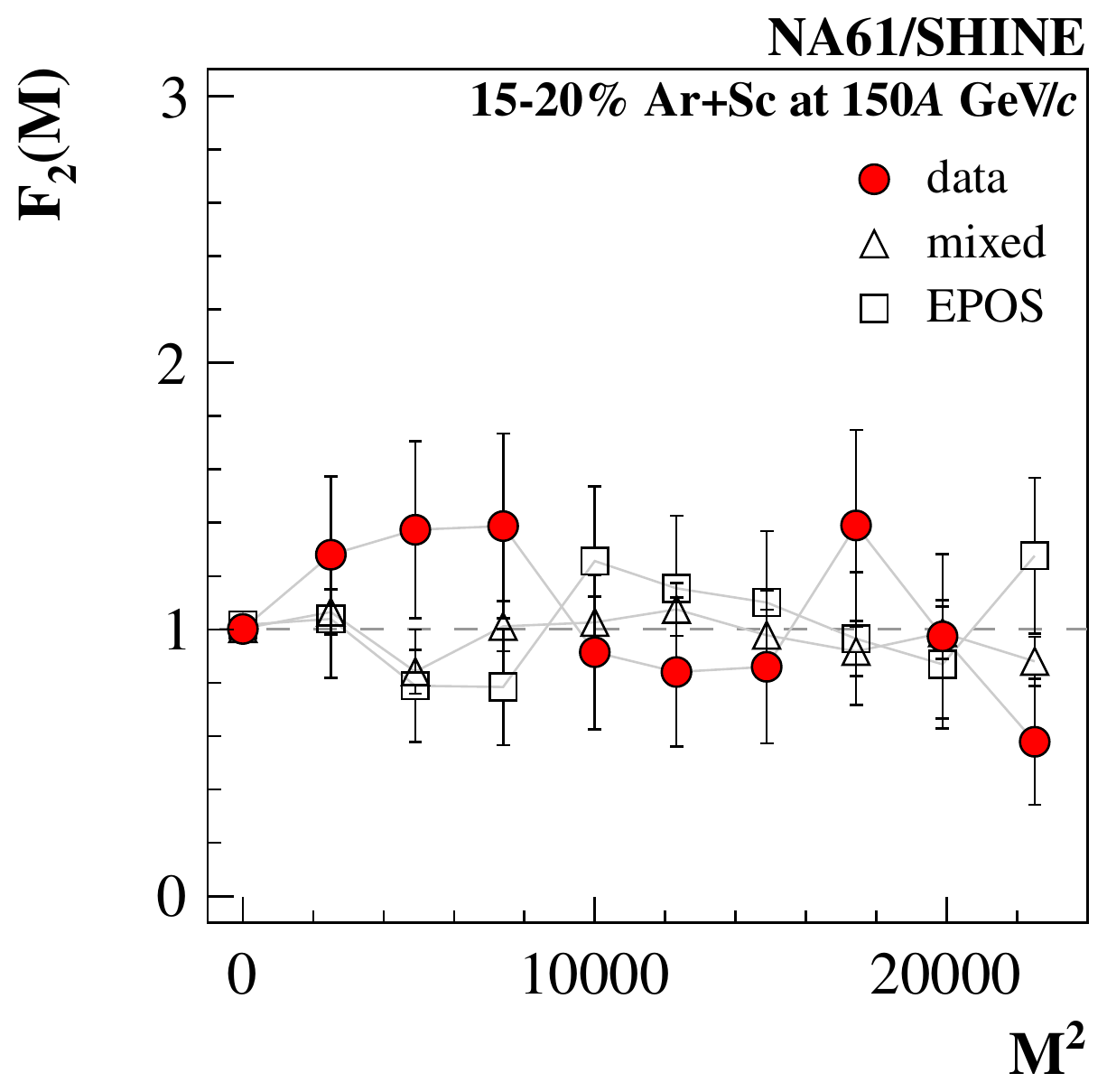}\qquad
    \includegraphics[width=.31\textwidth]{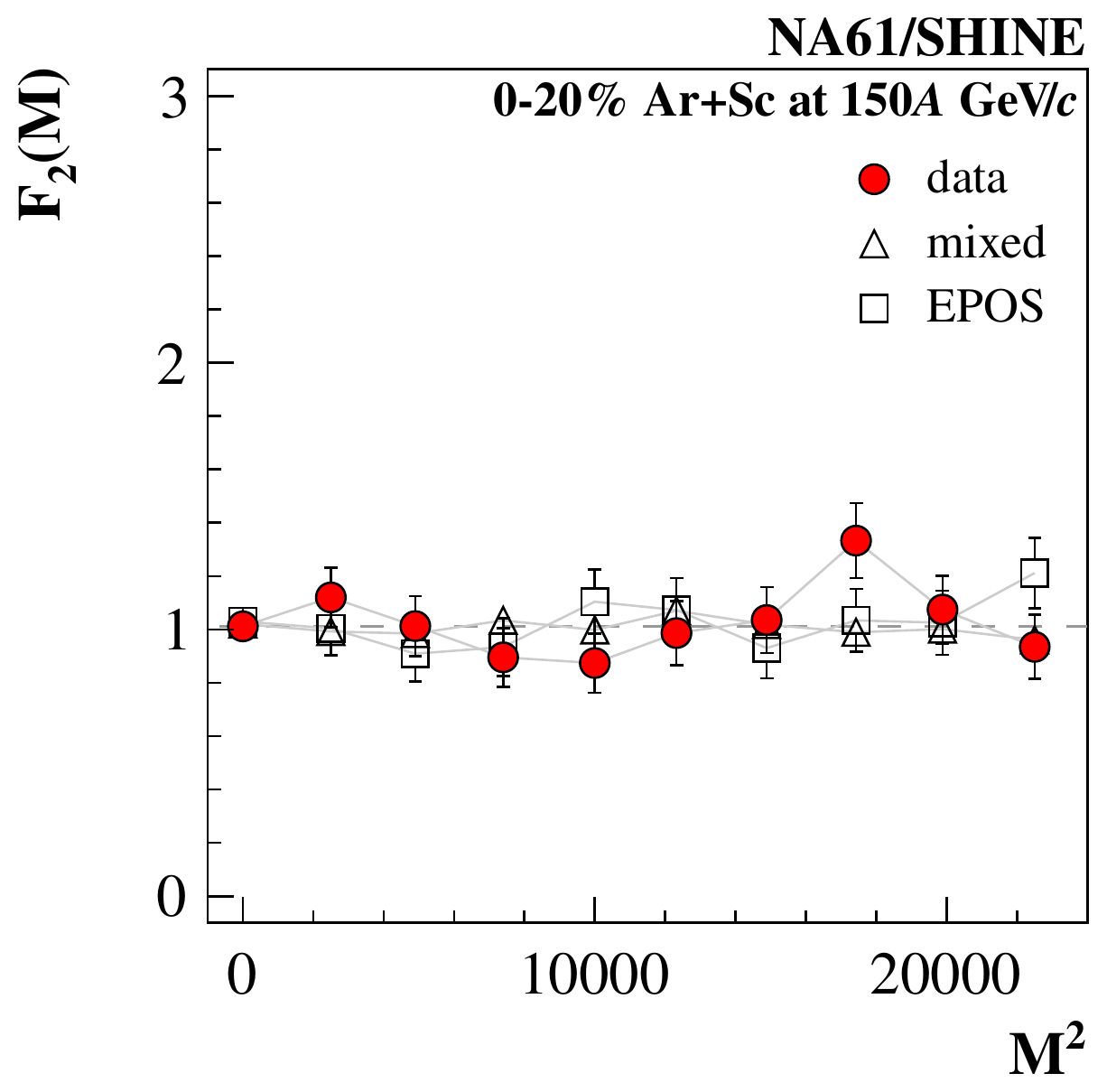}
    \caption{
        Results on the dependence of the scaled factorial moment of proton multiplicity distribution on the
        number of subdivisions in cumulative transverse momentum space $M^{2}$ for $1^{2} \leq M^{2} \leq 150^2$.
        Closed circles indicate the experimental data. For comparison, corresponding results for mixed events
        (open triangles) and the \EposLong model  (open squares) are also shown.
        Results for five centrality selections of events are presented in different panels. Only statistical
        uncertainties are indicated.
    }
    \label{fig:results-fine}
\end{figure}

\begin{figure}[!ht]
    \centering
    \includegraphics[width=.31\textwidth]{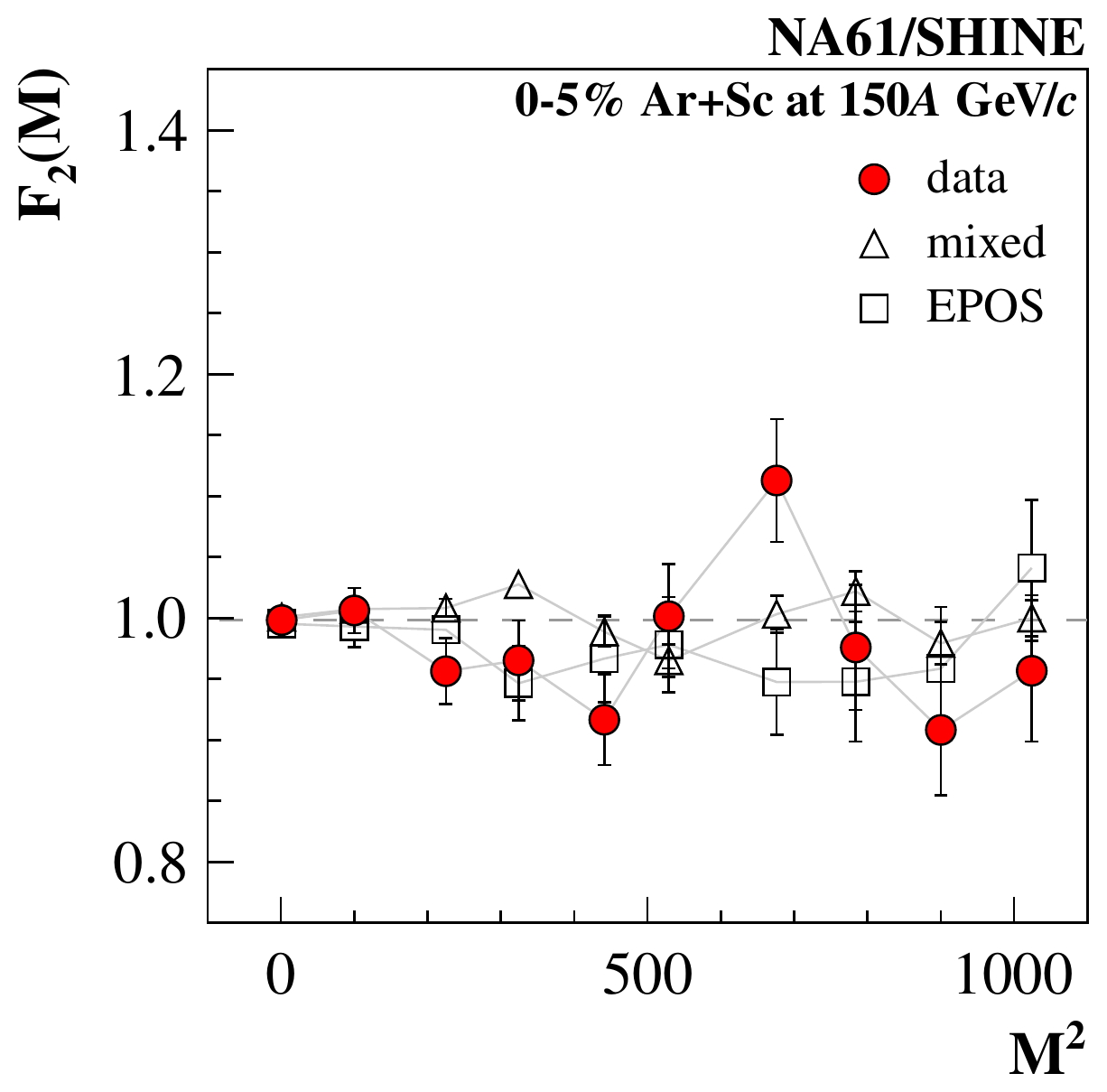}\hfill
    \includegraphics[width=.31\textwidth]{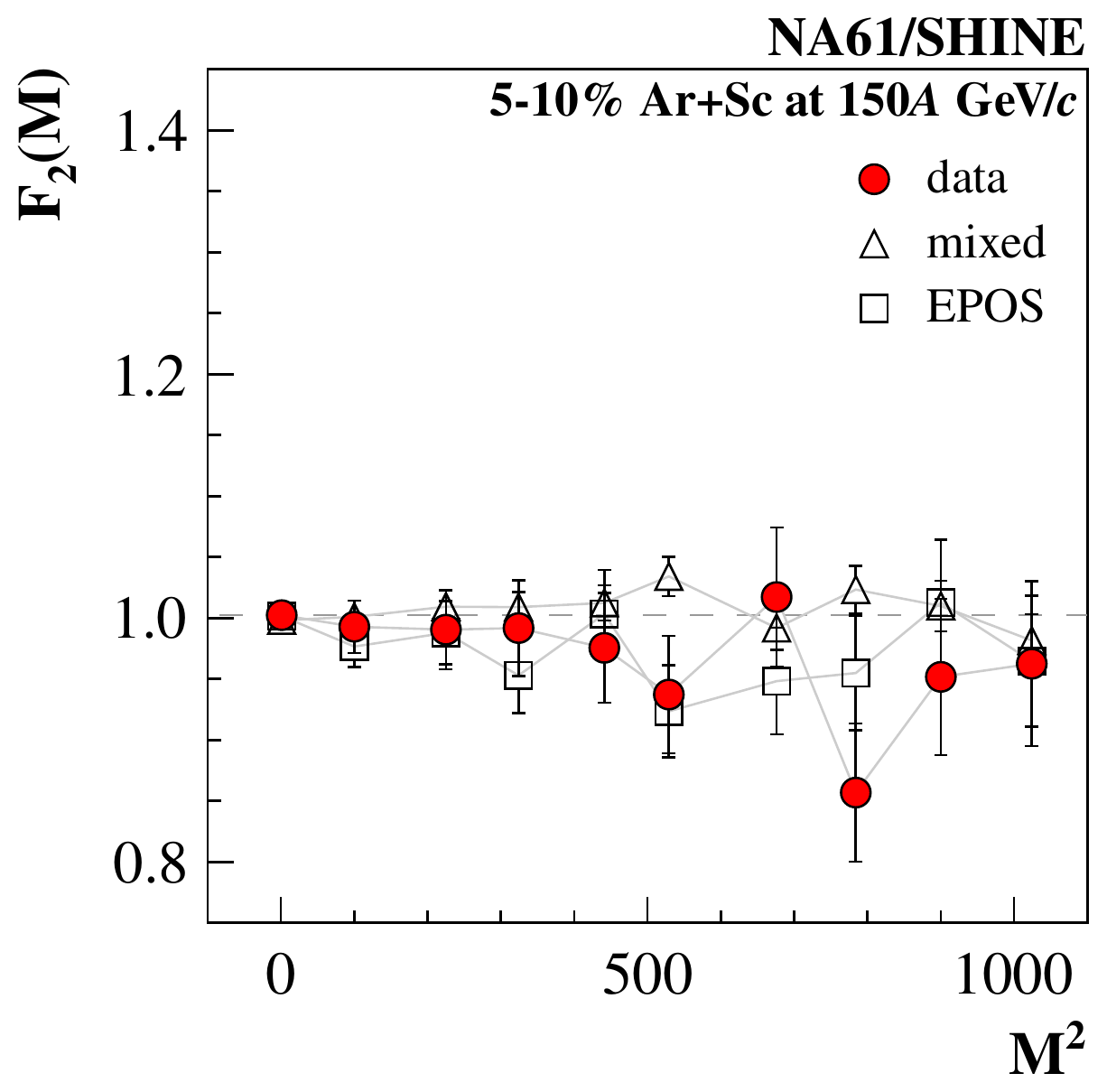}\hfill
    \includegraphics[width=.31\textwidth]{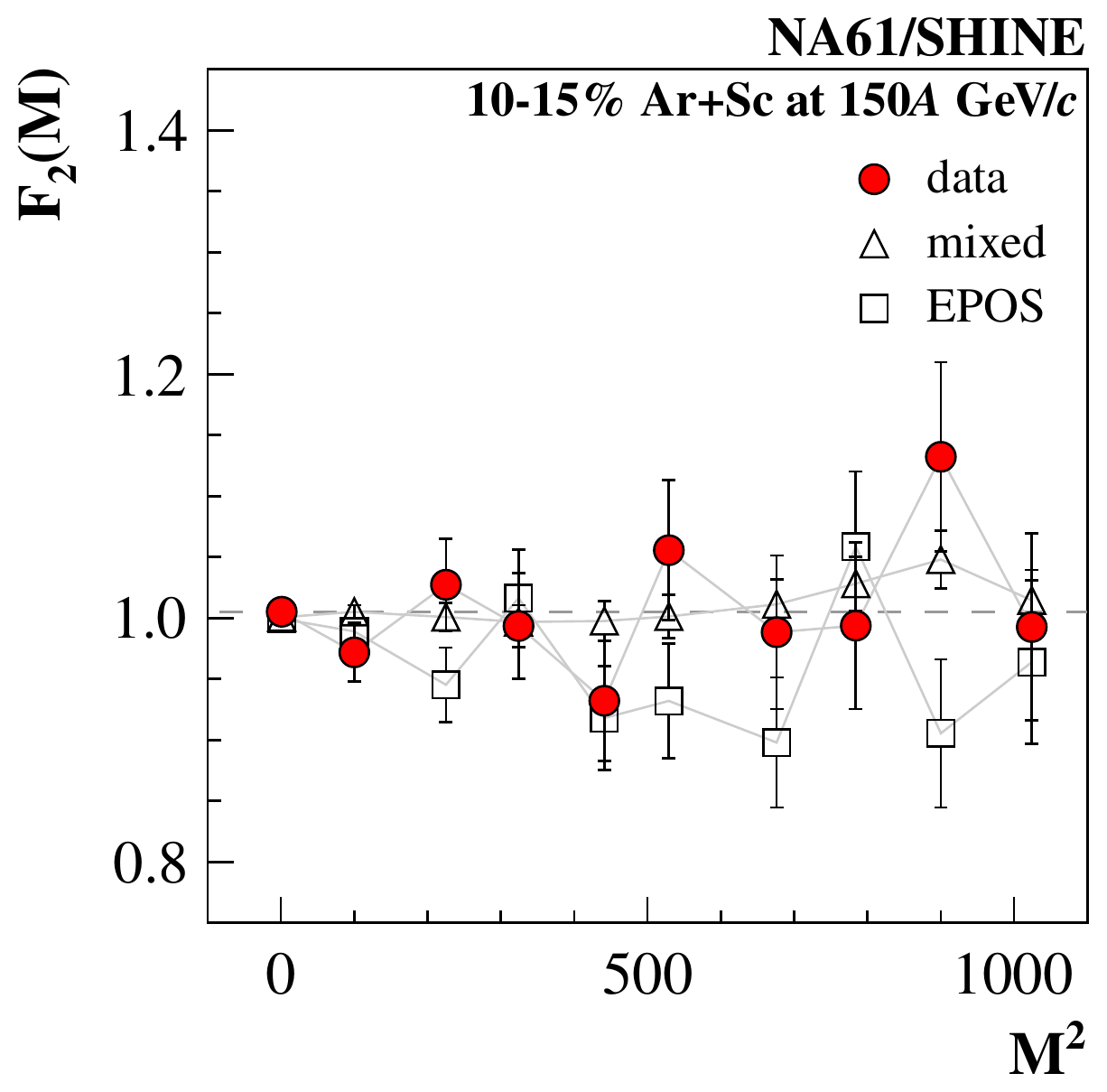}\\
    \includegraphics[width=.31\textwidth]{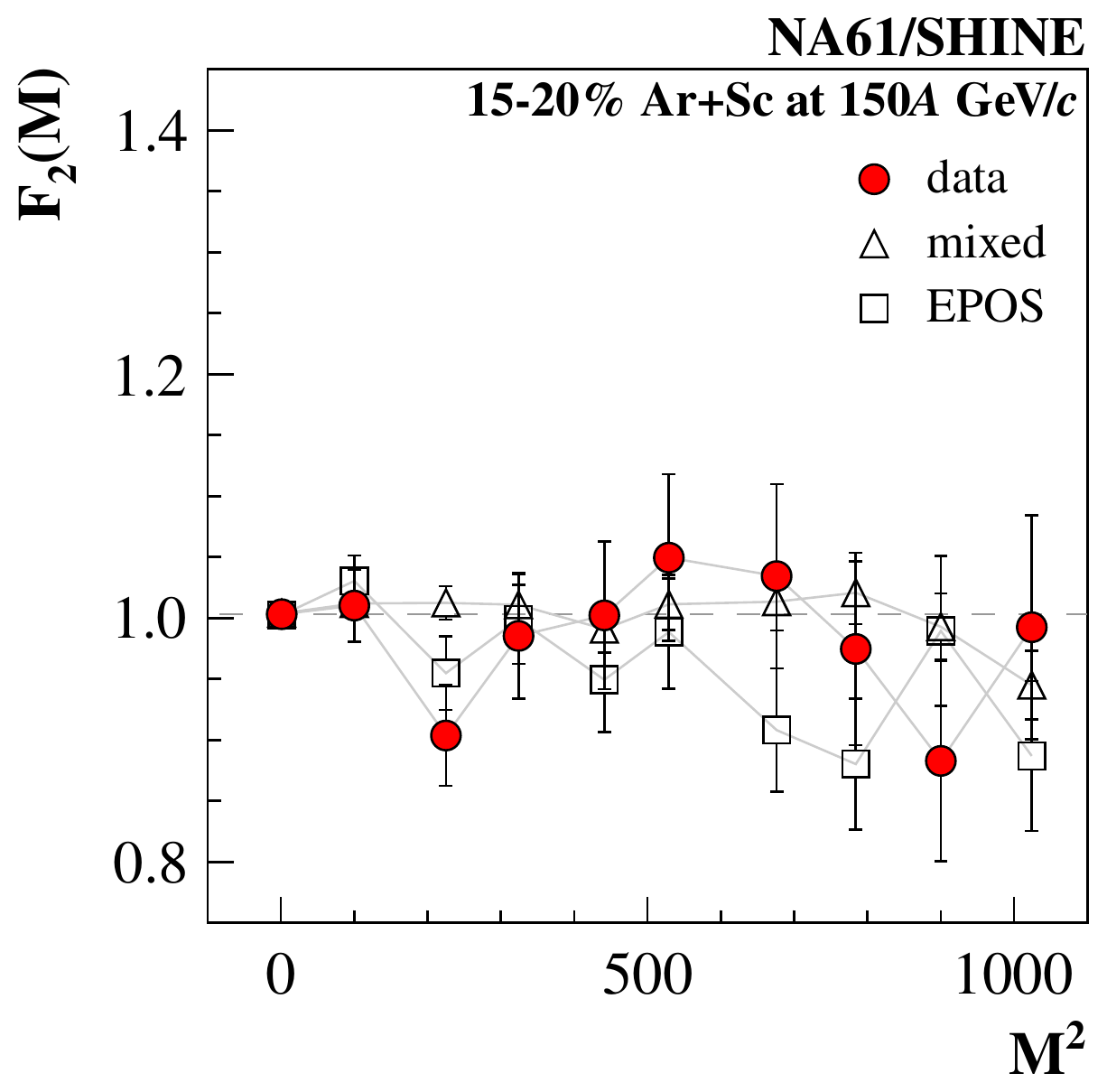}\qquad
    \includegraphics[width=.31\textwidth]{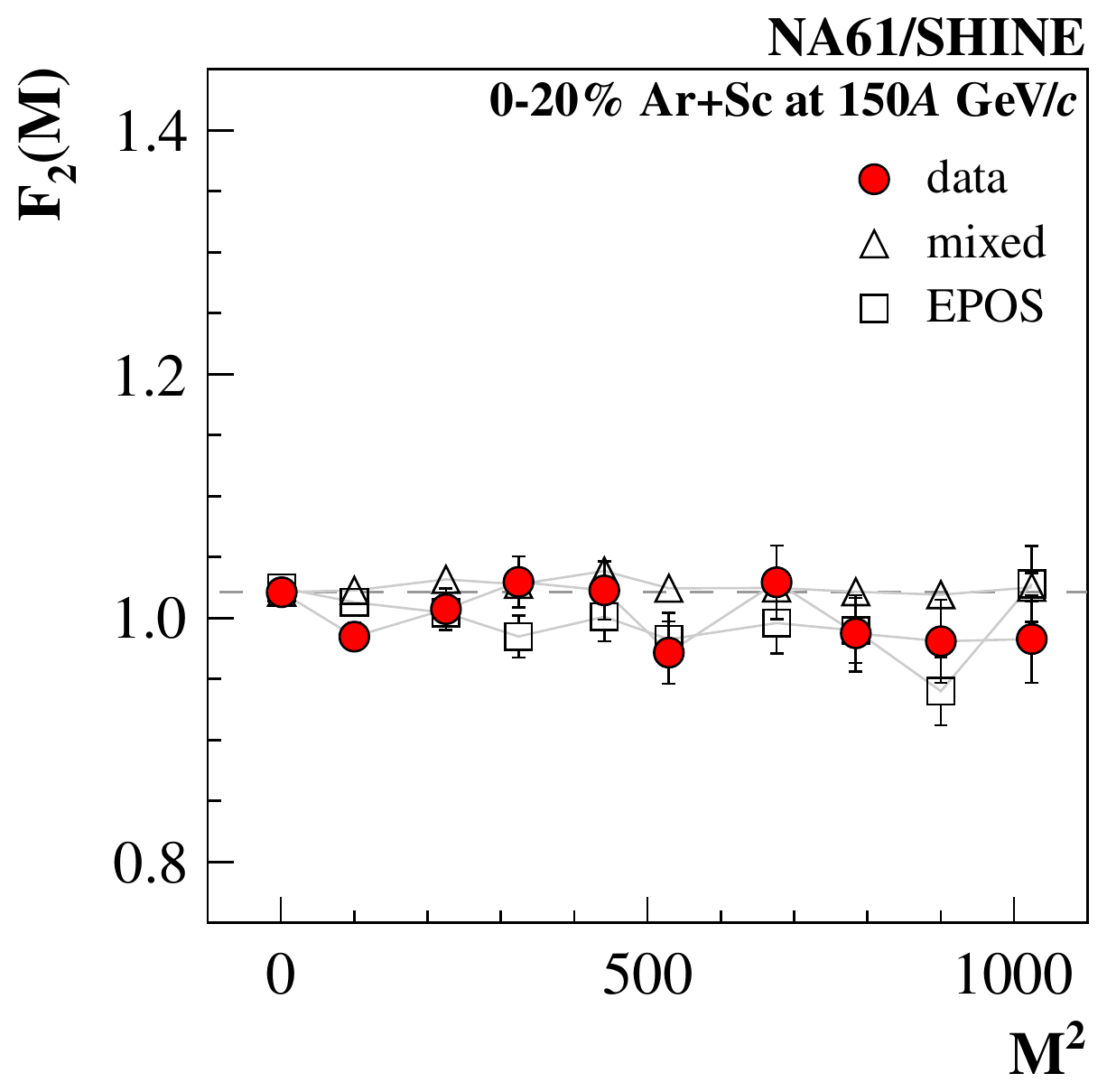}
    \caption{
        Results on the dependence of the scaled factorial moment of proton multiplicity distribution on
        the number of subdivisions in cumulative transverse momentum space $M^{2}$ for $1^{2} \leq M^{2} \leq 32^2$.
        Closed circles indicate the experimental data. For comparison, corresponding results for mixed events
        (open triangles) and the \EposLong model (open squares) are also shown.
        Results for five centrality selections of events are presented in different panels.
        Only statistical uncertainties are indicated.
    }
    \label{fig:results-coarse}
\end{figure}

Figures~\ref{fig:results-fine} and~\ref{fig:results-coarse} present the dependence of the factorial
moment on the number of subdivisions in cumulative-transverse momentum space for the maximum
subdivision number of $M^{2} = 150^2$ and $M^{2} = 32^2$, respectively. The latter, coarse subdivision,
was introduced to limit the effect of experimental momentum resolution;
see Ref.~\cite{Samanta:2021usk} and below for details.
The experimental results are shown for five different selections of events with respect to centrality.
As a reference, the corresponding results for mixed events are also shown.
The mixed data set is constructed by randomly swapping particles from different events such that each
particle in a mixed event comes from different recorded events.
Note that by construction,
the multiplicity distribution of protons in mixed events for $M^{2} = 1^2$ is equal to the corresponding
distribution for the data. In the mixed events, protons are uncorrelated in the transverse momentum space.
Therefore for them, the scaled factorial moment is independent of $M^{2}$, $F_2(M) = F_2(1^2)$.
The experimental results do not show any significant dependence on $M^{2}$.
The obtained values are consistent with the value of the first data point (dashed line)
with \chindf = 8.7/9 on average for the fine binning (Fig.~\ref{fig:results-fine})
and \chindf = 11.4/9 for the coarse binning (Fig.~\ref{fig:results-coarse}).
There is no indication of the critical fluctuations for selected protons.

\subsection{Subdivisions in transverse momentum space}

\begin{figure}[!ht]
    \centering
    \includegraphics[width=.31\textwidth]{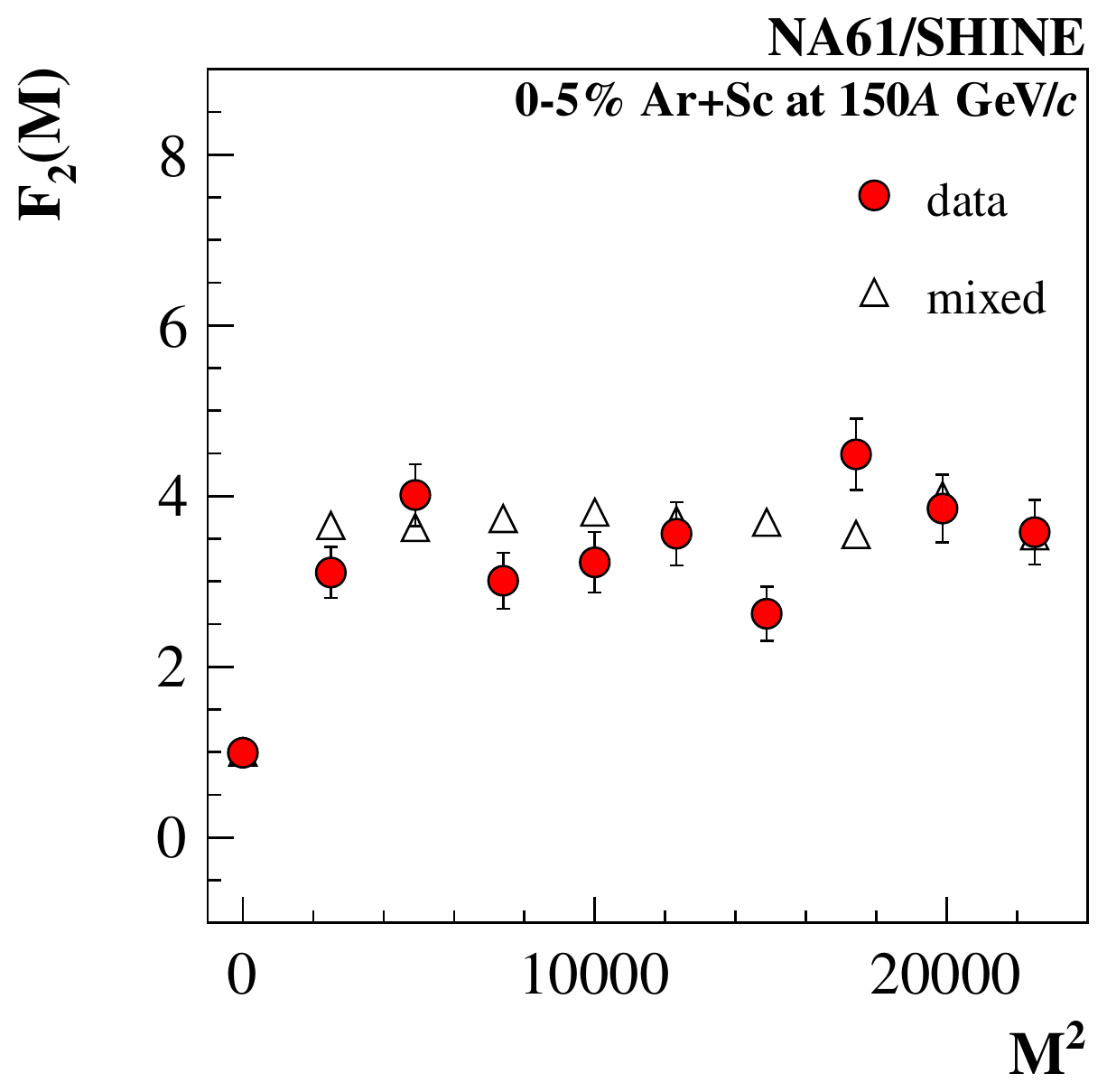}\hfill
    \includegraphics[width=.31\textwidth]{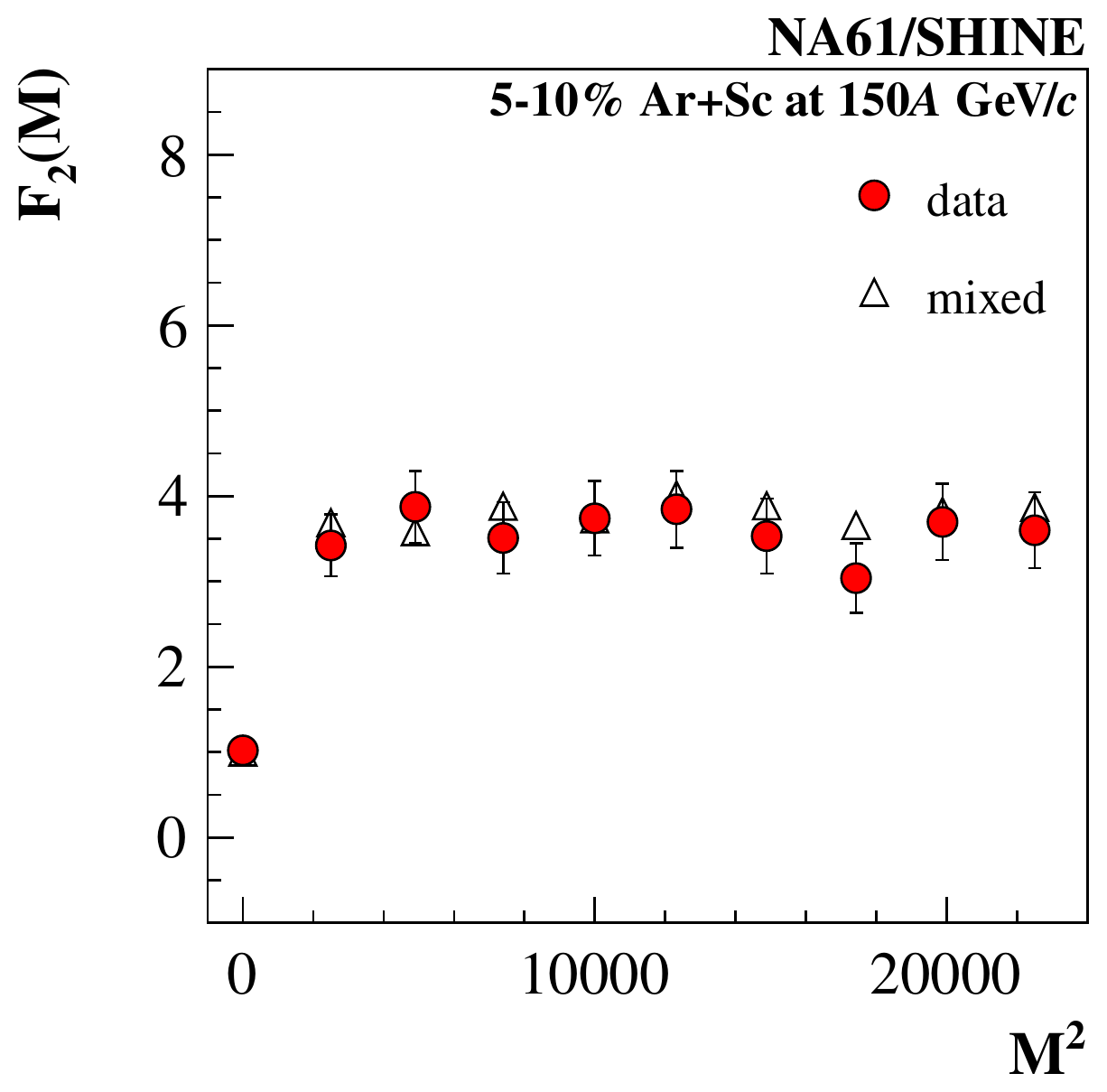}\hfill
    \includegraphics[width=.31\textwidth]{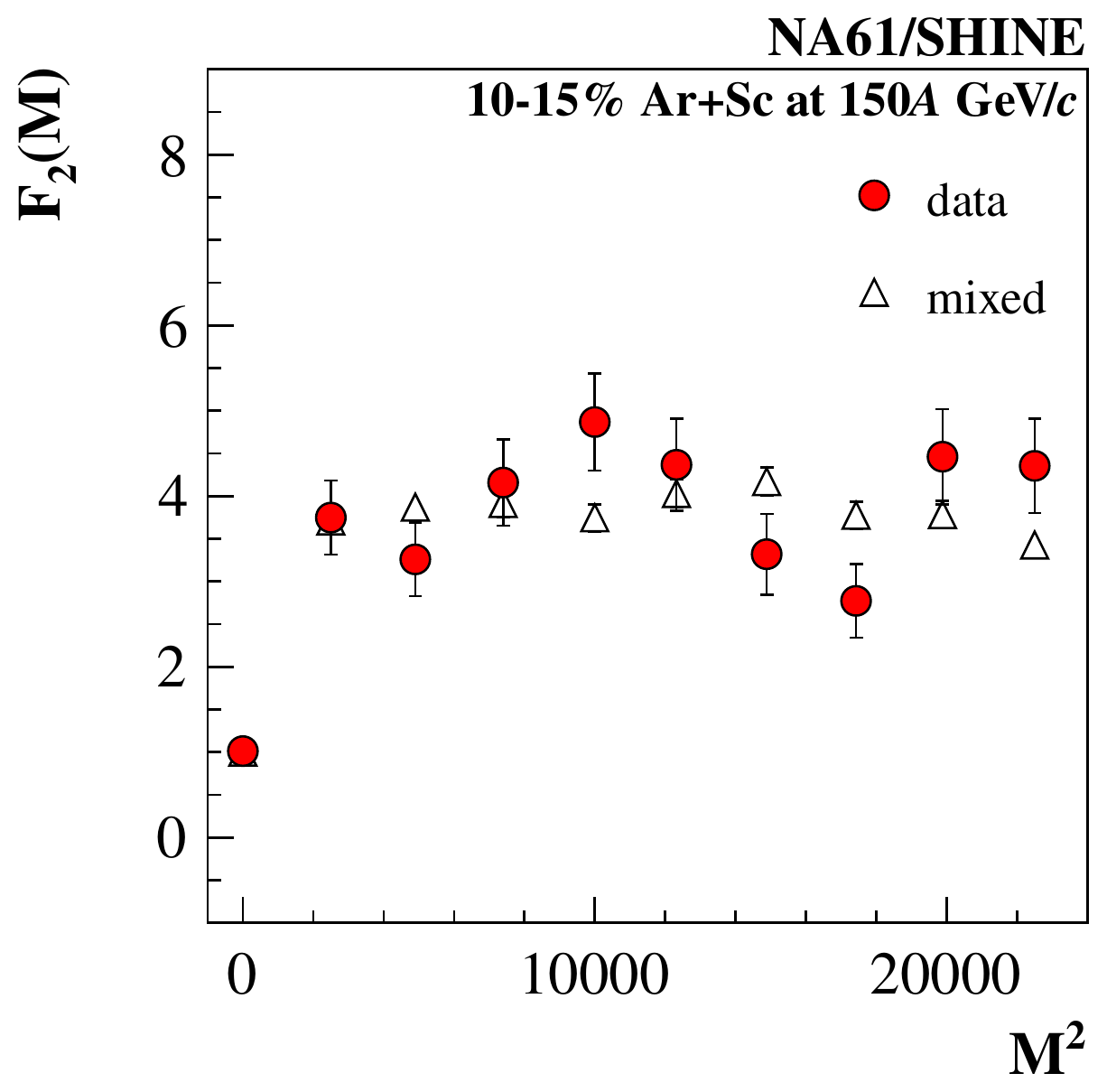}\\
    \includegraphics[width=.31\textwidth]{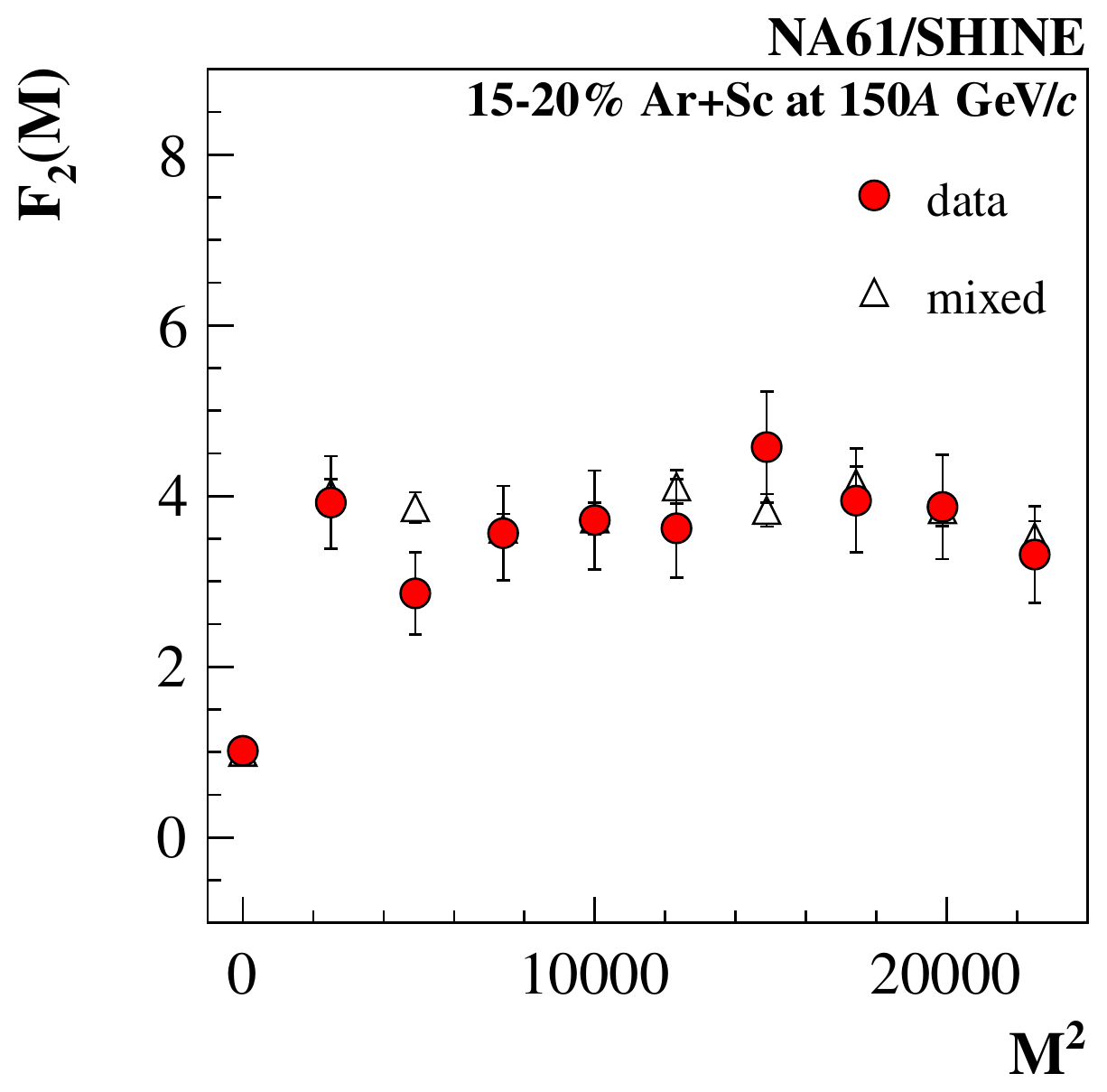}\qquad
    \includegraphics[width=.31\textwidth]{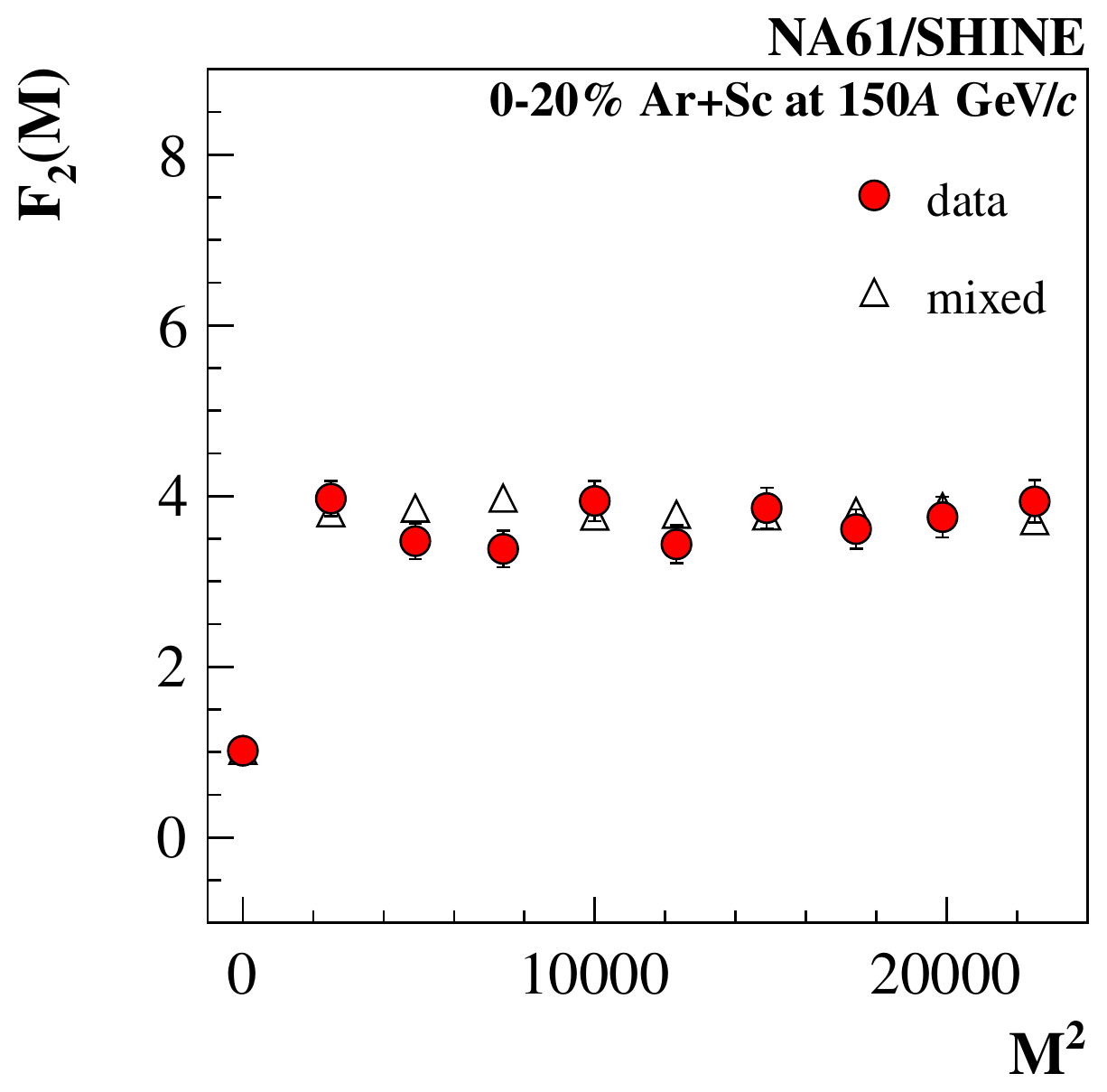}
    \caption{
       Results on the dependence of the scaled factorial moment of proton multiplicity distribution
       on the number of subdivisions in  transverse momentum space $M^{2}$ for $1^{2} \leq M^{2} \leq 150^2$.
        Closed circles indicate the experimental data. For comparison, corresponding results for mixed events
        (open triangles) are also shown.
        Results for five centrality selections of events are presented in different panels.
        Only statistical uncertainties are indicated.
    }
    \label{fig:results-noncum}
\end{figure}

Figure~\ref{fig:results-noncum} presents the results which
correspond to the results shown in Fig.~\ref{fig:results-fine}, but subdivisions are done in the
original transverse momentum space.
By construction, $F_2(1^2)$ values are equal for subdivisions in cumulative transverse momentum
space and transverse momentum space. But for the latter,
$F_2(M)$ strongly depends on $M^{2}$. This dependence is primarily due to
non-uniform shape of the single-particle transverse momentum distributions,
see Sec.~\ref{sec:sfm_cumulative}. It can be accounted for by comparing the results for
the experimental data with the corresponding results obtained for the mixed events.
There is no significant difference between the two, which confirms the previous conclusion of no
indication of significant critical fluctuations.

\clearpage

\section{Comparison with models}
\label{sec:models}

This section presents a comparison of the experimental results with
two models. The first one, \EposLong~\cite{Werner:2008zza}, takes into account numerous sources
of particle correlations, in particular, conservation laws and resonance decays,
but without critical fluctuations.
The second one, the Power-law Model~\cite{Czopowicz:2023}, produces particles correlated by the
power law together with fully uncorrelated particles.

\subsection{\Epos}
\label{sec:models-epos}

For comparison, almost $20\cdot10^{6}$ minimum bias \ArSc events have been generated with
\EposLong. Signals from the \NASixtyOne detector were simulated with \GeantThree software, and the
recorded events were reconstructed using the standard \NASixtyOne procedure. Number of analyzed
events is shown in Table~\ref{tab:events}.

To calculate model predictions (pure \Epos), all generated central events
were analyzed. Protons and proton pairs within the single-particle and
two-particle acceptance maps were selected. Moreover, 60\% of accepted protons were
randomly selected for the analysis to take into account the effect of the
proton misidentification.

Results for the reconstructed \Epos events were obtained as follows.
The model events were required to have the reconstructed primary vertex.
Selected protons and proton pairs (matching to the generated particles was used
for identification) were subject to the same cuts as used for the experimental
data analysis, see Sec.~\ref{sec:analysis}.

\begin{figure}[!ht]
    \centering
    \includegraphics[width=.4\textwidth]{figures/results/EPOS.pdf}\qquad
    \includegraphics[width=.4\textwidth]{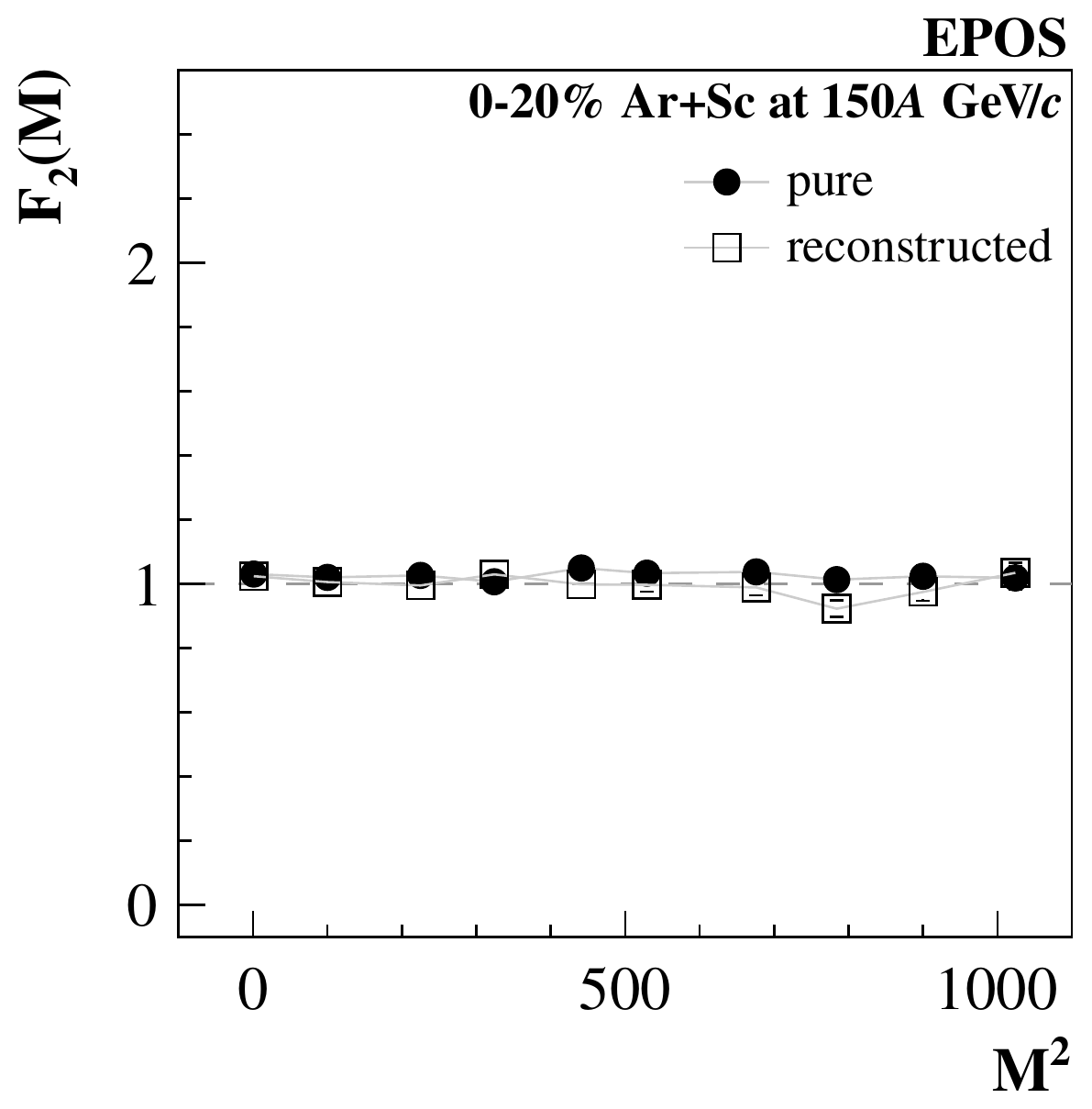}
    \caption{
        Results on the dependence of the scaled factorial moment of proton multiplicity distribution
        on the number of subdivisions in cumulative transverse momentum space
        for events generated with \EposLong for fine
        $M^{2} = 1^{2} ... 150^2$ (\emph{left}) and coarse,
        $M^{2} = 1^{2} ... 32^2$ (\emph{right}) subdivisions.
    }
    \label{fig:epos}
\end{figure}

The results for the pure and the reconstructed \Epos events are compared
in Fig.~\ref{fig:epos}. They agree for both fine and coarse subdivisions. As the
statistics of the \Epos events is several times higher than of the data, one concludes that
for the \Epos-like physics, the biases of
the experimental data are significantly smaller than the statistical uncertainties of the data.

Finally, the experimental results are compared with the pure \Epos predictions
in Figs.~\ref{fig:results-fine}, \ref{fig:results-coarse} and~\ref{fig:results-noncum}.
No significant differences are found.

\subsection{Power-law Model}
\label{sec:models-power}

Inspired by expectations of the power-law correlations between particles near
the critical point,
the Power-law Model was developed~\cite{Czopowicz:2023} to compare with the experimental result.
It generates momenta of uncorrelated and correlated protons with a given single-particle transverse
momentum distribution in events with a given multiplicity distribution. The model has two controllable
parameters:
\begin{enumerate}[(i)]
    \item fraction of correlated particles,
    \item strength of the correlation (the power-law exponent).
\end{enumerate}

The transverse momentum of particles is drawn from the input transverse momentum distribution.
Correlated-particle pairs' transverse momentum difference follows a power-law distribution:
\begin{equation}
    \rho(|\Delta\overrightarrow{p_{T}}|) \sim |\Delta\overrightarrow{p_{T}}|^{-\phi_{2}},
\end{equation}
where the exponent $\phi_{2} < 1$.
Azimuthal-angle distribution is assumed to be uniform.
The momentum component along the beamline, $p_{z}$, is calculated assuming a uniform rapidity
distribution from $-0.75$ to $0.75$ and proton mass.

\begin{figure}
    \centering
    \includegraphics[width=.4\textwidth]{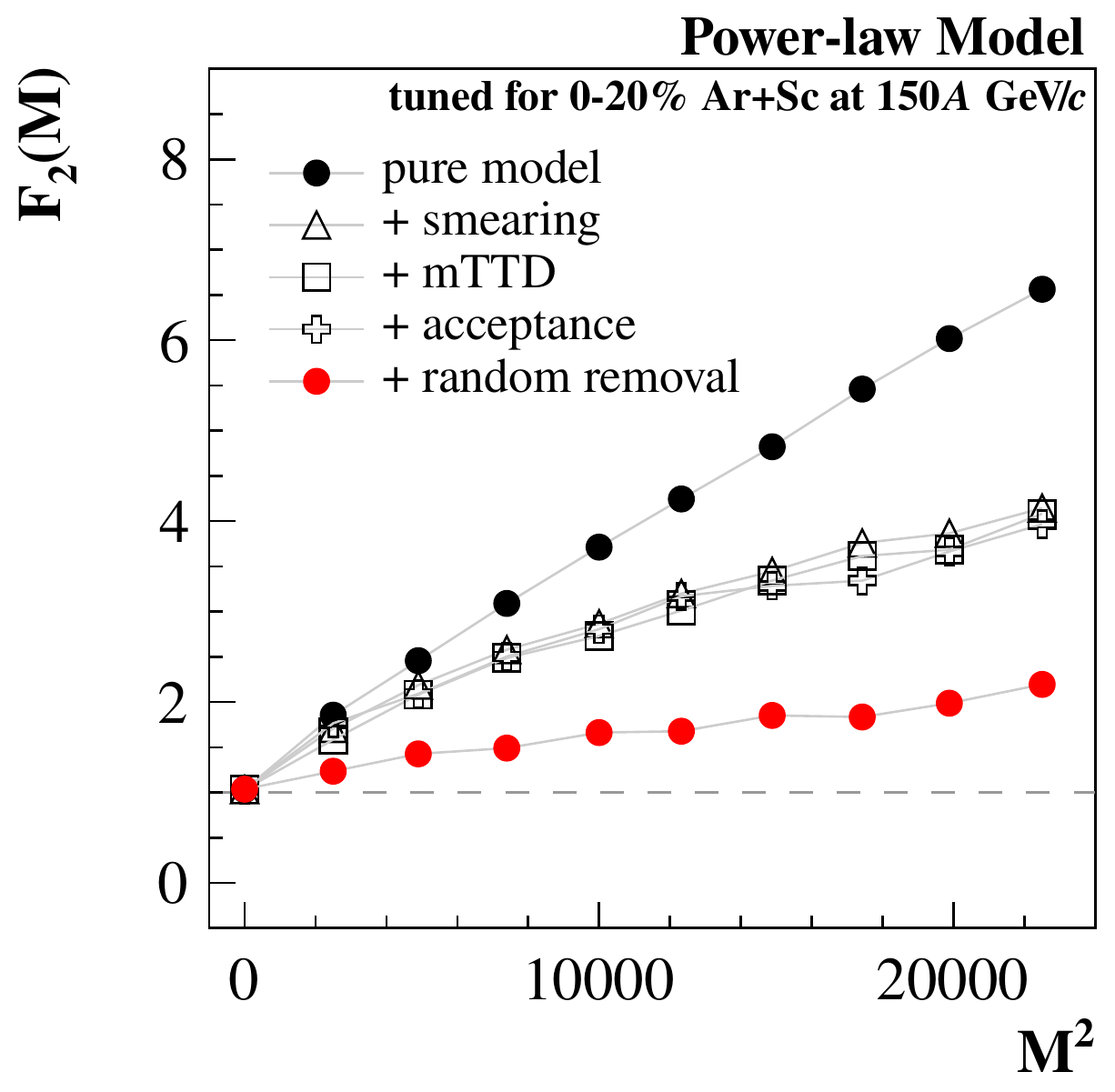}\qquad
    \includegraphics[width=.4\textwidth]{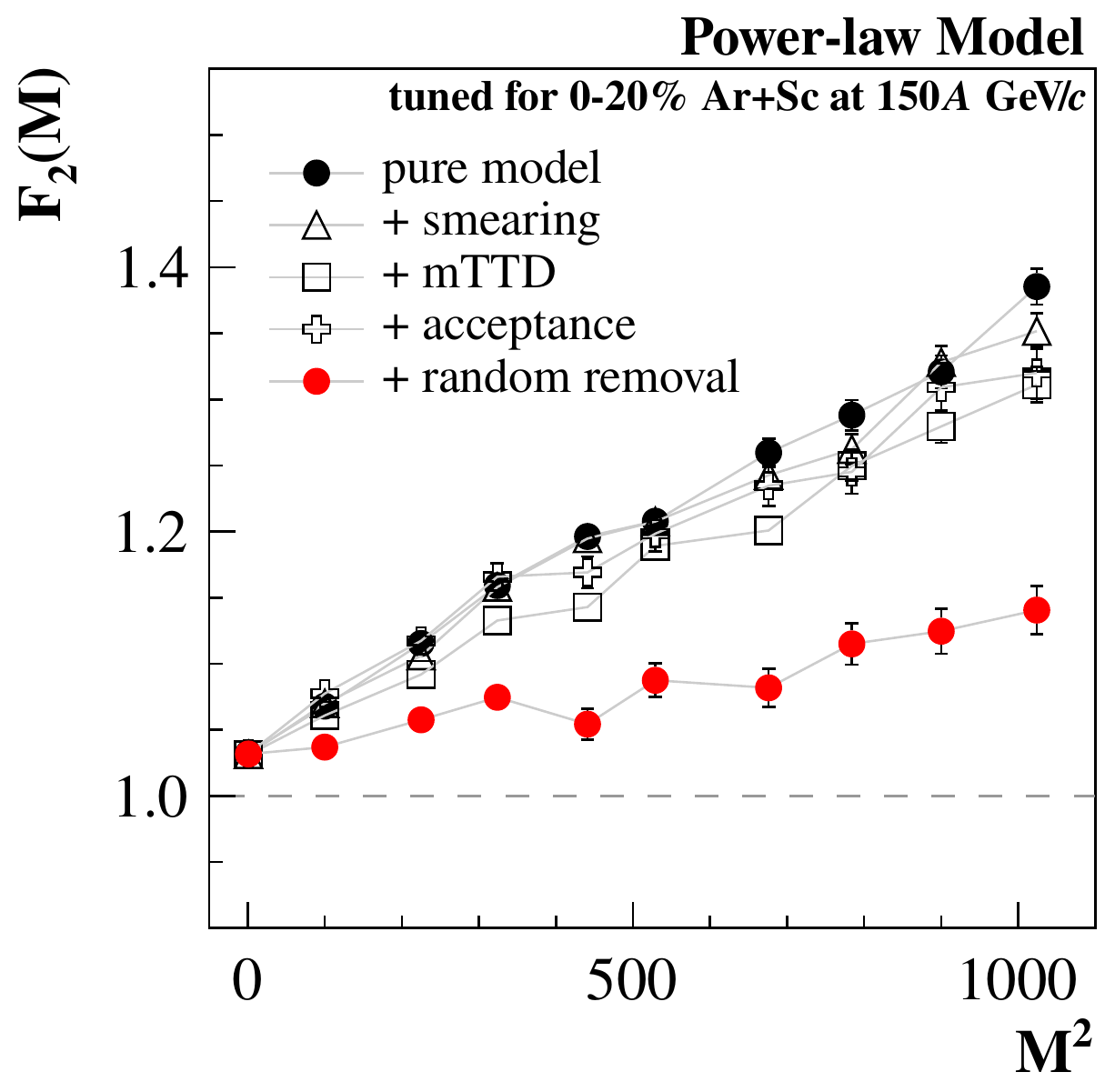}
    \caption{Dependence of the scaled factorial moment on the number of subdivisions in the
    cumulative transverse momentum for the Power-law Model with power-law exponent set to 0.80
    and fraction of correlated particles to 3\%.
    Each line presents a result with a different effect included separately, and the red
    circles all of them together.
    The results for the fine and coarse subdivisions are shown in the left and
    right panels, respectively.
    }
    \label{fig:model-biases}
\end{figure}

Many high-statistics data sets with multiplicity distributions identical to the experimental
data and similar inclusive transverse momentum distributions have been produced using the model.
Each data set has a different fraction of correlated particles (varying from 0 to 2\%) and a
different power-law exponent (varying from 0.00 to 0.95).
The following effects have been included:
\begin{enumerate}[(i)]
    \item Gaussian smearing of momentum components to mimic reconstruction resolution of the momentum,
    (see Eq.~\ref{eq:smearing}),
    \item random exchange of 40\% of correlated particles with uncorrelated ones to simulate 60\%
    acceptance of protons (preserves the desired multiplicity distribution, but requires generating more
    correlated pairs at the beginning),
    \item two-particle acceptance map, see Sec.~\ref{sec:maps},
    \item single-particle acceptance map, see Sec.~\ref{sec:maps}.
\end{enumerate}
The influence of each of the above effects separately and all of them applied together on
$F_{2}(M)$ is shown in Fig.~\ref{fig:model-biases} for
an example model parameters, and fine and coarse subdivisions.

\begin{figure}[!ht]
    \centering
    \includegraphics[width=.4\textwidth]{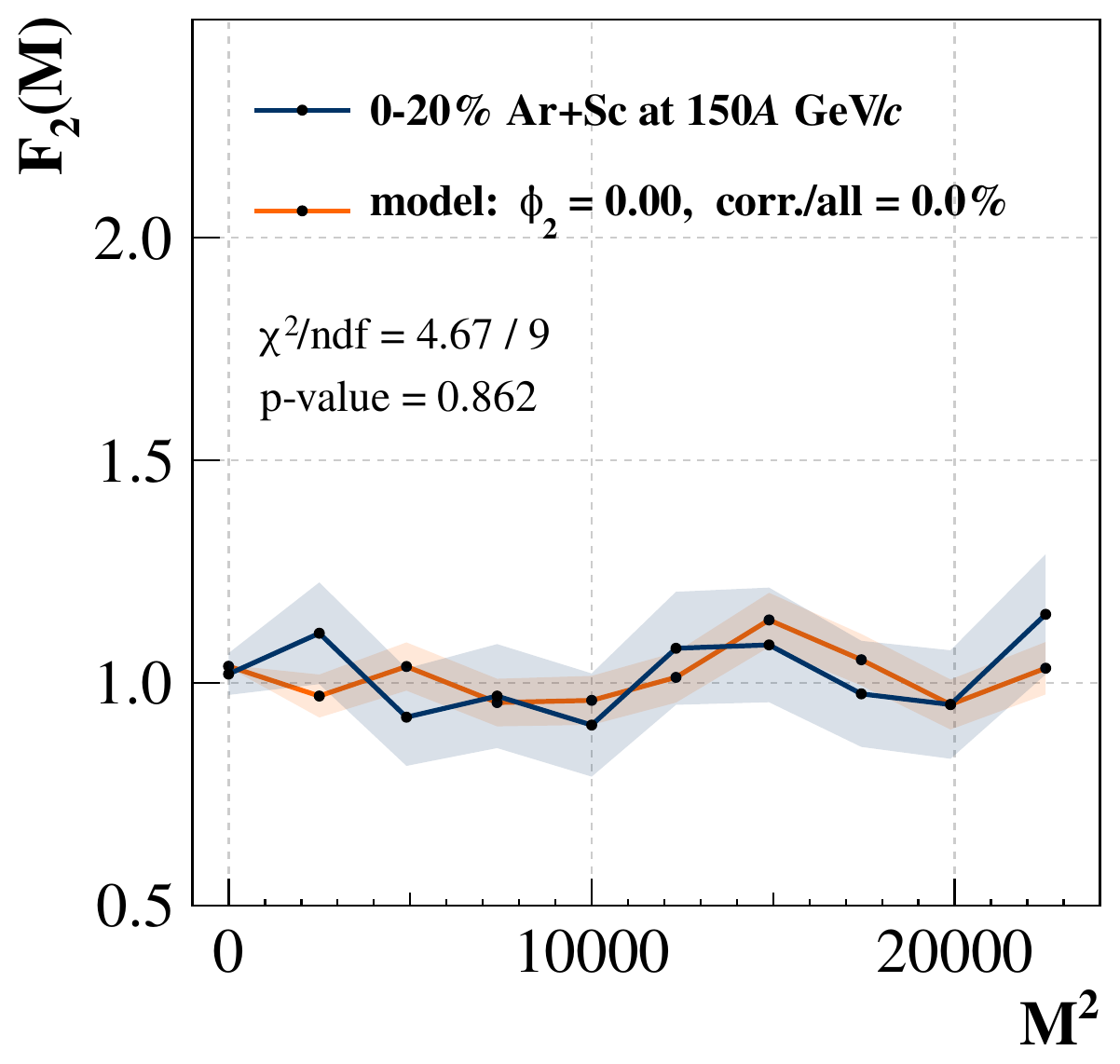}\qquad
    \includegraphics[width=.4\textwidth]{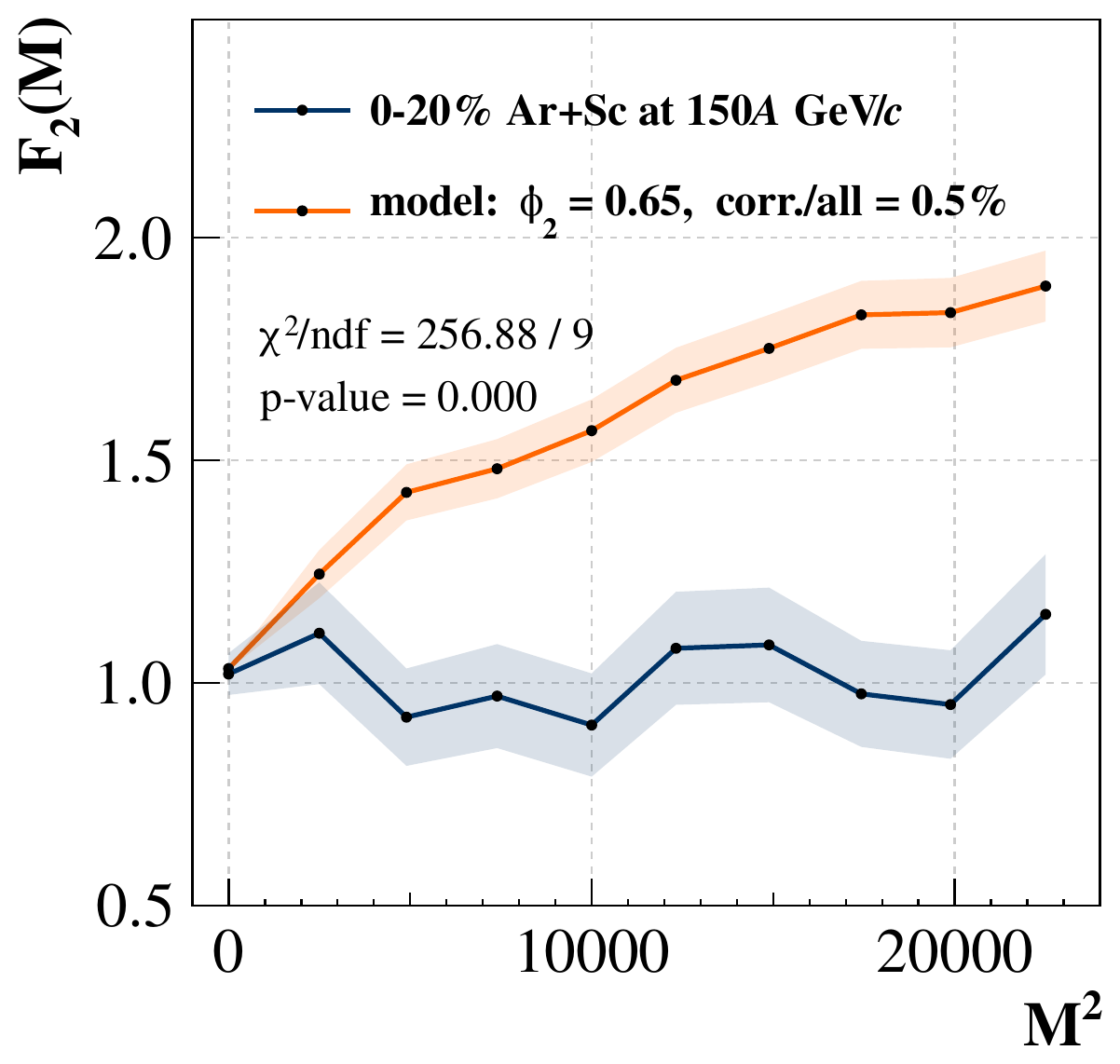}
    \caption{
        Examples of comparison of results for two Power-law Model data sets with the experimental data.
        The left panel includes model predictions assuming only
        uncorrelated protons, whereas the right one shows predictions
        for 0.5\% of correlated protons with power-law exponent $\phi_{2} = 0.65$.
    }
    \label{fig:model-results}
\end{figure}

Next, all generated data sets with all the above effects included have been analyzed the same
way as the experimental data. Obtained $F_{2}(M)$ results have been compared
with the corresponding experimental results and
$\chi^{2}$ and a p-value were calculated. For the calculation, statistical uncertainties from the
model with similar statistics to the data were used.
Examples of such comparison are presented in Fig.~\ref{fig:model-results}.

\begin{figure}
    \centering
    \includegraphics[width=.45\textwidth]{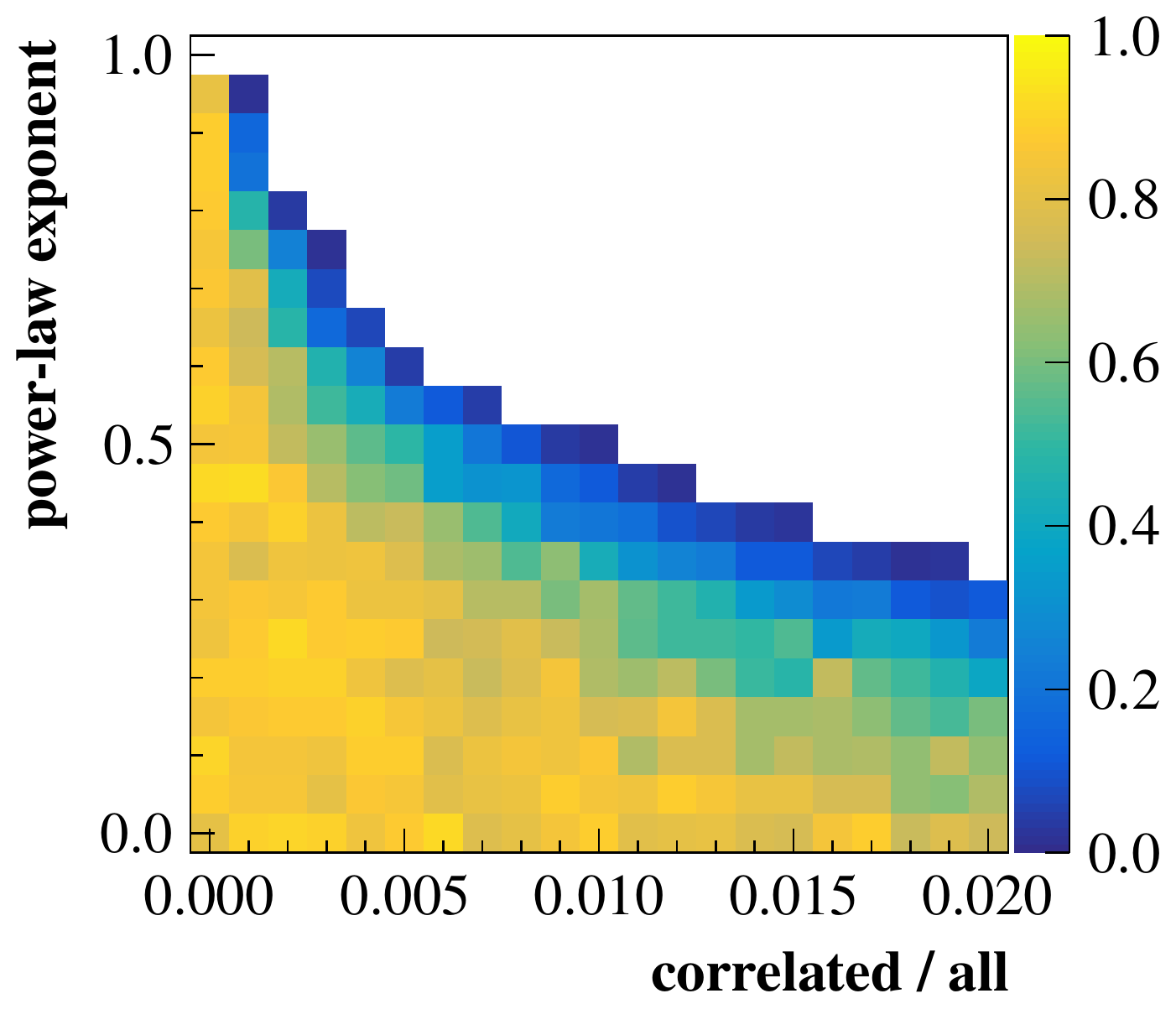}\hfill
    \caption{
        Exclusion plot, the p-values, for the Power-law Model parameters -- the fraction of correlated
        protons and the power-law exponent.
        The white areas correspond to p-values less than 1\%.
        The exclusion plot for the fine subdivisions.
    }
    \label{fig:exclusion-plot}
\end{figure}

Figure~\ref{fig:exclusion-plot} shows obtained p-values as a function of
the fraction of correlated protons and the power-law exponent.
White areas correspond to a p-value of less than 1\% and may be considered excluded (for this
particular model). Results for the coarse subdivision have low statistical uncertainties, thus
small deviations from the behavior expected for uncorrelated particle production due to non-critical
correlations (conservation laws, resonance decays, quantum statistics, ...), as well as possible
experimental biases may lead to significant decrease of the p-values.

The intermittency index $\phi_{2}$ for an infinite system at the QCD critical point is expected
to be $\phi_{2}$ = 5/6, assuming that the latter belongs to the 3-D Ising universality class.
If this value is set as the power-law exponent of the Power-law Model with coarse subdivisions
(Fig.~\ref{fig:exclusion-plot}), the \NASixtyOne data on central \ArSc collisions at
150\AGeVc exclude fractions of correlated protons larger than about 0.1\%.

\section{Summary}
\label{sec:summary}

This paper reports
on the search for the critical point of strongly interacting matter in central \ArSc collisions at
150\AGeVc. Results on second-order scaled factorial moments of proton multiplicity distribution at
mid-rapidity are presented.
Protons produced in strong and electromagnetic processes in \ArSc interactions
and selected by the single- and two-particle acceptance maps, as well as the identification cuts are
used.

The scaled factorial moments are shown as a function of the number of subdivisions of
transverse momentum space -- the so-called intermittency analysis.
The analysis was performed for cumulative and non-cumulative transverse momentum components.
Independent data sets were used to calculate results for each subdivision.
Influence of several experimental effects was discussed and quantified.
The results show no significant intermittency signal.

The experimental data are consistent with the mixed events and the \Epos model
predictions. An upper limit on the fraction of critical proton pairs and the power of the correlation
function was obtained based on a comparison with the Power-law Model.

The intermittency analysis of other reactions recorded within the \NASixtyOne
program on strong interactions is well advanced and the new final results should
be expected soon.

\clearpage

\section*{Acknowledgements}
We would like to thank the CERN EP, BE, HSE and EN Departments for the
strong support of NA61/SHINE.

This work was supported by
the Hungarian Scientific Research Fund (grant NKFIH 138136\slash138152),
the Polish Ministry of Science and Higher Education
(DIR\slash WK\slash\-2016\slash 2017\slash\-10-1, WUT ID-UB), the National Science Centre Poland (grants
2014\slash 14\slash E\slash ST2\slash 00018, %AR, settled
2016\slash 21\slash D\slash ST2\slash 01983, %MMP, settled
2017\slash 25\slash N\slash ST2\slash 02575, %AT, settled
2018\slash 29\slash N\slash ST2\slash 02595, %AM, completed, not settled
2018\slash 30\slash A\slash ST2\slash 00226, %MG, in progress
2018\slash 31\slash G\slash ST2\slash 03910, %SK, in progress
2019\slash 33\slash B\slash ST9\slash 03059, %DB, LT (astro), in progress
2020\slash 39\slash O\slash ST2\slash 00277), %MR, in progress
the Norwegian Financial Mechanism 2014--2021 (grant 2019\slash 34\slash H\slash ST2\slash 00585),
the Polish Minister of Education and Science (contract No. 2021\slash WK\slash 10),
%the Russian Science Foundation (grant 17-72-20045),
%the Russian Academy of Science and the
%Russian Foundation for Basic Research (grants 08-02-00018, 09-02-00664 and 12-02-91503-CERN),
%the Russian Foundation for Basic Research (RFBR) funding within the research project no. 18-02-40086,
%the Ministry of Science and Higher Education of the Russian Federation, Project "Fundamental properties of elementary particles and cosmology" No 0723-2020-0041,
the European Union's Horizon 2020 research and innovation programme under grant agreement No. 871072,
the Ministry of Education, Culture, Sports,
Science and Tech\-no\-lo\-gy, Japan, Grant-in-Aid for Sci\-en\-ti\-fic
Research (grants 18071005, 19034011, 19740162, 20740160 and 20039012),
the German Research Foundation DFG (grants GA\,1480\slash8-1 and project 426579465),
the Bulgarian Ministry of Education and Science within the National
Roadmap for Research Infrastructures 2020--2027, contract No. D01-374/18.12.2020,
Ministry of Education
and Science of the Republic of Serbia (grant OI171002), Swiss
Nationalfonds Foundation (grant 200020\-117913/1), ETH Research Grant
TH-01\,07-3 and the Fermi National Accelerator Laboratory (Fermilab), a U.S. Department of Energy, Office of Science, HEP User Facility managed by Fermi Research Alliance, LLC (FRA), acting under Contract No. DE-AC02-07CH11359 and the IN2P3-CNRS (France).\\

The data used in this paper were collected before February 2022.

\clearpage
\bibliographystyle{include/na61Utphys}
\bibliography{include/na61References.bib}

\clearpage
{\Large The \NASixtyOne Collaboration}
\bigskip
\begin{sloppypar}
  % based on XML DB with time Fri Apr 28 12:49:46 2023
% Authors in alphabetical order.

\noindent
\mbox{H.\;Adhikary\href{https://orcid.org/0000-0002-5746-1268}{\includegraphics[height=1.7ex]{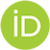}}\textsuperscript{\,14}},
\mbox{P.\;Adrich\href{https://orcid.org/0000-0002-7019-5451}{\includegraphics[height=1.7ex]{figures/orcid-logo.png}}\textsuperscript{\,16}},
\mbox{K.K.\;Allison\href{https://orcid.org/0000-0002-3494-9383}{\includegraphics[height=1.7ex]{figures/orcid-logo.png}}\textsuperscript{\,27}},
\mbox{N.\;Amin\href{https://orcid.org/0009-0004-7572-3817}{\includegraphics[height=1.7ex]{figures/orcid-logo.png}}\textsuperscript{\,5}},
\mbox{E.V.\;Andronov\href{https://orcid.org/0000-0003-0437-9292}{\includegraphics[height=1.7ex]{figures/orcid-logo.png}}\textsuperscript{\,23}},
\mbox{T.\;Anti\'ci\'c\href{https://orcid.org/0000-0002-6606-0191}{\includegraphics[height=1.7ex]{figures/orcid-logo.png}}\textsuperscript{\,3}},
\mbox{I.-C.\;Arsene\href{https://orcid.org/0000-0003-2316-9565}{\includegraphics[height=1.7ex]{figures/orcid-logo.png}}\textsuperscript{\,13}},
\mbox{M.\;Bajda\href{https://orcid.org/0009-0005-8859-1099}{\includegraphics[height=1.7ex]{figures/orcid-logo.png}}\textsuperscript{\,17}},
\mbox{Y.\;Balkova\href{https://orcid.org/0000-0002-6957-573X}{\includegraphics[height=1.7ex]{figures/orcid-logo.png}}\textsuperscript{\,19}},
\mbox{M.\;Baszczyk\href{https://orcid.org/0000-0002-2595-0104}{\includegraphics[height=1.7ex]{figures/orcid-logo.png}}\textsuperscript{\,18}},
\mbox{D.\;Battaglia\href{https://orcid.org/0000-0002-5283-0992}{\includegraphics[height=1.7ex]{figures/orcid-logo.png}}\textsuperscript{\,26}},
\mbox{A.\;Bazgir\href{https://orcid.org/0000-0003-0358-0576}{\includegraphics[height=1.7ex]{figures/orcid-logo.png}}\textsuperscript{\,14}},
\mbox{S.\;Bhosale\href{https://orcid.org/0000-0001-5709-4747}{\includegraphics[height=1.7ex]{figures/orcid-logo.png}}\textsuperscript{\,15}},
\mbox{M.\;Bielewicz\href{https://orcid.org/0000-0001-8267-4874}{\includegraphics[height=1.7ex]{figures/orcid-logo.png}}\textsuperscript{\,16}},
\mbox{A.\;Blondel\href{https://orcid.org/0000-0002-1597-8859}{\includegraphics[height=1.7ex]{figures/orcid-logo.png}}\textsuperscript{\,4}},
\mbox{M.\;Bogomilov\href{https://orcid.org/0000-0001-7738-2041}{\includegraphics[height=1.7ex]{figures/orcid-logo.png}}\textsuperscript{\,2}},
\mbox{Y.\;Bondar\href{https://orcid.org/0000-0003-2773-9668}{\includegraphics[height=1.7ex]{figures/orcid-logo.png}}\textsuperscript{\,14}},
\mbox{N.\;Bostan\href{https://orcid.org/0000-0002-1129-4345}{\includegraphics[height=1.7ex]{figures/orcid-logo.png}}\textsuperscript{\,26}},
\mbox{A.\;Brandin\textsuperscript{\,23}},
\mbox{W.\;Bryli\'nski\href{https://orcid.org/0000-0002-3457-6601}{\includegraphics[height=1.7ex]{figures/orcid-logo.png}}\textsuperscript{\,22}},
\mbox{J.\;Brzychczyk\href{https://orcid.org/0000-0001-5320-6748}{\includegraphics[height=1.7ex]{figures/orcid-logo.png}}\textsuperscript{\,17}},
\mbox{M.\;Buryakov\href{https://orcid.org/0009-0008-2394-4967}{\includegraphics[height=1.7ex]{figures/orcid-logo.png}}\textsuperscript{\,23}},
\mbox{A.F.\;Camino\textsuperscript{\,29}},
\mbox{P.\;Christakoglou\href{https://orcid.org/0000-0002-4325-0646}{\includegraphics[height=1.7ex]{figures/orcid-logo.png}}\textsuperscript{\,7}},
\mbox{M.\;\'Cirkovi\'c\href{https://orcid.org/0000-0002-4420-9688}{\includegraphics[height=1.7ex]{figures/orcid-logo.png}}\textsuperscript{\,24}},
\mbox{M.\;Csan\'ad\href{https://orcid.org/0000-0002-3154-6925}{\includegraphics[height=1.7ex]{figures/orcid-logo.png}}\textsuperscript{\,9}},
\mbox{J.\;Cybowska\href{https://orcid.org/0000-0003-2568-3664}{\includegraphics[height=1.7ex]{figures/orcid-logo.png}}\textsuperscript{\,22}},
\mbox{T.\;Czopowicz\href{https://orcid.org/0000-0003-1908-2977}{\includegraphics[height=1.7ex]{figures/orcid-logo.png}}\textsuperscript{\,14}},
\mbox{C.\;Dalmazzone\href{https://orcid.org/0000-0001-6945-5845}{\includegraphics[height=1.7ex]{figures/orcid-logo.png}}\textsuperscript{\,4}},
\mbox{N.\;Davis\href{https://orcid.org/0000-0003-3047-6854}{\includegraphics[height=1.7ex]{figures/orcid-logo.png}}\textsuperscript{\,15}},
\mbox{F.\;Diakonos\href{https://orcid.org/0000-0003-0142-9098}{\includegraphics[height=1.7ex]{figures/orcid-logo.png}}\textsuperscript{\,7}},
\mbox{A.\;Dmitriev\href{https://orcid.org/0000-0001-7853-0173}{\includegraphics[height=1.7ex]{figures/orcid-logo.png}}\textsuperscript{\,23}},
\mbox{P.~von\;Doetinchem\href{https://orcid.org/0000-0002-7801-3376}{\includegraphics[height=1.7ex]{figures/orcid-logo.png}}\textsuperscript{\,28}},
\mbox{W.\;Dominik\href{https://orcid.org/0000-0001-7444-9239}{\includegraphics[height=1.7ex]{figures/orcid-logo.png}}\textsuperscript{\,20}},
\mbox{P.\;Dorosz\href{https://orcid.org/0000-0002-8884-0981}{\includegraphics[height=1.7ex]{figures/orcid-logo.png}}\textsuperscript{\,18}},
\mbox{J.\;Dumarchez\href{https://orcid.org/0000-0002-9243-4425}{\includegraphics[height=1.7ex]{figures/orcid-logo.png}}\textsuperscript{\,4}},
\mbox{R.\;Engel\href{https://orcid.org/0000-0003-2924-8889}{\includegraphics[height=1.7ex]{figures/orcid-logo.png}}\textsuperscript{\,5}},
\mbox{G.A.\;Feofilov\href{https://orcid.org/0000-0003-3700-8623}{\includegraphics[height=1.7ex]{figures/orcid-logo.png}}\textsuperscript{\,23}},
\mbox{L.\;Fields\href{https://orcid.org/0000-0001-8281-3686}{\includegraphics[height=1.7ex]{figures/orcid-logo.png}}\textsuperscript{\,26}},
\mbox{Z.\;Fodor\href{https://orcid.org/0000-0003-2519-5687}{\includegraphics[height=1.7ex]{figures/orcid-logo.png}}\textsuperscript{\,8,21}},
\mbox{M.\;Friend\href{https://orcid.org/0000-0003-4660-4670}{\includegraphics[height=1.7ex]{figures/orcid-logo.png}}\textsuperscript{\,10}},
\mbox{M.\;Ga\'zdzicki\href{https://orcid.org/0000-0002-6114-8223}{\includegraphics[height=1.7ex]{figures/orcid-logo.png}}\textsuperscript{\,14,6}},
\mbox{O.\;Golosov\href{https://orcid.org/0000-0001-6562-2925}{\includegraphics[height=1.7ex]{figures/orcid-logo.png}}\textsuperscript{\,23}},
\mbox{V.\;Golovatyuk\href{https://orcid.org/0009-0006-5201-0990}{\includegraphics[height=1.7ex]{figures/orcid-logo.png}}\textsuperscript{\,23}},
\mbox{M.\;Golubeva\href{https://orcid.org/0009-0003-4756-2449}{\includegraphics[height=1.7ex]{figures/orcid-logo.png}}\textsuperscript{\,23}},
\mbox{K.\;Grebieszkow\href{https://orcid.org/0000-0002-6754-9554}{\includegraphics[height=1.7ex]{figures/orcid-logo.png}}\textsuperscript{\,22}},
\mbox{F.\;Guber\href{https://orcid.org/0000-0001-8790-3218}{\includegraphics[height=1.7ex]{figures/orcid-logo.png}}\textsuperscript{\,23}},
\mbox{S.N.\;Igolkin\textsuperscript{\,23}},
\mbox{S.\;Ilieva\href{https://orcid.org/0000-0001-9204-2563}{\includegraphics[height=1.7ex]{figures/orcid-logo.png}}\textsuperscript{\,2}},
\mbox{A.\;Ivashkin\href{https://orcid.org/0000-0003-4595-5866}{\includegraphics[height=1.7ex]{figures/orcid-logo.png}}\textsuperscript{\,23}},
\mbox{A.\;Izvestnyy\href{https://orcid.org/0009-0009-1305-7309}{\includegraphics[height=1.7ex]{figures/orcid-logo.png}}\textsuperscript{\,23}},
\mbox{K.\;Kadija\textsuperscript{\,3}},
\mbox{A.\;Kapoyannis\href{https://orcid.org/0000-0002-7732-8552}{\includegraphics[height=1.7ex]{figures/orcid-logo.png}}\textsuperscript{\,7}},
\mbox{N.\;Kargin\textsuperscript{\,23}},
\mbox{N.\;Karpushkin\href{https://orcid.org/0000-0001-5513-9331}{\includegraphics[height=1.7ex]{figures/orcid-logo.png}}\textsuperscript{\,23}},
\mbox{E.\;Kashirin\href{https://orcid.org/0000-0001-6062-7997}{\includegraphics[height=1.7ex]{figures/orcid-logo.png}}\textsuperscript{\,23}},
\mbox{M.\;Kie{\l}bowicz\href{https://orcid.org/0000-0002-4403-9201}{\includegraphics[height=1.7ex]{figures/orcid-logo.png}}\textsuperscript{\,15}},
\mbox{V.A.\;Kireyeu\href{https://orcid.org/0000-0002-5630-9264}{\includegraphics[height=1.7ex]{figures/orcid-logo.png}}\textsuperscript{\,23}},
\mbox{H.\;Kitagawa\textsuperscript{\,11}},
\mbox{R.\;Kolesnikov\href{https://orcid.org/0009-0006-4224-1058}{\includegraphics[height=1.7ex]{figures/orcid-logo.png}}\textsuperscript{\,23}},
\mbox{D.\;Kolev\href{https://orcid.org/0000-0002-9203-4739}{\includegraphics[height=1.7ex]{figures/orcid-logo.png}}\textsuperscript{\,2}},
\mbox{Y.\;Koshio\textsuperscript{\,11}},
\mbox{V.N.\;Kovalenko\href{https://orcid.org/0000-0001-6012-6615}{\includegraphics[height=1.7ex]{figures/orcid-logo.png}}\textsuperscript{\,23}},
\mbox{S.\;Kowalski\href{https://orcid.org/0000-0001-9888-4008}{\includegraphics[height=1.7ex]{figures/orcid-logo.png}}\textsuperscript{\,19}},
\mbox{B.\;Koz{\l}owski\href{https://orcid.org/0000-0001-8442-2320}{\includegraphics[height=1.7ex]{figures/orcid-logo.png}}\textsuperscript{\,22}},
\mbox{A.\;Krasnoperov\href{https://orcid.org/0000-0002-1425-2861}{\includegraphics[height=1.7ex]{figures/orcid-logo.png}}\textsuperscript{\,23}},
\mbox{W.\;Kucewicz\href{https://orcid.org/0000-0002-2073-711X}{\includegraphics[height=1.7ex]{figures/orcid-logo.png}}\textsuperscript{\,18}},
\mbox{M.\;Kuchowicz\href{https://orcid.org/0000-0003-3174-585X}{\includegraphics[height=1.7ex]{figures/orcid-logo.png}}\textsuperscript{\,21}},
\mbox{M.\;Kuich\href{https://orcid.org/0000-0002-6507-8699}{\includegraphics[height=1.7ex]{figures/orcid-logo.png}}\textsuperscript{\,20}},
\mbox{A.\;Kurepin\href{https://orcid.org/0000-0002-1851-4136}{\includegraphics[height=1.7ex]{figures/orcid-logo.png}}\textsuperscript{\,23}},
\mbox{A.\;L\'aszl\'o\href{https://orcid.org/0000-0003-2712-6968}{\includegraphics[height=1.7ex]{figures/orcid-logo.png}}\textsuperscript{\,8}},
\mbox{M.\;Lewicki\href{https://orcid.org/0000-0002-8972-3066}{\includegraphics[height=1.7ex]{figures/orcid-logo.png}}\textsuperscript{\,21}},
\mbox{G.\;Lykasov\href{https://orcid.org/0000-0002-1544-6959}{\includegraphics[height=1.7ex]{figures/orcid-logo.png}}\textsuperscript{\,23}},
\mbox{V.V.\;Lyubushkin\href{https://orcid.org/0000-0003-0136-233X}{\includegraphics[height=1.7ex]{figures/orcid-logo.png}}\textsuperscript{\,23}},
\mbox{M.\;Ma\'ckowiak-Paw{\l}owska\href{https://orcid.org/0000-0003-3954-6329}{\includegraphics[height=1.7ex]{figures/orcid-logo.png}}\textsuperscript{\,22}},
\mbox{Z.\;Majka\href{https://orcid.org/0000-0003-3064-6577}{\includegraphics[height=1.7ex]{figures/orcid-logo.png}}\textsuperscript{\,17}},
\mbox{A.\;Makhnev\href{https://orcid.org/0009-0002-9745-1897}{\includegraphics[height=1.7ex]{figures/orcid-logo.png}}\textsuperscript{\,23}},
\mbox{B.\;Maksiak\href{https://orcid.org/0000-0002-7950-2307}{\includegraphics[height=1.7ex]{figures/orcid-logo.png}}\textsuperscript{\,16}},
\mbox{A.I.\;Malakhov\href{https://orcid.org/0000-0001-8569-8409}{\includegraphics[height=1.7ex]{figures/orcid-logo.png}}\textsuperscript{\,23}},
\mbox{A.\;Marcinek\href{https://orcid.org/0000-0001-9922-743X}{\includegraphics[height=1.7ex]{figures/orcid-logo.png}}\textsuperscript{\,15}},
\mbox{A.D.\;Marino\href{https://orcid.org/0000-0002-1709-538X}{\includegraphics[height=1.7ex]{figures/orcid-logo.png}}\textsuperscript{\,27}},
\mbox{H.-J.\;Mathes\href{https://orcid.org/0000-0002-0680-040X}{\includegraphics[height=1.7ex]{figures/orcid-logo.png}}\textsuperscript{\,5}},
\mbox{T.\;Matulewicz\href{https://orcid.org/0000-0003-2098-1216}{\includegraphics[height=1.7ex]{figures/orcid-logo.png}}\textsuperscript{\,20}},
\mbox{V.\;Matveev\href{https://orcid.org/0000-0002-2745-5908}{\includegraphics[height=1.7ex]{figures/orcid-logo.png}}\textsuperscript{\,23}},
\mbox{G.L.\;Melkumov\href{https://orcid.org/0009-0004-2074-6755}{\includegraphics[height=1.7ex]{figures/orcid-logo.png}}\textsuperscript{\,23}},
\mbox{A.\;Merzlaya\href{https://orcid.org/0000-0002-6553-2783}{\includegraphics[height=1.7ex]{figures/orcid-logo.png}}\textsuperscript{\,13}},
\mbox{{\L}.\;Mik\href{https://orcid.org/0000-0003-2712-6861}{\includegraphics[height=1.7ex]{figures/orcid-logo.png}}\textsuperscript{\,18}},
\mbox{A.\;Morawiec\href{https://orcid.org/0009-0001-9845-4005}{\includegraphics[height=1.7ex]{figures/orcid-logo.png}}\textsuperscript{\,17}},
\mbox{S.\;Morozov\href{https://orcid.org/0000-0002-6748-7277}{\includegraphics[height=1.7ex]{figures/orcid-logo.png}}\textsuperscript{\,23}},
\mbox{Y.\;Nagai\href{https://orcid.org/0000-0002-1792-5005}{\includegraphics[height=1.7ex]{figures/orcid-logo.png}}\textsuperscript{\,9}},
\mbox{T.\;Nakadaira\href{https://orcid.org/0000-0003-4327-7598}{\includegraphics[height=1.7ex]{figures/orcid-logo.png}}\textsuperscript{\,10}},
\mbox{M.\;Naskr\k{e}t\href{https://orcid.org/0000-0002-5634-6639}{\includegraphics[height=1.7ex]{figures/orcid-logo.png}}\textsuperscript{\,21}},
\mbox{S.\;Nishimori\href{https://orcid.org/~0000-0002-1820-0938}{\includegraphics[height=1.7ex]{figures/orcid-logo.png}}\textsuperscript{\,10}},
\mbox{V.\;Ozvenchuk\href{https://orcid.org/0000-0002-7821-7109}{\includegraphics[height=1.7ex]{figures/orcid-logo.png}}\textsuperscript{\,15}},
\mbox{A.D.\;Panagiotou\textsuperscript{\,7}},
\mbox{O.\;Panova\href{https://orcid.org/0000-0001-5039-7788}{\includegraphics[height=1.7ex]{figures/orcid-logo.png}}\textsuperscript{\,14}},
\mbox{V.\;Paolone\href{https://orcid.org/0000-0003-2162-0957}{\includegraphics[height=1.7ex]{figures/orcid-logo.png}}\textsuperscript{\,29}},
\mbox{O.\;Petukhov\href{https://orcid.org/0000-0002-8872-8324}{\includegraphics[height=1.7ex]{figures/orcid-logo.png}}\textsuperscript{\,23}},
\mbox{I.\;Pidhurskyi\href{https://orcid.org/0000-0001-9916-9436}{\includegraphics[height=1.7ex]{figures/orcid-logo.png}}\textsuperscript{\,14,6}},
\mbox{R.\;P{\l}aneta\href{https://orcid.org/0000-0001-8007-8577}{\includegraphics[height=1.7ex]{figures/orcid-logo.png}}\textsuperscript{\,17}},
\mbox{P.\;Podlaski\href{https://orcid.org/0000-0002-0232-9841}{\includegraphics[height=1.7ex]{figures/orcid-logo.png}}\textsuperscript{\,20}},
\mbox{B.A.\;Popov\href{https://orcid.org/0000-0001-5416-9301}{\includegraphics[height=1.7ex]{figures/orcid-logo.png}}\textsuperscript{\,23,4}},
\mbox{B.\;P\'orfy\href{https://orcid.org/0000-0001-5724-9737}{\includegraphics[height=1.7ex]{figures/orcid-logo.png}}\textsuperscript{\,8,9}},
\mbox{M.\;Posiada{\l}a-Zezula\href{https://orcid.org/0000-0002-5154-5348}{\includegraphics[height=1.7ex]{figures/orcid-logo.png}}\textsuperscript{\,20}},
\mbox{D.S.\;Prokhorova\href{https://orcid.org/0000-0003-3726-9196}{\includegraphics[height=1.7ex]{figures/orcid-logo.png}}\textsuperscript{\,23}},
\mbox{D.\;Pszczel\href{https://orcid.org/0000-0002-4697-6688}{\includegraphics[height=1.7ex]{figures/orcid-logo.png}}\textsuperscript{\,16}},
\mbox{S.\;Pu{\l}awski\href{https://orcid.org/0000-0003-1982-2787}{\includegraphics[height=1.7ex]{figures/orcid-logo.png}}\textsuperscript{\,19}},
\mbox{J.\;Puzovi\'c\textsuperscript{\,24}\textsuperscript{\dag}},
\mbox{R.\;Renfordt\href{https://orcid.org/0000-0002-5633-104X}{\includegraphics[height=1.7ex]{figures/orcid-logo.png}}\textsuperscript{\,19}},
\mbox{L.\;Ren\href{https://orcid.org/0000-0003-1709-7673}{\includegraphics[height=1.7ex]{figures/orcid-logo.png}}\textsuperscript{\,27}},
\mbox{V.Z.\;Reyna~Ortiz\href{https://orcid.org/0000-0002-7026-8198}{\includegraphics[height=1.7ex]{figures/orcid-logo.png}}\textsuperscript{\,14}},
\mbox{D.\;R\"ohrich\textsuperscript{\,12}},
\mbox{E.\;Rondio\href{https://orcid.org/0000-0002-2607-4820}{\includegraphics[height=1.7ex]{figures/orcid-logo.png}}\textsuperscript{\,16}},
\mbox{M.\;Roth\href{https://orcid.org/0000-0003-1281-4477}{\includegraphics[height=1.7ex]{figures/orcid-logo.png}}\textsuperscript{\,5}},
\mbox{{\L}.\;Rozp{\l}ochowski\href{https://orcid.org/0000-0003-3680-6738}{\includegraphics[height=1.7ex]{figures/orcid-logo.png}}\textsuperscript{\,15}},
\mbox{B.T.\;Rumberger\href{https://orcid.org/0000-0002-4867-945X}{\includegraphics[height=1.7ex]{figures/orcid-logo.png}}\textsuperscript{\,27}},
\mbox{M.\;Rumyantsev\href{https://orcid.org/0000-0001-8233-2030}{\includegraphics[height=1.7ex]{figures/orcid-logo.png}}\textsuperscript{\,23}},
\mbox{A.\;Rustamov\href{https://orcid.org/0000-0001-8678-6400}{\includegraphics[height=1.7ex]{figures/orcid-logo.png}}\textsuperscript{\,1,6}},
\mbox{M.\;Rybczynski\href{https://orcid.org/0000-0002-3638-3766}{\includegraphics[height=1.7ex]{figures/orcid-logo.png}}\textsuperscript{\,14}},
\mbox{A.\;Rybicki\href{https://orcid.org/0000-0003-3076-0505}{\includegraphics[height=1.7ex]{figures/orcid-logo.png}}\textsuperscript{\,15}},
\mbox{K.\;Sakashita\href{https://orcid.org/0000-0003-2602-7837}{\includegraphics[height=1.7ex]{figures/orcid-logo.png}}\textsuperscript{\,10}},
\mbox{K.\;Schmidt\href{https://orcid.org/0000-0002-0903-5790}{\includegraphics[height=1.7ex]{figures/orcid-logo.png}}\textsuperscript{\,19}},
\mbox{A.Yu.\;Seryakov\href{https://orcid.org/0000-0002-5759-5485}{\includegraphics[height=1.7ex]{figures/orcid-logo.png}}\textsuperscript{\,23}},
\mbox{P.\;Seyboth\href{https://orcid.org/0000-0002-4821-6105}{\includegraphics[height=1.7ex]{figures/orcid-logo.png}}\textsuperscript{\,14}},
\mbox{U.A.\;Shah\href{https://orcid.org/0000-0002-9315-1304}{\includegraphics[height=1.7ex]{figures/orcid-logo.png}}\textsuperscript{\,14}},
\mbox{Y.\;Shiraishi\textsuperscript{\,11}},
\mbox{A.\;Shukla\href{https://orcid.org/0000-0003-3839-7229}{\includegraphics[height=1.7ex]{figures/orcid-logo.png}}\textsuperscript{\,28}},
\mbox{M.\;S{\l}odkowski\href{https://orcid.org/0000-0003-0463-2753}{\includegraphics[height=1.7ex]{figures/orcid-logo.png}}\textsuperscript{\,22}},
\mbox{P.\;Staszel\href{https://orcid.org/0000-0003-4002-1626}{\includegraphics[height=1.7ex]{figures/orcid-logo.png}}\textsuperscript{\,17}},
\mbox{G.\;Stefanek\href{https://orcid.org/0000-0001-6656-9177}{\includegraphics[height=1.7ex]{figures/orcid-logo.png}}\textsuperscript{\,14}},
\mbox{J.\;Stepaniak\href{https://orcid.org/0000-0003-2064-9870}{\includegraphics[height=1.7ex]{figures/orcid-logo.png}}\textsuperscript{\,16}},
\mbox{M.\;Strikhanov\textsuperscript{\,23}},
\mbox{H.\;Str\"obele\textsuperscript{\,6}},
\mbox{T.\;\v{S}u\v{s}a\href{https://orcid.org/0000-0001-7430-2552}{\includegraphics[height=1.7ex]{figures/orcid-logo.png}}\textsuperscript{\,3}},
\mbox{L.\;Swiderski\href{https://orcid.org/0000-0001-5857-2085}{\includegraphics[height=1.7ex]{figures/orcid-logo.png}}\textsuperscript{\,16}},
\mbox{J.\;Szewi\'nski\href{https://orcid.org/0000-0003-2981-9303}{\includegraphics[height=1.7ex]{figures/orcid-logo.png}}\textsuperscript{\,16}},
\mbox{R.\;Szukiewicz\href{https://orcid.org/0000-0002-1291-4040}{\includegraphics[height=1.7ex]{figures/orcid-logo.png}}\textsuperscript{\,21}},
\mbox{A.\;Taranenko\href{https://orcid.org/0000-0003-1737-4474}{\includegraphics[height=1.7ex]{figures/orcid-logo.png}}\textsuperscript{\,23}},
\mbox{A.\;Tefelska\href{https://orcid.org/0000-0002-6069-4273}{\includegraphics[height=1.7ex]{figures/orcid-logo.png}}\textsuperscript{\,22}},
\mbox{D.\;Tefelski\href{https://orcid.org/0000-0003-0802-2290}{\includegraphics[height=1.7ex]{figures/orcid-logo.png}}\textsuperscript{\,22}},
\mbox{V.\;Tereshchenko\textsuperscript{\,23}},
\mbox{A.\;Toia\href{https://orcid.org/0000-0001-9567-3360}{\includegraphics[height=1.7ex]{figures/orcid-logo.png}}\textsuperscript{\,6}},
\mbox{R.\;Tsenov\href{https://orcid.org/0000-0002-1330-8640}{\includegraphics[height=1.7ex]{figures/orcid-logo.png}}\textsuperscript{\,2}},
\mbox{L.\;Turko\href{https://orcid.org/0000-0002-5474-8650}{\includegraphics[height=1.7ex]{figures/orcid-logo.png}}\textsuperscript{\,21}},
\mbox{T.S.\;Tveter\href{https://orcid.org/0009-0003-7140-8644}{\includegraphics[height=1.7ex]{figures/orcid-logo.png}}\textsuperscript{\,13}},
\mbox{M.\;Unger\href{https://orcid.org/0000-0002-7651-0272~}{\includegraphics[height=1.7ex]{figures/orcid-logo.png}}\textsuperscript{\,5}},
\mbox{M.\;Urbaniak\href{https://orcid.org/0000-0002-9768-030X}{\includegraphics[height=1.7ex]{figures/orcid-logo.png}}\textsuperscript{\,19}},
\mbox{F.F.\;Valiev\href{https://orcid.org/0000-0001-5130-5603}{\includegraphics[height=1.7ex]{figures/orcid-logo.png}}\textsuperscript{\,23}},
\mbox{M.\;Vassiliou\textsuperscript{\,7}},
\mbox{D.\;Veberi\v{c}\href{https://orcid.org/0000-0003-2683-1526}{\includegraphics[height=1.7ex]{figures/orcid-logo.png}}\textsuperscript{\,5}},
\mbox{V.V.\;Vechernin\href{https://orcid.org/0000-0003-1458-8055}{\includegraphics[height=1.7ex]{figures/orcid-logo.png}}\textsuperscript{\,23}},
\mbox{V.\;Volkov\href{https://orcid.org/0000-0002-4785-7517}{\includegraphics[height=1.7ex]{figures/orcid-logo.png}}\textsuperscript{\,23}},
\mbox{A.\;Wickremasinghe\href{https://orcid.org/0000-0002-5325-0455}{\includegraphics[height=1.7ex]{figures/orcid-logo.png}}\textsuperscript{\,25}},
\mbox{K.\;W\'ojcik\href{https://orcid.org/0000-0002-8315-9281}{\includegraphics[height=1.7ex]{figures/orcid-logo.png}}\textsuperscript{\,19}},
\mbox{O.\;Wyszy\'nski\href{https://orcid.org/0000-0002-6652-0450}{\includegraphics[height=1.7ex]{figures/orcid-logo.png}}\textsuperscript{\,14}},
\mbox{A.\;Zaitsev\href{https://orcid.org/0000-0003-4711-9925}{\includegraphics[height=1.7ex]{figures/orcid-logo.png}}\textsuperscript{\,23}},
\mbox{E.D.\;Zimmerman\href{https://orcid.org/0000-0002-6394-6659}{\includegraphics[height=1.7ex]{figures/orcid-logo.png}}\textsuperscript{\,27}},
\mbox{A.\;Zviagina\href{https://orcid.org/0009-0007-5211-6493}{\includegraphics[height=1.7ex]{figures/orcid-logo.png}}\textsuperscript{\,23}}, and
\mbox{R.\;Zwaska\href{https://orcid.org/0000-0002-4889-5988}{\includegraphics[height=1.7ex]{figures/orcid-logo.png}}\textsuperscript{\,25}}
\\\rule{2cm}{.5pt}\\[-.5ex]\textit{\textsuperscript{\dag} \footnotesize deceased}

\end{sloppypar}
% based on XML DB with time Fri Apr 28 12:49:46 2023
% Institutes in alphabetical order.

\noindent
\textsuperscript{1}~National Nuclear Research Center, Baku, Azerbaijan\\
\textsuperscript{2}~Faculty of Physics, University of Sofia, Sofia, Bulgaria\\
\textsuperscript{3}~Ru{\dj}er Bo\v{s}kovi\'c Institute, Zagreb, Croatia\\
\textsuperscript{4}~LPNHE, University of Paris VI and VII, Paris, France\\
\textsuperscript{5}~Karlsruhe Institute of Technology, Karlsruhe, Germany\\
\textsuperscript{6}~University of Frankfurt, Frankfurt, Germany\\
\textsuperscript{7}~University of Athens, Athens, Greece\\
\textsuperscript{8}~Wigner Research Centre for Physics, Budapest, Hungary\\
\textsuperscript{9}~E\"otv\"os Lor\'and University, Budapest, Hungary\\
\textsuperscript{10}~Institute for Particle and Nuclear Studies, Tsukuba, Japan\\
\textsuperscript{11}~Okayama University, Japan\\
\textsuperscript{12}~University of Bergen, Bergen, Norway\\
\textsuperscript{13}~University of Oslo, Oslo, Norway\\
\textsuperscript{14}~Jan Kochanowski University, Kielce, Poland\\
\textsuperscript{15}~Institute of Nuclear Physics, Polish Academy of Sciences, Cracow, Poland\\
\textsuperscript{16}~National Centre for Nuclear Research, Warsaw, Poland\\
\textsuperscript{17}~Jagiellonian University, Cracow, Poland\\
\textsuperscript{18}~AGH - University of Science and Technology, Cracow, Poland\\
\textsuperscript{19}~University of Silesia, Katowice, Poland\\
\textsuperscript{20}~University of Warsaw, Warsaw, Poland\\
\textsuperscript{21}~University of Wroc{\l}aw,  Wroc{\l}aw, Poland\\
\textsuperscript{22}~Warsaw University of Technology, Warsaw, Poland\\
\textsuperscript{23}~Affiliated with an institution covered by a cooperation agreement with CERN\\
\textsuperscript{24}~University of Belgrade, Belgrade, Serbia\\
\textsuperscript{25}~Fermilab, Batavia, USA\\
\textsuperscript{26}~University of Notre Dame, Notre Dame, USA\\
\textsuperscript{27}~University of Colorado, Boulder, USA\\
\textsuperscript{28}~University of Hawaii at Manoa, Honolulu, USA\\
\textsuperscript{29}~University of Pittsburgh, Pittsburgh, USA\\

\end{document}